\documentclass[reprint,amsmath,longbibliography,amssymb,floatfix,aps,pra]{revtex4-1}
\usepackage{graphics}
\usepackage{graphicx}
\usepackage{amssymb}
\usepackage{amsthm}
\usepackage{amsmath}
\usepackage{bm}
\usepackage{bbm}
\usepackage{xcolor}
\usepackage{comment}
\newcommand{\be}{\begin{equation}}
\newcommand{\ben}{\begin{eqnarray}}
\newcommand{\een}{\end{eqnarray}}
\newcommand{\ee}{\end{equation}}
\newcommand{\ps}{+\!\!}
\newcommand{\ms}{-\!\!}
\newcommand{\daa}{\rho^{(2)}_{\uparrow\uparrow}}
\newcommand{\dab}{\rho^{(2)}_{\uparrow\downarrow}}

\newcommand{\nud}{{\cal N}_{\uparrow\downarrow}}
\newcommand{\iud}{I_{\uparrow\downarrow}}
\newcommand{\nuu}{{\cal N}_{\uparrow\uparrow}}

\usepackage{braket}
\usepackage{multirow}
\usepackage{lipsum}
\begin{document}
\title{\vspace*{-0.5cm} 
Fermionic entanglement and quantum correlation measures in molecules}
\author{J.\ Garcia$^1$, J.A. Cianciulli$^1$, R. Rossignoli$^{1,2}$}
\affiliation{$^1$ Instituto de F\'{\i}sica La Plata, CONICET and 
 Departamento\ de F\'{\i}sica, Universidad Nacional de La Plata,
C.C. 67, La Plata (1900), Argentina\\
$^{2}$ Comisi\'on de Investigaciones Cient\'{\i}ficas (CIC), La Plata (1900), Argentina}
\begin{abstract} 
We analyze fermionic entanglement and correlation measures in the ground and the low temperature thermal state  of the water molecule as a function of the internuclear distance in the context of the full configuration interaction approach. The aim is to obtain a  general entanglement based characterization of the electronic eigenstates.  We consider first the spin-up -- spin-down partition and the associated Schmidt decomposition, 
 examining the total up-down entanglement of the electronic wave function. We then consider the one- and two-body entanglement derived from the one- and two-body reduced density matrices (DMs), which measure the deviation of the state from a Slater Determinant (SD).  All blocks of these DMs are examined.  We also introduce and analyze new measures like the up-down two-body mutual information,  which measures the up-down correlation at the two-body level,  and two types of two-body negativities, the latter detecting and measuring the ``inner'' entanglement of the reduced two-body DMs, i.e., their deviation from a convex mixture of SDs.  Finally, the dissociation limit is also analyzed, considering both the exact ground state (GS) as well as the thermal state in the zero temperature limit, representing the projector onto the ``GS band'' of almost degenerate lowest lying eigenstates. 
\end{abstract}
\maketitle
\vspace*{-1.5cm}

\section{Introduction}

The quantitative characterization of electronic correlations remains one of the most significant challenges in Quantum Chemistry (QC) and condensed matter physics. While traditional energy-based criteria successfully categorize correlations into dynamic and static types \cite{Wigner_1934, L_wdin_1955_b}, the perspective of Quantum Information Theory (QIT) \cite{NielsenBook,HHHH.09} has emerged as a powerful complementary framework, offering rigorous tools to quantify the complexity of many-body wavefunctions \cite{A.08, Aliverti.24}. In particular, the study of entanglement in systems of indistinguishable fermions has garnered significant attention, providing structural insights into many-body states that go beyond standard energetic descriptors \cite{BFFM.20,SC.01,ES.02,Za.02,FL.13,SSG.18,GR.15}.

Unlike systems of distinguishable components, where entanglement is defined via the tensor product structure of the Hilbert space, fermionic systems require a formulation that explicitly respects the antisymmetry principle. In this context, mode-independent fermionic entanglement \cite{SC.01,ES.02,GR.15,GDR.20,GDR.21} can be  rigorously defined as the correlations existing ``beyond antisymmetrization,'' where a single SD  represents the unentangled, ``mean-field''  reference state. Any deviation from this reference manifests as mixedness in the reduced density matrices (RDMs), allowing for a basis-independent quantification of correlation effects \cite{GR.15,GDR.20,GDR.21,Iem.14,MB.16,GR.17,SD.18,DGR.18,DRCG.19}. This formalism has been generalized to higher orders and bosonic systems, establishing a hierarchy of $M$-body entanglement measures based on the spectral properties of $M$-body reduced DMs and associated bipartite representations \cite{GDR.21,CR.24}.

Typical studies of fermionic correlations in quantum chemistry involve the computation of the von Neumann entropy of the 
one-particle RDM \cite{Wang_2007}, and of higher order RDMs \cite{Luzanov_2007}, which were first applied to small molecular systems. 
Other QIT techniques include the 
Shannon entropy of the Configuration Interaction (CI) expansion coefficients \cite{Ivanov_2005, Alcoba_2016}. 
The cumulants of the RDMs (i.e., the parts of higher-order RDMs that do not depend on the lower order ones) have also been used to describe electron correlation \cite{Juh_sz_2006, Alcoba_2010, Li_2021}.
Even so, most of the discussion of electron correlation in QC is still performed in terms of static and dynamic correlations  
\cite{Benavides_Riveros_2017, Izs_k_2023, _ulka_2023, Ganoe_2024}.
A different approach relies on measures of modal fermionic entanglement \cite{Boguslawski_2012, Boguslawski_2013, Boguslawski_2014_a, DS.21, Ding_2022, Ding_2023, Ding_2023_a, Ratini_2024}; in this regard, the quintessential tool is the Mutual Information (MI) of pairs of orbitals. 

Our investigation adopts two complementary perspectives. First, we examine the bipartite entanglement between the spin-up and spin-down electronic sectors. Since electrons with opposite spin projections are distinguishable within the spatial orbital basis, this partition admits a standard Schmidt decomposition \cite{NielsenBook}, providing a global measure of up-down correlations. Secondly, we analyze fermionic entanglement through the one- and two-body RDMs \cite{GDR.20,GDR.21,CR.24}. We explicitly consider the blocked structure of these matrices  arising from spin conservation and evaluate their entropies to quantify the departure from the independent-particle picture.

A central contribution of this study is the introduction and analysis of novel correlation measures tailored to the two-body level, based on the mutual information and negativity \cite{VW.02, P.05}.  We examine the up-down two-body mutual information and introduce two types of two-body negativities. 
The latter are measures of the ``inner'' entanglement of the reduced two-body DMs, vanishing for convex mixtures of SDs and providing a sensitive probe for non-classical correlations in the two-body sector.

Finally, we address the dissociation limit, where the GS becomes degenerate \cite{DS.20}. To provide a physically meaningful characterization, we analyze not only the exact GS but also the thermal state in the zero-temperature limit. The latter captures the manifold of the lowest  closely-lying eigenstates, naturally revealing the local nature of the system in this limit. Additional  features  concerning pertinent entanglement witnesses and results for distinct  basis sets are provided in the Appendices \ref{apA} and \ref{apB}. 

\section{Formalism}
\subsection{Electronic Hamiltonian and eigenstates} \label{sec:hamiltonian_eigenstates}
The Born-Oppenheimer approximation allows us
to approximate the molecular Hamiltonian as the sum of a nuclear Hamiltonian and an electronic one.
Within the second-quantization formalism, the electronic Hamiltonian can be written in terms of fermionic creation and annihilation operators $c^\dagger_i$ and $c_i$, which  satisfy the usual fermionic anticommutation relations, 
\begin{equation}
    \{c_i, c_j\} = \{c^\dagger_i, c^\dagger_j\} = 0, \quad \{c_i, c_j^\dagger\} = \delta_{ij}\,,
\end{equation}
and create or destroy an electron in  a spin-orbital $\ket{\phi_i}$. These single particle (sp) states  form an orthonormal set, $\braket{\phi_i|\phi_j} = \delta_{ij}$.
Although the dimension of the sp space is in principle infinite, for practical purposes a finite set of $d$ sp spin-orbitals are taken  such that computations are feasible.
Typically, these sets involve linear combinations of contracted Gaussian functions centered around the atoms that comprise the molecule.

The electronic Hamiltonian can be written as
\begin{equation}
    H = \sum_{i,j}^{d} h_{ij} c^\dagger_i c_j + \tfrac{1}{2}\sum_{i,j,k,l}^{d} R_{ij,kl} c^\dagger_i c^\dagger_j c_l c_k,
    \label{eq:hamiltonian}
\end{equation}
where $h_{ij} = \braket{\phi_i|h|\phi_j}$, with $h = t + v$ the sp operator containing the kinetic energy operator $t$ plus its Coulombic interaction $v$ with the nuclei, and $R_{ij,kl} \propto\braket{\phi_i\phi_j| r_{12}^{-1} | \phi_k\phi_l}$ are the matrix elements of the  Coulombic interaction between a pair of electrons.  Its eigenstates $\ket{\Psi}$ such that $H\ket{\Psi}=E\ket{\Psi}$, can be written as  linear combinations of SDs,  
\begin{equation}
    \begin{aligned}
        \ket{\Psi} &= \sum_{\bm\gamma} \Gamma_{\bm\gamma} \ket{\bm\gamma},\\
        \ket{\bm\gamma} &= C_{\bm\gamma}^\dagger \ket{0}, \quad C_{\bm\gamma}^\dagger = c_1^{\dagger\,n_1}\ldots c_d^{\dagger\,n_d},
\end{aligned}
    \label{eq:sds}
\end{equation}
where $n_i = 0,1$ are the occupation numbers of each spin-orbital in the SD $\ket{\boldsymbol{\gamma}}$  and $\ket{0}$ is the vacuum ($c_i|0\rangle=0$). 
For a  fixed number $N$ of electrons there is a total of $\binom{d}{N}$ orthogonal  SDs. The eigenenergies $E$  and the $\Gamma_{\bm{\gamma}}$ coefficients are then obtained through diagonalization of the matrix of elements $H_{\boldsymbol{\gamma\gamma^\prime}} = \braket{\boldsymbol{\gamma}|H|\boldsymbol{\gamma^\prime}}$. 
In QC, this procedure is usually referred to as Full Configuration Interaction (FCI) \cite{GSB.25}. 

The  Hamiltonian \eqref{eq:hamiltonian}  commutes not only with the particle number operator $N$, but also with the total spin operator $\bm S$ (as spin-orbit coupling is omitted).  Therefore, its eigenstates  can be  chosen as eigenstates of $S^2=\bm{S}\cdot\bm{S}$ and the total spin component $S_z$ along the $z$ axis. We may  write $S_z=\frac{1}{2}(N_{\uparrow}-N_{\downarrow})=M_S$ and $N=N_{\uparrow}+ N_{\downarrow}$, with $N_{\uparrow}$, $N_{\downarrow}$ the total spin-up and spin-down particle number operators, then satisfying 
 $[H,N_{\uparrow}]=[H,N_{\downarrow}]=0$. 

The eigenstates $|\Psi\rangle$ of $H$ can therefore be all chosen to be eigenstates of both $N_{\uparrow}$ and $N_{\downarrow}$.  The  
$N$-particle creation operator $C^\dagger_{\bm{\gamma}}$ in \eqref{eq:sds} can then be written as a product of $N_\uparrow$- and $N_\downarrow$-particle creation operators 
$C^\dagger_{\bm\alpha},   C^\dagger_{\bar{\bm\beta}}$:
\begin{equation}\ket{\boldsymbol{\gamma}}= C^\dagger_{\bm\alpha} C^\dagger_{\bm{\bar\beta}}|0\rangle = \ket{\bm\alpha\bar{\bm\beta}}
\label{ab0}\,,\end{equation}
where in what follows the bar will
denote spin-down states and its absence spin-up states. This representation 
 allows for a reduction in the total number of SDs required to describe an eigenstate $\ket{\Psi}$, since, for a given $M_S$, only $\binom{d/2}{N_\uparrow}\binom{d/2}{N_\downarrow}$ of them are needed. Notice that 
in spite of having good $M_S$, these SDs are not necessarily eigenstates of the total spin operator $S^2$. 
Nevertheless, since $H$ commutes with $S^2$,  its exact eigenstates  will directly be eigenstates of $S^2$ if non-degenerate.  

The exact eigenstates of $H$ can then be written as  
\begin{subequations}
\label{psig}
\begin{eqnarray}
|\Psi\rangle&=&\sum_{\bm\alpha,{\bm\beta}}\Gamma_{\bm\alpha\bm\beta}\,|\bm\alpha\bar{\bm\beta}\rangle\label{psiga}\\&=&\sum_{\nu=1}^{n_s}\Gamma_\nu\,|\nu\bar\nu\rangle\,,\;\;\;
|\nu\bar\nu\rangle=A^{\dag}_{\nu} B^{\dag}_{\bar\nu}|0\rangle\,,\label{Scd}
\end{eqnarray}
\end{subequations}
where \eqref{Scd}  is the up-down {\it Schmidt decomposition} \cite{NielsenBook} of $|\Psi\rangle$. Here $\Gamma_\nu$ are the singular values of the tensor ${\bf \Gamma}$ of elements $\Gamma_{\bm\alpha\bm\beta}$ (i.e.\ the square root of the nonzero eigenvalues of ${\bf \Gamma}{\bf \Gamma}^\dag$ or equivalently ${\bf \Gamma}^\dag{\bf \Gamma}$), satisfying $\Gamma_\nu>0$, $\sum_{\nu}\Gamma_\nu^2={\rm Tr}\,{\bf \Gamma}^\dag{\bf \Gamma}=1$, 
and $A^{\dag}_{\nu}=
\sum_{\bm\alpha}U_{\bm\alpha\nu}C^{\dag}_{\bm\alpha}$,  $B^{\dag}_{\bar\nu}=
\sum_{\bm\beta}V^*_{\bm\beta\nu}C^{\dag}_{\bm{\bar\beta}}$,   the normal $N_\uparrow$- and $N_\downarrow$- 
particle creation operators respectively,  with ${\bf U}$, ${\bf V}$ the unitary matrices of the singular value decomposition (SVD) ${\bf \Gamma}={\bf U}{\bf \Gamma}^{D}{\bf V}^\dag$ and  ${\bf \Gamma}^D$ a diagonal matrix of elements $\Gamma^D_{\nu\nu'}=\delta_{\nu\nu'}\Gamma_\nu$. These normal operators create orthogonal $N_\uparrow$- and $N_\downarrow$-particle states $|\nu\rangle=A^\dag_{\nu}|0\rangle$,  $|\bar\nu\rangle=B^\dag_{\bar\nu}|0\rangle$, with $\langle \nu'|\nu\rangle=\delta_{\nu\nu'}=\langle \bar\nu'|\bar\nu\rangle$, which in general are not SDs. The number of nonzero singular values $\Gamma_\nu$ is the Schmidt rank $n_s$ and is just the rank of the tensor ${\bf \Gamma}$. 

\subsection{Entanglement and correlation measures}
We now discuss the entanglement and correlation measures  employed in this work. They are such that they all vanish for SDs of the form \eqref{ab0}, hence measuring  the deviation of the actual eigenstates of $H$ from such SDs. 

\subsubsection{Total up-down entanglement of eigenstates}
Starting from the expansion \eqref{psig} of $\ket{\Psi}$,  the total bipartite up-down entanglement in these eigenstates is determined by the mixedness of the RDMs $\rho_\uparrow$, $\rho_\downarrow$  of the up and down electrons, which are isospectral. Setting 
\be \rho\equiv|\Psi\rangle\langle\Psi|,\ee they are given by the partial traces 
\begin{subequations}
    \label{rup}
\begin{eqnarray}
\rho_{\uparrow}
&=&
{\rm Tr}_{\downarrow}\, \rho=\sum_{\bm\beta}C_{\bar{\bm\beta}}|\Psi\rangle\langle\Psi|C^\dag_{\bar{\bm\beta}}\\
&=&\sum_{\bm\alpha,\bm\alpha'}({\bf\Gamma\Gamma}^\dag)_{\bm\alpha\bm\alpha'}|\bm\alpha\rangle\langle\bm\alpha'|
=\sum_\nu \Gamma_\nu^2\,|\nu\rangle\langle\nu|\,,
\end{eqnarray}
\end{subequations}
and similarly, $\rho_{\downarrow}={\rm Tr}_{\uparrow}\,\rho=\sum_{\bm\alpha}C_{\bm\alpha}|\Psi\rangle\langle\Psi|C_{\bm\alpha}^\dag=\sum_{\bm\beta,\bm\beta'}({\bf \Gamma^\dag \Gamma})_{\bm\beta'\bm\beta}|\bar{\bm\beta}\rangle\langle\bar{\bm\beta'}|=\sum_\nu \Gamma_\nu^2|\bar\nu\rangle\langle\bar\nu|$, where $|\bm\alpha\rangle=C^\dag_{\bm \alpha}|0\rangle$, $|\bm{\bar\beta}\rangle=C^\dag_{\bm {\bar\beta}}|0\rangle$.
Hence, the squared singular values $\Gamma_\nu^2$ are their eigenvalues while the states $|\nu\rangle$, $|\bar\nu\rangle$ of the Schmidt decomposition \eqref{Scd} the corresponding eigenvectors. They  satisfy ${\rm Tr}\,\rho_\uparrow={\rm Tr}\,\rho_{\downarrow}=1$ and   determine the averages of any observable concerning just the up (or down) electrons 
through $\langle \Psi|O_{\uparrow}|\Psi\rangle={\rm Tr}\rho_{\uparrow}O_{\uparrow}$, $\langle \Psi|O_{\downarrow}|\Psi\rangle={\rm Tr}\rho_{\downarrow}O_{\downarrow}$. 
  
Their common entropy is the  total {\it up-down entanglement entropy}: 
\begin{equation}
E_{\uparrow\downarrow}=
S(\rho_{\,\uparrow})=S(\rho_{\,\downarrow})=-\sum_\nu \Gamma^2_\nu\log \Gamma^2_\nu\label{Eud}\,,
\end{equation}
where we have set $S(\rho)=-{\rm Tr}\,\rho\log\rho$ as the von Neumann entropy. Eq.\ \eqref{Eud} vanishes if and only if $|\Psi\rangle$ is any up-down  ``product'' state  $|\nu\bar\nu\rangle=A^\dag_{\nu}B^\dag_{\bar\nu}|0\rangle$, i.e.\ $n_s=1$ in \eqref{Scd}  (hence vanishing in any SD of the form \eqref{ab0}, though being a SD is not a necessary condition for its vanishing),  
and is maximum for maximally mixed reduced states with $\Gamma_\nu^2=1/D$ $\forall\,\nu$ 
(with $D={\rm Min}[\binom{d/2}{N_\uparrow},\binom{d/2}{N_\downarrow}])$, for which $E_{\uparrow\downarrow}=\log D$.  

The associated mutual information, which is a measure of total (classical plus quantum) up-down correlations, is$\!\!$ 
\begin{subequations}\label{Itud}\begin{eqnarray}   I_{\uparrow\downarrow}&=&
S(\rho||\rho_{\uparrow}\otimes\rho_{\downarrow})=
S(\rho_{\uparrow})+S(\rho_{\downarrow})-S(\rho)\label{iud0}\\&=&2E_{\uparrow\downarrow}
\label{Iud}\,,
\end{eqnarray}
\end{subequations}
since $S(\rho)=0$. Here $S(\rho||\sigma)=-{\rm Tr}\rho\,(\log\sigma\,-\log\rho)$ is the relative entropy ($S(\rho||\sigma)\geq 0$, with $S(\rho||\sigma)=0$ iff $\rho=\sigma$ \cite{NielsenBook}), such that $I_{\uparrow\downarrow}=0$  iff $\rho=\rho_{\uparrow}\otimes\rho_{\downarrow}$. 

Thus, in pure up-down entangled eigenstates, $I_{\uparrow\downarrow}$  violates the classical upper bound  $I_{\uparrow\downarrow}\leq {\rm Min}[S(\rho_\uparrow),S(\rho_{\downarrow})]$,  which would hold if $\rho$ were a classical random variable (with $\rho_{\uparrow}$, $\rho_{\downarrow}$ its marginals). 
In the quantum case such upper bound  
 still  holds  for  all {\it separable} up-down states (pure or mixed) $\rho=\sum_{n}p_n\rho_{n\uparrow}\otimes\rho_{n\downarrow}$  with $p_{n}>0$,  i.e.\ convex mixtures of up-down product DMs,   
  which lead to $S(\rho)\geq {\rm Max}[S(\rho_\uparrow),S(\rho_{\downarrow})]$  \cite{NK.01,RC.02}. Thus, 
  though not a  measure of entanglement, $I_{\uparrow\downarrow}$ can  reveal entanglement through violation of this upper bound. In the general quantum case it satisfies instead $I_{\uparrow\downarrow}\leq 2{\rm Min}[S(\rho_{\uparrow}),S(\rho_{\downarrow})]$, according to the Araki-Lieb inequality 
  \cite{AL.70}.  

\subsubsection{One-body entanglement}
We now consider  fermionic entanglement  measures, which quantify the deviation of $|\Psi\rangle$ from a single SD, irrespective of the choice of sp modes. They are based on the reduced $M$-body DMs, which are idempotent in any SD \cite{SC.01,ES.02,GR.15,GDR.20,GDR.21}. 

We start with the one-body DM $\rho^{(1)}$, whose elements in a pure state $|\Psi\rangle$ are defined as 
\begin{equation}
    \rho^{(1)}_{ij}=\langle\Psi|c^\dag_j c_i|\Psi\rangle\,,\label{rh1}
\end{equation}
and satisfies $(\rho^{(1)})^2=\rho^{(1)}$ iff $|\Psi\rangle$ is a SD.  Since  the number of spin-up and spin-down electrons is fixed in each eigenstate $|\Psi\rangle$, it becomes here blocked, \begin{equation}\rho^{(1)}=\begin{pmatrix}\rho^{
(1)}_{\uparrow}&0\\0&\rho^{(1)}_{\downarrow}\end{pmatrix}\label{blrho1}\end{equation}
in a basis of sp orbitals with definite $s_z$, as $\langle c^\dag_i c_{\bar j}\rangle=0$. The blocks can be calculated as 
$\rho^{(1)}_{\uparrow_{ij}}=
{\rm Tr}\,\rho_\uparrow c^\dag_{j}c_{i}$, $\rho^{(1)}_{\downarrow_{ij}}=
{\rm Tr}\,\rho_\downarrow  c^\dag_{\bar j}c_{\bar i}$, in terms of the reduced up and down DMs \eqref{rup},   
satisfying ${\rm Tr}\,\rho^{(1)}_\uparrow=N_\uparrow$, ${\rm Tr}\,\rho^{(1)}_\downarrow=N_\downarrow$. 
In a SD its eigenvalues are obviously $\lambda^{(1)}_k=1$ ($0$) for occupied (empty) natural orbitals, but otherwise  $\lambda_k^{(1)}\in(0,1)$ for ``active'' natural orbitals. 

The associated {\it one-body entanglement} is determined by the ``mixedness'' of the one-body DM, and can be quantified by its entropy 
\begin{equation}
    E^{(1)}=S(\rho^{(1)})=
S(\rho^{(1)}_\uparrow)+S(\rho^{(1)}_\downarrow)\,.\label{E11}
\end{equation}
We will use here the von Neumann entropy 
though other entropies can also be employed \cite{GR.15,GDR.20,GDR.21}. 
Then $E^{(1)}\geq 0$, with $E^{(1)}=0$ iff
$\ket{\Psi}$ is a SD. $E^{(1)}$  is not affected by empty or fully occupied (i.e.\ ``core'') sp levels ($\lambda_k^{(1)}=0$ or $1$),    
and it is maximum when all sp levels have the same average occupation $2N_\mu/d$ ($\mu=\uparrow,\downarrow$), in which case $S(\rho^{(1)}_\mu)=-N_\mu\log(2N_\mu/d)$. 

In  pure states with definite particle number $N$, $\rho^{(1)}$  is isospectral with the $(N\!-1\!)$-body DM $\rho^{(N-1)}$ (and $\rho^{(M)}$ is isospectral with $\rho^{(N-M)}$ \cite{GDR.20,GDR.21}). Accordingly,  the one-body entanglement \eqref{E11} can also be viewed as the $(1,N\!-\!1)$-particle entanglement, within a generalized $(1,N-1)$ bipartite representation of the state \cite{GDR.20,GDR.21}. 

An important remark is that in order to have {\it nonzero total up-down entanglement} $E_{\uparrow\downarrow}>0$ {\it with fixed} up and down particle number $N_\uparrow$, $N_{\downarrow}$, {\it it is necessary to have one-body entanglement} $E^{(1)}>0$, i.e., $|\Psi\rangle$ cannot be a SD \cite{GDR.20}.  If it were zero, $\rho^{(1)}$, and hence both $\rho^{(1)}_{\uparrow}$ and $\rho^{(1)}_{\downarrow}$, should be idempotent, since the blocked form \eqref{blrho1} holds if $N_{\uparrow}$ and $N_{\downarrow}$ are fixed, implying that  both $\rho_{\uparrow}$ and $\rho_{\downarrow}$  should be SDs,  then leading to a SD $|\Psi\rangle$ of the form \eqref{ab0}, 
 which has  no up-down entanglement. 
 On the other hand, $E^{(1)}>0$ is not sufficient, since a mixed $\rho^{(1)}_{\uparrow}$ or $\rho^{(1)}_{\downarrow}$ could also stem from a ``product'' state $A^\dag_\nu B^\dag_{\bar\nu}|0\rangle$ with $A^\dag_\nu |0\rangle$ and/or $B^\dag_{\bar\nu}|0\rangle$ not SDs. 

\subsubsection{Two-body entanglement}\label{IIC2}
Further analysis of the state can be obtained through the two-body DM, of elements $\rho^{(2)}_{ij,kl}=\langle \Psi|c^\dag_k c^\dag_l c_j c_i|\Psi\rangle$ (with $i<j$, $k<l$).  For  present eigenstates $|\Psi\rangle$, it will contain three blocks \cite{CR.24}:
\begin{equation}
\rho^{(2)}=\begin{pmatrix}\rho^{(2)}_{\uparrow\uparrow}&0&0\\0&\rho^{(2)}_{\uparrow\downarrow}&0\\
0&0&\rho^{(2)}_{\downarrow\downarrow}\end{pmatrix}\,,
\label{rho2}
\end{equation}
where $\rho^{(2)}_{\uparrow\uparrow_{ij,kl}}={\rm Tr}\,[\rho_{\uparrow} c^\dag_kc^\dag_l c_jc_i]$,  $\rho^{(2)}_{\downarrow\downarrow_{ij,kl}}=
{\rm Tr}\,[\rho_{\downarrow} c^\dag_{\bar k}c^\dag_{\bar l} c_{\bar j}c_{\bar i}]$, with $i<j$, $k<l$ and  ${\rm Tr}\,\rho^{(2)}_{\uparrow\uparrow}=\binom{N_\uparrow}{2}$, 
${\rm Tr}\,\rho^{(2)}_{\downarrow\downarrow}=\binom{N_\downarrow}{2}$, 
while  
\begin{equation}   \rho^{(2)}_{\uparrow\downarrow_{ij,kl}}=
\langle\Psi|c^\dag_kc^\dag_{\bar l}c_{\bar j}c_i|\Psi\rangle \label{r2ud}\,,
\end{equation}
with  $i,j,k,l$ unrestricted and  
 ${\rm Tr}\,\rho^{(2)}_{\uparrow\downarrow}=
N_\uparrow N_\downarrow$, such that  ${\rm Tr}\,\rho^{(2)}=
\binom{N_\uparrow+N_\downarrow}{2}$. Remaining elements vanish for fixed $N_{\uparrow}$, $N_{\downarrow}$. 
The associated total two-body  entropy   is 
\begin{equation}    E^{(2)}=S(\rho^{(2)})=S(\rho^{(2)}_{\uparrow\uparrow})+S(\rho^{(2)}_{\uparrow\downarrow})+S(\rho^{(2)}_{\downarrow\downarrow})\,,
\end{equation}
where we will use again the von Neumann entropy. 
It can be considered  a measure of the total $(2,N-2)$ particle entanglement  within a general $(2,N-2)$ bipartite representation of $|\Psi\rangle$ \cite{GDR.21,CR.24}, vanishing  
 in a SD of the form \eqref{ab0}, where all eigenvalues of 
 these blocks are $1$ or $0$.  
It should be noticed, however, that in other states the eigenvalues of these two-body DMs can not only be  smaller, but also larger than $1$, due to the approximate bosonic character of collective  pair creation operators. Such large eigenvalues in $\rho^{(2)}$  reflect  the presence of pairing-type correlations \cite{Yang.62,GDR.21,SS.21,CR.24}.  

On the other hand, in any  {\it pure two-fermion  state},  
$\rho^{(2)}$ has just a single nonzero eigenvalue equal to $1$,  
with the pair operator $A^\dag$ creating the state as eigenvector, implying $E^{(2)}=0$ \cite{GDR.21}.  Explicitly, 
any such state  can be written in the natural sp basis diagonalizing $\rho^{(1)}$  as \cite{SC.01,ES.02} 
\be|\Psi\rangle=A^\dag|0\rangle\,,\;\;\;A^\dag=\sum_{k=1}^{n_s}\sigma_k c^\dag_{k}c^\dag_{\bar k}\label{psi2}\,,\ee
with $\sigma_k$ real and $\sum_k \sigma_k^2=1$. In this representation (corresponding to  $N_\uparrow=N_{\downarrow}=1$ and hence to the Schmidt decomposition \eqref{Scd}), just the central block $\rho^{(2)}_{\uparrow\downarrow}$ is nonzero in \eqref{rho2}, with $\langle c^\dag_{k}c^\dag_{\bar k}c_{\bar k'}c_{k'}\rangle=\sigma_k\sigma_{k'}$ its only nonzero elements. This implies rank $\rho^{(2)}_{\uparrow\downarrow}=1$, i.e., a single nonzero eigenvalue $\lambda_1=\sum_k\sigma_k^2=1$, with eigenvector $A^\dag$ ($\langle A^\dag A\rangle=1)$.  
In contrast, 
$\rho^{(1)}_{\uparrow}$   and $\rho^{(1)}_{\downarrow}$ have identical eigenvalues $\sigma_k^2$, leading to $E^{(1)}=-2\sum_k \sigma_k^2\log \sigma_k^2=2E_{\uparrow\downarrow}=I_{\uparrow\downarrow}$. The separable case corresponds to a single term in \eqref{psi2}, i.e.\ just a single nonzero $\sigma_k=1$.   

\subsubsection{Reduced two-body  up-down mutual information}
From $\rho^{(2)}_{\uparrow\downarrow}$ we can recover the one-body DM blocks  as
$\rho^{(1)}_{\uparrow}={\rm Tr}_{\downarrow}\rho^{(2)}_{\uparrow\downarrow}/N_{\downarrow}$, 
$\rho^{(1)}_{\downarrow}={\rm Tr}_{\uparrow}\rho^{(2)}_{\uparrow\downarrow}/N_\uparrow$, 
where ${\rm Tr}_{\mu}$ denotes the partial trace over $\mu=\,\uparrow$ or $\downarrow$ modes. 
Then we can also examine the total (classical plus quantum) up-down correlations at the level of two particles through the mutual information associated with $\rho^{(2)}_{\uparrow\downarrow}$, defined as 
\begin{subequations}
\label{iupd2}
\begin{eqnarray}    I^{(2)}_{\uparrow\downarrow}&=&S(\rho^{(2)}_{\uparrow\downarrow}\,||\,\rho^{(1)}_{\uparrow}\otimes\rho^{(1)}_{\downarrow})\\
&=&
   N_{\downarrow}\,S(\rho^{(1)}_{\uparrow})
     +N_{\uparrow}\,S(\rho^{(1)}_{\downarrow})
     -S(\rho^{(2)}_{\uparrow\downarrow})\label{Imud}\,,
\end{eqnarray}\end{subequations}
where $N_{\downarrow}={\rm Tr}\,\rho^{(1)}_{\downarrow}$, $N_{\uparrow}={\rm Tr}\,\rho^{(1)}_{\uparrow}$.  For a two-particle state with $N_\uparrow=N_\downarrow=1$,  Eq.\ \eqref{iupd2} becomes  identical with \eqref{Iud} (and hence with $E^{(1)}$, Eq.\ \eqref{E11}).  We show below its main general properties:

{\bf 1}. {\it $I^{(2)}_{\uparrow\downarrow}\geq 0$, with $I^{(2)}_{\uparrow\downarrow}=0$ iff  $\rho^{(2)}_{\uparrow\downarrow}=\rho^{(1)}_{\uparrow}\otimes \rho^{(1)}_{\downarrow}$}.\\
This last equality is obviously  satisfied  in {\it any} ``product'' state $|\Psi\rangle=A^\dag_{\nu}B^\dag_{\bar\nu}|0\rangle$, including (but not limited to)  SDs \eqref{ab0} with fixed $N_\uparrow,N_\downarrow$. 

{\it Proof:} We note that 
$I^{(2)}_{\uparrow\downarrow}=N_{\uparrow}N_{\downarrow}S(\rho^{(2)}_{n\uparrow\downarrow}||\rho^{(1)}_{n\uparrow}\otimes\rho^{(1)}_{n\downarrow})$, where $\rho^{(1)}_{n\uparrow}:=\rho^{(1)}_{\uparrow}/N_\uparrow={\rm Tr}_{\downarrow}\rho^{(2)}_{n\uparrow\downarrow}$, 
$\rho^{(1)}_{n\downarrow}=\rho^{(1)}_{\downarrow}/N_\downarrow={\rm Tr}_{\uparrow}\rho^{(2)}_{n\uparrow\downarrow}$ and 
$\rho^{(2)}_{n\uparrow\downarrow}=\rho^{(2)}_{\uparrow\downarrow}/(N_\uparrow N_\downarrow)$ are all normalized densities with unit trace. Then, from the basic properties of the relative entropy, it follows that $S(\rho^{(2)}_{n\uparrow\downarrow}||\rho^{(1)}_{n\uparrow}\otimes\rho^{(1)}_{n\downarrow})\geq 0$, vanishing iff 
$\rho^{(2)}_{n\uparrow\downarrow}=\rho^{(1)}_{n\uparrow}\otimes \rho^{(1)}_{n\downarrow}$, i.e.,  
$\rho^{(2)}_{\uparrow\downarrow}=\rho^{(1)}_\uparrow\otimes \rho^{(1)}_\downarrow$ (we assume $N_\uparrow N_\downarrow>0$), which leads to {\bf 1.} \qed

{\bf 2}. {\it $I^{(2)}_{\uparrow\downarrow}$ is  independent of the number of up and down ``core'' fermions.} \\
In other words [and in contrast with $I^{(2)}_{\uparrow\downarrow}/(N_\uparrow N_\downarrow)$] it does not depend on the 
number of sp levels having fixed occupation $1$, and obviously nor on those with null occupation,  such that just ``active'' electrons  need be considered. 

{\it Proof:} Addition of a core with $N^c_\uparrow$ and $N^c_\downarrow$ fully occupied orbitals leads to $\rho^{(1)}_{\mu}\rightarrow\rho^{(1)}_{\mu}\oplus\mathbbm 1^c_\mu$ and $N_\mu\rightarrow N_\mu+N^c_\mu$ for $\mu=\uparrow,\downarrow$, implying   $\rho^{(2)}_{\uparrow\downarrow}\rightarrow
\rho^{(2)}_{\uparrow\downarrow}\oplus (\rho^{(1)}_{\uparrow}\otimes \mathbbm 1^c_\downarrow)\oplus(\mathbbm 1^c_\uparrow\otimes \rho^{(1)}_{\downarrow})\oplus(\mathbbm 1^c_\uparrow\otimes\mathbbm 1^c_\downarrow)$ and hence 
$S(\rho^{(2)}_{\uparrow\downarrow})\rightarrow S(\rho^{(2)}_{\uparrow\downarrow})+N^c_\downarrow S(\rho^{(1)}_{\uparrow})+N^c_\uparrow S(\rho^{(1)}_{\downarrow})$, such that Eq.\ \eqref{Imud} remains invariant.\qed 

{\bf 3.} {\it If 
$|\Psi\rangle=
A^\dag_I A^\dag_{II}|0\rangle$, 
where $A^\dag_I$, $A^\dag_{II}$ create respectively $N^I_\uparrow$, 
$N^I_\downarrow$ and $N^{II}_\uparrow$, $N^{II}_\downarrow$  fermions  in fully orthogonal sp subspaces $I$ and $II$,  then 
\begin{equation}
I^{(2)}_{\uparrow\downarrow}=
I^{(2)}_{\uparrow\downarrow,I}+I^{(2)}_{\uparrow\downarrow,II}\,,\label{pr2}\end{equation} 
where $I^{(2)}_{\uparrow\downarrow,X}$ is the mutual information \eqref{iupd2} of $A^\dag_X|0\rangle$.} 

{\it Proof}: In this case $\rho^{(1)}_\mu=\rho^{(1)}_{\mu,I}\oplus\rho^{(1)}_{\mu,II}$, $N_\mu=N_\mu^I+N_\mu^{II}$ and also $\rho^{(2)}_{\uparrow\downarrow}=\rho^{(2)}_{\uparrow\downarrow,I}\oplus\rho^{(2)}_{\uparrow\downarrow,II}\oplus
(\rho^{(1)}_{\uparrow,I}\otimes\rho^{(1)}_{\downarrow,II})\oplus (\rho^{(1)}_{\uparrow,II}\otimes\rho^{(1)}_{\downarrow,I})$, implying $S(\rho^{(2)}_{\uparrow\downarrow})=S(\rho^{(2)}_{\uparrow\downarrow,I})+S(\rho^{(2)}_{\uparrow\downarrow,II})+N_\downarrow^{II}S(\rho^{(1)}_{\uparrow,I})+
N_\uparrow^I S(\rho^{(1)}_{\downarrow,II})+
N_\downarrow^I S(\rho^{(1)}_{\uparrow,II})+
N_\uparrow^{II}S(\rho^{(1)}_{\downarrow,I})$ and $S(\rho^{(1)}_{\mu})=S(\rho^{(1)}_{\mu,I})+
S(\rho^{(1)}_{\mu,II})$, which leads to \eqref{pr2}. \qed 

In particular, if $A^\dag_{II}|0\rangle$  corresponds to a ``core''  (all sp levels fully occupied), $I^{(2)}_{\uparrow\downarrow,II}=0$ and we recover from \eqref{pr2} the preceding property {\bf 2}.

{\bf 4.} {\it In the ``classically correlated'' case}
\begin{equation}\rho^{(2)}_{\uparrow\downarrow}=\sum_{\nu}p_\nu\,\rho^{(1)}_{\nu_\uparrow}\otimes\rho^{(1)}_{\nu_\downarrow}\,, 
\label{sep}\end{equation} 
{\it where $p_\nu> 0$, $\sum_\nu p_\nu=1$ and  the $\rho^{(1)}_{\nu_\mu}$  are assumed to have mutually orthogonal supports for distinct $\nu$'s and fixed $\nu$-independent traces ${\rm Tr}\,\rho^{(1)}_{\nu_\mu}=N_\mu$ for $\mu=\uparrow,\downarrow$, then 
 \be I^{(2)}_{\uparrow\downarrow}=N_\uparrow N_\downarrow S(\bm p)\,\label{Iclas}\,,\ee
where $S(\bm p)=-\sum_\nu p_\nu \log p_\nu$ is the  Shannon entropy of the 
probability distribution $\{p_\nu\}$.}

{\it Proof}: Under previous assumptions   
we obtain $S(\rho^{(2)}_{\uparrow\downarrow})=\sum_{\nu} p_\nu\,[N_{\downarrow}S(\rho^{(1)}_{\nu_\uparrow})+N_\uparrow S(\rho^{(1)}_{\nu_\downarrow})]+N_\uparrow N_\downarrow S(\bm p)$ while $S(\rho^{(1)}_{\mu})=\sum_\nu p_\nu S(\rho^{(1)}_{\nu_\mu})+N_\mu S(\bm p)$ for $\mu=\uparrow,\downarrow$.  
Then \eqref{Imud} leads to  \eqref{Iclas}.\qed

An example of an entangled global state leading to the classically correlated two-body DM  \eqref{sep}  is 
 \be|\Psi\rangle=\sum_\nu \sqrt{p_\nu} A^\dag_{\nu}B^\dag_{\bar\nu}|0\rangle\label{exmp}\ee 
 where $A^\dag_{\nu}$, $B^\dag_{\bar\nu}$ create up and down states of  $N_\uparrow\geq 2$ and $N_\downarrow\geq 2$ electrons respectively with fully orthogonal sp supports, such that 
 $\rho^{(2)}_{\uparrow\downarrow}$ becomes the average \eqref{sep} (as all cross terms  involving distinct $\nu$'s vanish). Eq.\ \eqref{exmp} is for this case the Schmidt decomposition \eqref{Scd} of $|\Psi\rangle$, with $\sqrt{p_\nu}=\Gamma_\nu$. The up-down entanglement of the state \eqref{exmp} is precisely $E_{\uparrow\downarrow}=S(\bm p)$, with 
 $I_{\uparrow\downarrow}=2E_{\uparrow\downarrow}$.

\subsubsection{Total up-down Negativity}
We first recall that the negativity  \cite{VW.02,P.05}    of a bipartite mixed state $\rho_{AB}$ of two distinguishable components,  is defined as minus the sum of the negative eigenvalues of its partial transpose  $\rho_{AB}^{t_B}$ (which are the same as those of $\rho_{AB}^{t_A}$). Since ${\rm Tr}\,\rho_{AB}^{t_B}={\rm Tr}\,\rho_{AB}=1$, it can be expressed as 
\begin{equation}{\cal N}_{AB}=\tfrac{1}{2}[{\rm Tr}\,|\rho_{AB}^{t_B}|-1]\,.\label{neg0}\end{equation}
Based on the well-known positive partial transpose necessary criterion for separability \cite{P.96,HHH.96}, this non-negative quantity vanishes in any  separable state 
$\rho_{AB}=\sum_\nu p_\nu \rho_{\nu A}\otimes\rho_{\nu B}$  
where $p_\nu>0$, $\sum_{\nu}p_\nu=1$ 
  \cite{W.89}, i.e.\ for any convex mixture of product densities (as $\sum_\nu p_\nu \rho_{\nu A}\otimes\rho_{\nu B}^t$ is positive semidefinite),  but 
can  be positive in entangled mixed states (and is always positive in entangled pure states, see Eq.\ \eqref{negp} below), providing a simple computable indicator of entanglement for mixed states: ${\cal N}(\rho_{AB})>0\Rightarrow \rho_{AB}$ is entangled. Moreover, it is an entanglement monotone \cite{VW.02,P.05}: it does not increase under local operations and classical communication. 

Here we use this measure  to detect up-down entanglement in convex mixtures of eigenstates,  
\begin{equation}\rho=\sum_n p_n\ket{\Psi_n}\bra{\Psi_n}\,,
\label{com}
\end{equation}
where $H\ket{\Psi_n}=E_n\ket{\Psi_n}$ and $p_n\geq 0$.    In particular, for analyzing the dissociation limit we will consider thermal-like states with fixed $M_S$,  where $p_n\propto \delta_{M_{S_n},M_S}\,e^{-\beta(E_n-E_0)}$  with $\beta=1/kT>0$, 
such that all states  in the mixture have the same  $N_{\uparrow}$, $N_{\downarrow}$. 

In such a case we can directly apply Eq.\ \eqref{neg0} for $A$, $B$ identified with the set of spin-up and spin-down fermions, since  they can be considered as distinguishable due to their distinct spin quantum number. We then define the total up-down negativity as 
\begin{equation}
{\cal N}_{\uparrow\downarrow}=\tfrac{1}{2}[{\rm Tr}\,|\rho^{t_\downarrow}|-1]\label{neg}\,,
\end{equation}
where $\rho^{t_\downarrow}_{\bm\alpha\bar{\bm \beta},\bm\alpha'\bar{\bm\beta'}}=\rho_{\bm\alpha\bar{\bm\beta'},\bm\alpha'\bar{\bm\beta}}$. It satisfies 
${\cal N}_{\uparrow\downarrow}=0$ for any up-down separable state  
$\rho=\sum_\nu p_\nu\rho_{\nu\uparrow}\otimes \rho_{\nu\downarrow}$ ($p_\nu\geq 0$), i.e., which can be written as a convex mixture of product up-down states (in particular for any mixture of SDs with definite $N_\uparrow$, $N_\downarrow$). Hence, ${\cal N}_{\uparrow\downarrow}>0$ ensures that $\rho$ is up-down entangled, i.e.\ not of the previous separable form. We recall  that an up-down separable mixed state will still normally have $I_{\uparrow\downarrow}>0$ (it can be classically correlated), as $I_{\uparrow\downarrow}=0$ just for single products  $\rho_{\nu\uparrow}\otimes \rho_{\nu\downarrow}$. $I_{\uparrow\downarrow}$ is then clearly nonequivalent to ${\cal N}_{\uparrow\downarrow}$ for mixed states.

Nonetheless, in the case of a pure state $\rho=|\Psi\rangle\langle\Psi|$, from the up-down Schmidt decomposition \eqref{Scd} we obtain 
\be {\cal N}_{\uparrow\downarrow}=\sum_{\nu<\nu'}\Gamma_\nu\Gamma_{\nu'}=\tfrac{1}{2}[({\rm Tr}\sqrt{\rho_{\mu}})^2-1]\label{negp}\,,\ee
where $\rho_\mu$, $\mu=\uparrow$ or $\downarrow$, 
is the total up or down DM (Eq.\ \eqref{rup}). Hence, in the pure  case ${\cal N}_{\uparrow\downarrow}$ becomes  another entropic measure of total up-down entanglement (it is an increasing function of the $\alpha=1/2$  Renyi   entropy $\frac{1}{1-\alpha}\log {\rm Tr}\rho_\mu^{\alpha}$ of the reduced state),  vanishing just if $n_s=1$ in \eqref{Scd} and becoming maximum in the maximally entangled case.  Thus, in pure states  it always detects entanglement if present,  becoming equivalent to $E_{\uparrow\downarrow}$.  

\subsubsection{Reduced two-body up-down Negativity}
Similarly, in order to obtain  an indicator of the ``inner'' entanglement of the  up-down block of the two-body DM  
$\rho^{(2)}_{\uparrow\downarrow}$,  we can define its negativity  again as minus the sum of the negative eigenvalues of its partial transpose. This yields, noting that ${\rm Tr}\,\rho^{(2)\,t_\downarrow}_{\uparrow\downarrow}={\rm Tr}\,\rho^{(2)}_{\uparrow\downarrow}=N_\uparrow N_\downarrow$,
\be {\cal N}^{(2)}_{\uparrow\downarrow}=
\tfrac{1}{2}[{\rm Tr}\,|\rho^{(2)\,t_\downarrow}_{\uparrow\downarrow}|-N_\uparrow N_\downarrow]\,,\label{neg2}
\ee
where $(\rho^{(2)\,t_\downarrow}_{\uparrow\downarrow})_{i\bar j,k\bar l}=(\rho^{(2)}_{\uparrow\downarrow})_{i\bar l,k\bar j}$. This non-negative quantity vanishes for any {\it separable} $\rho^{(2)}_{\uparrow\downarrow}$: 
\be \rho^{(2)}_{\uparrow\downarrow}=\sum_\nu p_\nu\rho^{(1)}_{\nu\uparrow}\otimes\rho^{(1)}_{\nu\downarrow}\;\Rightarrow\;{\cal N}_{\uparrow\downarrow}^{(2)}=0\label{neg20}\ee 
where $p_\nu>0$, $\sum_\nu p_\nu=1$ and 
$\rho^{(1)}_{\nu\mu}$ are arbitrary  one-body densities 
satisfying ${\rm Tr}\,\rho^{(1)}_{\nu\mu}=N_\mu$ for $\mu=\uparrow$ or $\downarrow$. 
The proof is obvious since $(\rho^{(1)}_{\nu\uparrow}\otimes\rho^{(1)}_{\nu\downarrow})^{t_\downarrow}=\rho^{(1)}_{\nu\uparrow}\otimes\rho^{(1)t}_{\nu\downarrow}$ is positive semidefinite, as   $\rho^{(1)t}_{\alpha\downarrow}$ has the same eigenvalues as $\rho^{(1)}_{\alpha\downarrow}$ and the sum of positive semidefinite operators is positive semidefinite.  

Hence, a positive ${\cal N}^{(2)}_{\uparrow\downarrow}>0$ indicates an entangled  (i.e., nonseparable) $\rho^{(2)}_{\uparrow\downarrow}$, in the sense that it cannot be written as a convex mixture  \eqref{neg20}. 
For example, in the  two-fermion state \eqref{psi2} with $N_\uparrow=N_\downarrow=1$, it is easily seen that ${\cal N}^{(2)}_{\uparrow\downarrow}=\sum_{k<k'}\sigma_k\sigma_{k'}>0$ if $n_s\geq 2$, i.e.\ if the state is up-down entangled, coinciding in this case with the full  up-down negativity ${\cal N}_{\uparrow\downarrow}$, Eqs.\ \eqref{neg}--\eqref{negp}.  It is also apparent that the negative eigenvalues of $\rho^{(2)t_{\downarrow}}_{\uparrow\downarrow}$, if existent, are not affected by the presence of ``core'' fermions, since the latter  just add separable blocks to $\rho^{(2)}_{\uparrow\downarrow}$, such that they do not alter the value of ${\cal N}^{(2)}_{\uparrow\downarrow}$.

Moreover, if the many-electron state  is a convex mixture $\rho=\sum_{\nu}p_\nu|\Psi_\nu\rangle\langle\Psi_\nu|$ of up-down `product' states $|\Psi_\nu\rangle=A^\dag_\nu B^\dag_{\bar{\nu}}|0\rangle$, with $p_\nu>0$,  $\rho^{(2)}_{\uparrow\downarrow}$ will also have the separable form \eqref{neg20}, as each of these  $|\Psi_\nu\rangle$ leads to an up-down product $\rho^{(2)}_{\uparrow\downarrow}$,  implying ${\cal N}^{(2)}_{\uparrow\downarrow}=0$.  Such product states include all   SDs with definite $N_\uparrow,N_\downarrow$, but are not restricted to them.  Thus, ${\cal N}^{(2)}_{\uparrow\downarrow}>0$  {\it already indicates up-down entanglement of the full state}    using just two-body information,  ensuring $\rho$ is not a convex mixture of such up-down product states without requiring the evaluation of the full ${\cal N}_{\uparrow\downarrow}$ of Eq.\ \eqref{neg}. Other witnesses and examples of up-down entanglement are discussed in Appendix \ref{apA}.

On the other hand, as in the previous case ${\cal N}^{(2)}_{\uparrow\downarrow}>0$ is in general a sufficient yet not necessary condition for up-down entanglement, both at the total and two-body level. And as previously discussed, a separable $\rho^{(2)}_{\uparrow\downarrow}$ of the form \eqref{neg20} can also emerge from an up-down entangled  state  at the full level, like e.g.\  the pure state \eqref{exmp} with  orthogonal sp supports for distinct $\nu$'s.  

\subsubsection{
Two-body fermionic Negativity for real representations} 

In order to obtain an analogous measure of the ``inner'' entanglement  of a general $\rho^{(2)}$ or of the blocks $\rho^{(2)}_{\uparrow\uparrow}$ or $\rho^{(2)}_{\downarrow\downarrow}$,  we now introduce a two-body negativity for real states (in some fixed sp basis), which vanishes in any convex mixture of real SDs,  but can be otherwise positive, being always positive for real 
entangled pure two-particle states. Such real states can arise  e.g.\ as eigenstates of a  Hamiltonian with real representation in a given  sp basis, as occurs in the present work, such that its  eigenvectors can be always chosen as real in this basis. The present negativity is not associated to any a priori partition of the sp space, thus differing from the previous negativities.  

Setting as before $\rho^{(2)}_{ij,kl}=\langle c^\dag_k c^\dag_l c_j c_i\rangle$, 
we define  an antisymmetrized partial transpose of elements 
\begin{equation}    \rho^{(2)t_p}_{ij,kl}=\rho^{(2)}_{il,kj}-\rho^{(2)}_{ik,lj}\,,\label{rtp}
\end{equation}
where here we consider unrestricted sp labels (i.e.,  $d^2\times d^2$ matrices) and accordingly, an   antisymmetrized $\rho^{(2)}$ ($\rho^{(2)}_{ij,kl}=-\rho^{(2)}_{ij,lk}=-\rho^{(2)}_{ji,kl}$). 

The matrix \eqref{rtp} then has the same trace as  $\rho^{(2)}$, $\frac{1}{2}{\rm Tr}\,\rho^{(2)t_p}=\frac{1}{2}{\rm Tr}\,\rho^{(2)}=\frac{1}{2}\langle N^2- N\rangle$,  and is hermitian (i.e.\  symmetric in the present real case) and antisymmetrized ($\rho^{(2)t_p}_{ij,kl}=-\rho^{(2)t_p}_{ij,lk}=-\rho^{(2)t_p}_{ji,kl}$) for an hermitian real  $\rho^{(2)}$. And 
 if averages are taken with respect to a real SD or a real fermionic gaussian state commuting with $N$, Wick's theorem holds, i.e., 
$\rho^{(2)}_{ij,kl}=\rho^{(1)}_{ik}\rho^{(1)}_{jl}-\rho^{(1)}_{il}\rho^{(1)}_{jk}$ and hence, 
\[\rho^{(2)t_p}_{ij,kl}=\rho^{(1)}_{ik}\rho^{(1)}_{lj}-\rho^{(1)}_{ij}\rho^{(1)}_{lk}-\rho^{(1)}_{il}\rho^{(1)}_{kj}+
\rho^{(1)}_{ij}\rho^{(1)}_{kl}=\rho^{(2)}_{ij,kl}\]
since $\rho^{(1)}_{ij}=\rho^{(1)}_{ji}\,\forall\,i,j$ when $\rho^{(1)}$ is real. This implies  $\rho^{(2)t_p}$ positive semidefinite for any real SD or gaussian state in the given basis. Hence, it will remain so for any convex  mixture of SDs or gaussian states, since a sum of positive semidefinite matrices is positive semidefinite. 

Besides, under real unitary sp transformations $c_i\rightarrow W^\dagger  c_i W=\sum_{i'} U_{ii'}c_{i'}$, 
with $W=e^{c^\dag h c}$, $U=e^{h}$ and $h$ a real antisymmetric matrix, such that  $U^t U=\mathbbm 1$, both $\rho^{(2)}$ and $\rho^{(2)t_p}$ undergo a real unitary transformation, which leaves their eigenvalues unchanged: $\rho^{(2)}\rightarrow U^{(2)}\rho^{(2)}U^{(2)t}$, $\rho^{(2)t_p}\rightarrow U^{(2)}\rho^{(2)t_p}U^{(2)t}$,
with $U^{(2)}_{ij,i'j'}=U_{ii'}U_{jj'}$.  

On the other hand, for a general real two-fermion state, which can be always written as in Eq.\ \eqref{psi2} after a suitable choice of sp basis, 
we have 
$\rho^{(2)}_{k\bar k,k'\bar k'}=\sigma_k\sigma_{k'}$, while all other elements vanish (except those obtained by permutations  $k\leftrightarrow \bar k$ and/or $k'\leftrightarrow \bar k'$). Then, 
while $\frac{1}{2}\rho^{(2)}$ has  a  single nonzero eigenvalue equal to $1$ (for unrestricted labels), $\rho^{(2)t_p}$ will have nonzero elements $\rho^{(2)t_p}_{k\bar k',k'\bar k}$ and $\rho^{(2)t_p}_{kk',\bar k'\bar k}$ (along with the corresponding permutations), 
which lead to negative eigenvalues $-\sigma_k\sigma_{k'}$ for $k\neq k'$ if $n_s\geq 2$ in \eqref{psi2} 
 (together with positive eigenvalues $\sigma_k\sigma_{k'}$ $\forall\,k,k'$). 
 This leads to a negativity 
\begin{equation} {\cal N}^{(2)}=2\sum_{k<k'}\sigma_k\sigma_{k'}\,,\end{equation} which is strictly positive iff $n_s\geq 2$,  where 
\begin{equation} {\cal N}^{(2)}=\tfrac{1}{2}[{\rm Tr}\,|\tfrac{1}{2}\rho^{(2)t_p}|-{\rm Tr}\,(\tfrac{1}{2}\rho^{(2)})]\label{Neg2}\,,
\end{equation}
is minus the sum of the negative eigenvalues of $\tfrac{1}{2}\rho^{(2)t_p}$.
For this state it is just twice the value of ${\cal N}^{(2)}_{\uparrow\downarrow}$, though present negativity does not assume an a priori knowledge of the up-down nature of the correlations. The two-fermion case is relevant since any $\rho^{(2)}$ can be formally considered as  a (generally mixed) two-fermion state which leads to the same two-body averages as the full original state \cite{CR.24}.
Hence, for real states, ${\cal N}^{(2)}>0$ already indicates that both $\rho^{(2)}$ and  the full $\rho$ cannot be written as a convex mixture of real SDs. Other witnesses of this entanglement and its implications are discussed in Appendix\ \ref{apA}.

 In this work we will actually apply the negativity \eqref{Neg2} to the first and third blocks of $\rho^{(2)}$, for which there is no a priori partition, defining  
\begin{equation}{\cal N}^{(2)}_{\uparrow\uparrow}:=\tfrac{1}{2}[{\rm Tr}\,|\tfrac{1}{2}\rho^{(2)t_p}_{\uparrow\uparrow}|-
\tfrac{1}{2}N_{\uparrow}(N_{\uparrow}-1)]\,,
\label{Neg4}\end{equation} 
and similarly  ${\cal N}^{(2)}_{\downarrow\downarrow}$. Hence, for real states, ${\cal N}^{(2)}_{\uparrow\uparrow}\geq 0$, with ${\cal N}^{(2)}_{\uparrow\uparrow}>0$ already ensuring that  $\rho^{(2)}_{\uparrow\uparrow}$ cannot be written as a convex mixture of up-up two-fermion real SDs, and hence that the full up RDM $\rho_{\uparrow}$ is not a convex mixture of real $N_{\uparrow}$-fermion   SDs. Analogous results hold for ${\cal N}^{(2)}_{\downarrow\downarrow}$.   

\section{Application}
\begin{figure}
    \begin{center}
        \includegraphics[width=4.5cm]{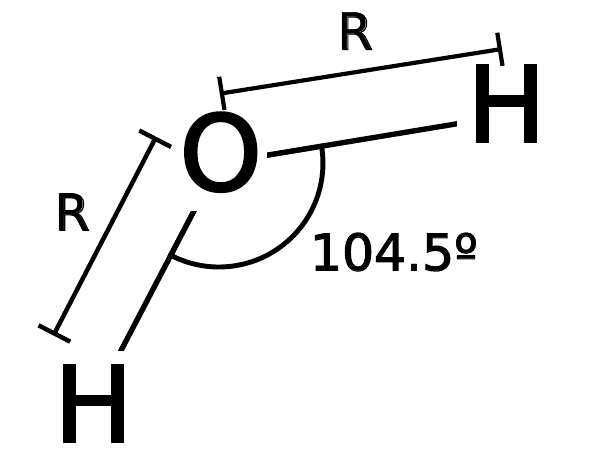}
        \caption{\label{fig:water} Water molecule as defined in our computations.}
    \end{center} 
\end{figure}

We will now study the GS  entanglement and correlations along the dissociation curve of the water molecule, keeping the H--O--H angle at 104.5\textdegree, and varying the O--H distance $R$ between 0.4 and 4 \AA, 
as indicated in Fig.\ \ref{fig:water}. This problem is usually referred to as the double dissociation of the water molecule, and is a classic test platform for various wavefunction methods \cite{Brown_1984, Olsen_1996, Li_1998, Ma_2005, Lee_2018}. 
We will focus our  discussion on states with $N_\uparrow = N_\downarrow$ ($M_S=0$). 

\subsection{Computational details}
The matrix elements of the electronic Hamiltonian were computed using  OpenFermion \cite{mcclean2019openfermionelectronicstructurepackage}, with the PySCF plugin \cite{Sun_2015, Sun_2017, Sun_2020}.
The sp space used was defined as in the STO-3G basis set \cite{hehre1969a}, which is a minimal basis set that contains 7 sp orbitals (for each spin).
Unless otherwise stated, the orbitals employed in the exact diagonalization are those coming from the Restricted Hartree-Fock  (RHF) method, i.e.,  those  that minimize the energy of a single SD, and are labeled from 0 to 6 ranging from the lowest to highest orbital energies \footnote{All entanglement measures were computed using our own programs \cite{fermionic-mbody, q-chemistry}, which rely on numpy and scipy for the linear-algebraic operations, and are made available online.}.
Fig.\ \ref{fig:energies} shows the energies of the lowest-lying states for this molecule. Some illustrative results for H$_2$O in the larger 6-31G basis, as well as for the simpler H$_2$ molecule in bases of increasing dimension, are provided in Appendix \ref{apB}.  

We have centered our analysis on  two states: 

\begin{figure}
    \begin{center}
        \includegraphics[width=1.02\linewidth]{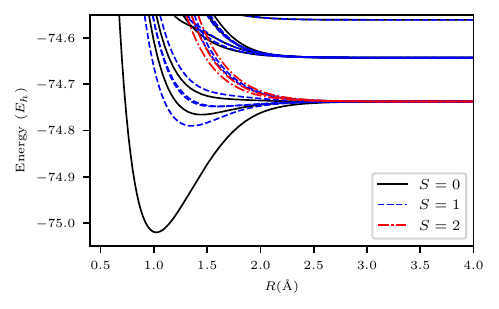}
    \end{center}
    \vspace*{-0.5cm}
    
    \caption{\label{fig:energies} Total energy of the lowest-lying states for the water molecule  (electronic plus nuclear repulsion), computed with the STO-3G basis set, as a function of the O--H distance. They are distinguished by color according to their total spin quantum number $S$.}
\end{figure}

i) The GS $\ket{\Psi}\equiv \ket{\Psi_0}$, 
i.e.\ the eigenfunction of $H$ with the lowest eigenvalue $E_0$.
It is a singlet state (total spin $S=0$), which evolves smoothly from a near SD at the equilibrium distance  ($R_{\rm eq}\approx 1.025${\AA} in present configuration space, slightly larger than the actual value $R^{\rm ex}_{\rm eq}\approx 0.96$\AA), to a  correlated state for larger distances, 
due to the fixed total spin  and the inner correlations at the O atom. 

ii) The $M_S=0$ thermal state, defined as
\begin{equation}
    \rho_0(\beta) = \frac{\Pi_0\,e^{-\beta (H - E_0)}}{{\rm Tr}\,[\Pi_0\, e^{-\beta (H - E_0)} ]} = \sum_{n} p_n \ket{\Psi_n}\bra{\Psi_n}\,,
    \label{eq:thermal}
\end{equation}
where $\Pi_0$ is the projector onto $M_S=0$ ($N_\uparrow=N_\downarrow=\frac{1}{2}N$) and $\ket{\Psi_n}$ the $M_S=0$ eigenstates: $H|\Psi_n\rangle=E_n|\Psi_n\rangle$,  $S_z|\Psi_n\rangle=0$, with  
$p_n=q_n/\sum_n q_n$ and $q_n=e^{-\beta(E_n-E_0)}$. We are actually interested in the low temperature limit $\beta\rightarrow\infty$ ($T=\frac{1}{k\beta}\rightarrow 0^+$). Although in the absence of GS degeneracy this limit leads again to the GS case, in the presence of GS degeneracy, as in the dissociation limit, $\rho_0(\beta)$ approaches the normalized projector onto the full $M_S=0$ GS subspace, then leading to averages over all eigenstates which  become degenerate with the GS. It then provides  a more real  picture of the actual correlations in this limit if individual quasidegenerate  levels cannot be resolved.
\vspace*{-.5cm}

\subsection{Ground State} \label{sec:GS}
In the natural orbital sp basis set, the GS of many molecules (such as the water molecule) approximately adopts the form
\begin{equation}
    \ket{\Psi} \approx C^\dagger_{\rm active}\,C^\dagger_{\rm core} \ket{0},
    \label{eq:core_act}
\end{equation}
where 
$C_{\rm core}^\dagger = \prod_{k = 0}^{n_{\rm core}} c^\dagger_k c^\dag_{\bar{k}}$ and $n_{\rm core}$ is the number of orbitals that are considered to be part of the core i.e., the number of natural orbitals with occupation number $1$. 
Splitting the sp space into core and active parts is not strictly exact, since  typically, even the highest occupation numbers are not exactly 1, but it is a good approximation for the levels with highest occupancy, and in this work  almost no detail is lost with it, with one notable exception that will be discussed later.

For the water molecule GS, $n_{\rm core} = 3$ along the whole dissociation curve, with the lowest occupation number within the core subspace being $\approx 0.9987$. 
The quality of the frozen-core approximation improves towards the dissociation limit, where the natural orbitals  become coincident with the RHF orbitals, and Eq.\ \eqref{eq:core_act} becomes exact (within the limits of the minimal sp basis set employed). In this limit, the core is formed by the $1s$, $2s$ and $2p_z$ orbitals centered around the O atom; the latter is perpendicular to the molecular plane.
The remaining orbitals form the active space, with occupation numbers 1/2.
These include the $2p_y$ orbital, centered around the O atom, 
perpendicular to the line that connects the two H atoms, the symmetric and antisymmetric linear combinations of the two $1s$ orbitals centered around the $H$ atoms, and the $2p_x$ orbital, which is parallel to the line that connects the H atoms. We label these orbitals from 0 to 6, in the order they were presented above.

\begin{figure}
    \begin{center}
    \includegraphics[width=.95\linewidth]{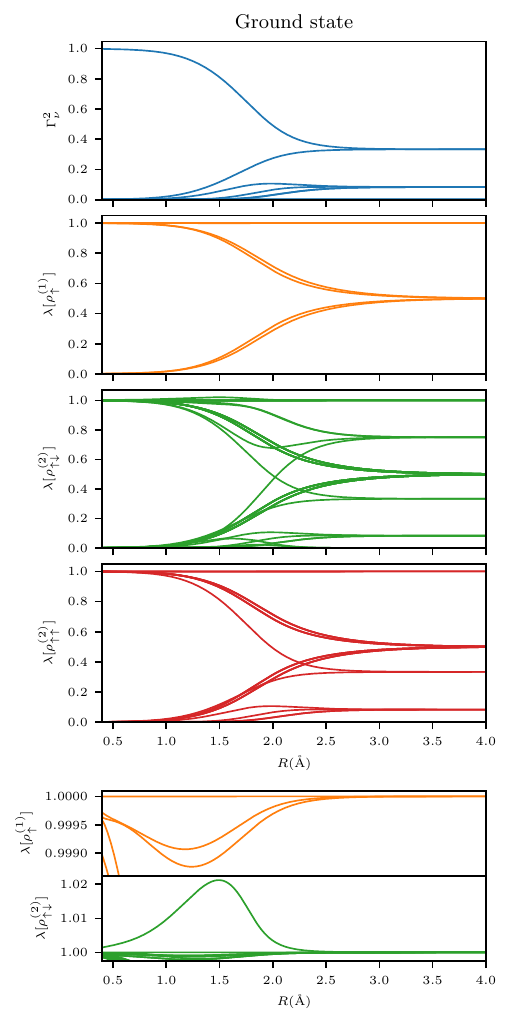}
    \vspace*{-0.5cm}
        \caption{\label{fig:dms_gs} Eigenvalues of the total spin-up reduced state $\rho_{\uparrow}$  (identical with those of $\rho_{\downarrow}$), top panel, and those of the blocks of the one- and two- body reduced density matrices (next three panels),  for the GS, as a function of the O--H distance.  In the dissociation limit  they approach the rational values summarized in Table \ref{tab:eigenvalues_dms}. The bottom panel shows a blow up of the largest eigenvalues of $\rho^{(1)}_\uparrow$ and $\rho^{(2)}_{\uparrow\downarrow}$. }
    \end{center}
\end{figure}

In Fig.\ \ref{fig:dms_gs}, we show the eigenvalues of $\rho_\uparrow$, $\rho^{(1)}_\uparrow$, $\rho^{(2)}_{\uparrow\uparrow}$, and $\rho^{(2)}_{\uparrow\downarrow}$ as a function of $R$ for the GS. 
They evolve from approximately $0$ or $1$ for low $R$, implying that the system can be approximated in this sector by a SD, to fractional numbers, that are detailed in Table \ref{tab:eigenvalues_dms}. They satisfy ${\rm Tr}\,\rho_{\uparrow}=1$, ${\rm Tr}\,\rho^{(1)}_{\uparrow}=5$, ${\rm Tr}\,\rho^{(2)}_{\uparrow\uparrow}=10$ and ${\rm Tr}\,\rho^{(2)}_{\uparrow\downarrow}=25$ $\forall\,R,\beta$.  Results for those of $\rho_{\downarrow}$, $\rho^{(1)}_{\downarrow}$ and $\rho^{(2)}_{\downarrow\downarrow}$ are fully identical in this system with those for the corresponding up RDMs when $M_S=0$.

\begin{table}
\begin{tabular}{c|c|llllll}
\hline
\multirow{4}{*}{$\;\;\ket{\Psi_0}\;\;$}             & $\rho^{(1)}_\uparrow$               & $\frac{1}{2}\;(4)$  & $1\;(3)$            &                      &                     &                     & \\        
\multicolumn{1}{r|}{\multirow{4}{*}{}}                  & $\;\rho^{(2)}_{\uparrow\uparrow}$\;     & $\frac{1}{12}\;(4)$ & $\frac{1}{3}\;(2)$  & $\frac{1}{2}\;(12)$  & $1\;(3)$            &                     & \\        
\multicolumn{1}{r|}{\multirow{4}{*}{}}                  & $\rho^{(2)}_{\uparrow\downarrow}$   & $\frac{1}{12}\;(4)$ & $\frac{1}{3}\;(2)$  & $\frac{1}{2}\;(24)$  & $\frac{3}{4}\;(4)$  & $1\;(9)$            & \\        
\multicolumn{1}{r|}{\multirow{4}{*}{}}                  & $\rho_\uparrow$                   & $\frac{1}{12}\;(4)$ & {$\frac{1}{3}\;(2)$\vspace*{0.05cm}}  &                      &                     &                     & \\        
\hline
\multirow{4}{*}{$\;\;\rho_0(\beta)\;\;$} & $\rho^{(1)}_\uparrow$               & $\frac{1}{2}\;(2)$  & $\frac{2}{3}\;(3)$  & $1\;(2)$             &                     &                     & \\        
\multicolumn{1}{r|}{\multirow{4}{*}{}}                  & $\rho^{(2)}_{\uparrow\uparrow}$     & $\frac{1}{4}\;(7)$  & $\frac{5}{12}\;(3)$ & $\frac{1}{2}\;(4)$   & $\frac{2}{3}\;(6)$  & $1\;(1)$            & \\        
\multicolumn{1}{r|}{\multirow{4}{*}{}}                  & $\rho^{(2)}_{\uparrow\downarrow}$   & $\frac{1}{4}\;(2)$  & $\frac{1}{3}\;(6)$  & $\frac{5}{12}\;(12)$ & $\frac{1}{2}\;(11)$ & $\frac{2}{3}\;(12)$ & $1\;(4)$\\
\multicolumn{1}{r|}{\multirow{4}{*}{}}                  & $\rho_\uparrow$                   & $\frac{1}{12}\;(9)$ & $\frac{1}{4}\;(1)$  &                      &                     &                     & \\        

\end{tabular}
\caption{\label{tab:eigenvalues_dms} Eigenvalues of the one- and two-body RDM blocks, as well as the total $\uparrow$ DM,  for the GS  $\ket{\Psi_0}$ and the thermal state \eqref{eq:thermal}, with $\beta = 1000 E_h^{-1}$, in the dissociation limit.}
\end{table}

We focus first on the eigenvalues of the reduced state $\rho_\uparrow$, i.e. the entanglement spectrum associated to the up-down partition of $\ket{\Psi}$, which are just the square of the Schmidt coefficients $\Gamma_\nu$ \eqref{Scd}. For low $R$ it has nearly rank one, in agreement with  the GS being here close to a SD with definite $N_\uparrow$ and $N_\downarrow$ electrons (though deviations are still visible, see bottom panel). However, as $R$ increases, further nonzero eigenvalues emerge, having  essentially $\binom{4}{2}=6$ non-zero eigenvalues 
(the remaining ones are less than $10^{-3}$), in 
agreement with the core of six electrons.  In the dissociation limit, they collapse into two values: 1/3, with degeneracy 2, and 1/12, with degeneracy 4, corresponding to the limit Schmidt decomposition of $\ket{\Psi}$,
\begin{equation}
    \begin{split}
        \!\!\!\!\!\ket{\Psi} =  & \left[ \sqrt{\tfrac{1}{12}} (C^\dagger_{34} C^\dagger_{\bar 5\bar 6} + C^\dagger_{5 6} C^\dagger_{\bar 3 \bar 4} - C^\dagger_{35}C^\dagger_{\bar 4 \bar 6} - C^\dagger_{46} C^\dagger_{\bar 3 \bar 5})  
        \right . \\
            & - 
            \left .\sqrt{\tfrac{1}{3}} (C^\dagger_{36} C^\dagger_{\bar 4\bar 5} + C^\dagger_{45} C^\dagger_{\bar 3 \bar 6}) \right ]C^\dagger_{\rm core}\ket{0} ,
    \end{split}
    \label{eq:psi_dissoc_up_down}
\end{equation}
where $C^\dag_{ij}:=c^\dag_i c^\dag_j$ and $C^\dag_{\rm core}=\prod_{k=0}^2 C^\dag_{k\bar k}=C^\dag_{012}C^\dag_{\bar 0 \bar 1\bar 2}$.  
Each $\Gamma_\nu$, and its associated normal $A^\dagger_{\nu}$,  $B^\dagger_{\bar\nu}$ operators are specified in Table \ref{tab:schmidtud}.  Here $\ket{\nu}$ and $\ket{\bar \nu}$, $\nu = 0, \ldots, 5$ can be taken as SDs, entailing that the GS reduced states $\rho_\uparrow$ and $\rho_\downarrow$ become in this limit convex mixtures of orthogonal SDs and are hence ``separable''. This is strictly true only in the dissociation limit.
The justification of the limit \eqref{eq:psi_dissoc_up_down} is provided in the next subsection 
[Eqs.\ \eqref{Astates}, \eqref{asint}]. 

\begin{table}
\begin{tabular}{lrrrrrr}
\multicolumn{1}{c|}{$\Gamma_\nu$}          &                         \multicolumn{2}{c|}{$\sqrt{\frac{1}{3}}$}                          &                                                             \multicolumn{4}{c}{$\sqrt{\frac{1}{12}}$}\\                                                            
\hline
\multicolumn{1}{l|}{$A^\dagger_{\nu_\ps}$} & \multicolumn{1}{r|}{$\,C_{36}^\dagger\,$}        & \multicolumn{1}{r|}{$\,-C_{45}^\dagger\,$}      & \multicolumn{1}{r|}{$\,C_{34}^\dagger\,$}       & \multicolumn{1}{r|}{$\,C_{56}^\dagger\,$}       & \multicolumn{1}{r|}{$\,-C_{35}^\dagger\,$}      & $\,C_{46}^\dagger\,$\\       
\multicolumn{1}{l|}{$B^\dagger_{\nu_-\!\!}$} & \multicolumn{1}{r|}{$\,-C_{\bar 4 \bar 5}^\dagger\,$} & \multicolumn{1}{r|}{$\,C_{\bar 3 \bar 6}^\dagger\,$} & \multicolumn{1}{r|}{$\,C_{\bar 5 \bar 6}^\dagger\,$} & \multicolumn{1}{r|}{$\,C_{\bar 3 \bar 4}^\dagger\,$} & \multicolumn{1}{r|}{$\,C_{\bar 4 \bar 6}^\dagger\,$} & $\,-C_{\bar 3 \bar 5}^\dagger\,$\\
\end{tabular}
\caption{Square root of the eigenvalues of $\rho_{\uparrow}$ (labeled as $\Gamma_\nu$), which are the coefficients of the Schmidt decomposition \eqref{Scd}, and the associated normal creation operators,  
for the GS in the dissociation limit. They are paired in such a way that they form the up-down Schmidt expansion \eqref{eq:psi_dissoc_up_down} of the active part of the GS in this limit. \label{tab:schmidtud}}
\end{table}

We now focus on the analysis of the one- and two-body RDMs. In the case of $\rho^{(1)}_{\uparrow}$, second panel from top in Fig.\ \ref{fig:dms_gs},  it has essentially (see below) just $4$ ``active'' eigenvalues $\in(0,1)$ (nondegenerate but  coming in almost degenerate pairs which sum to $1$), which approach $1/2$ in the dissociation limit, as can be derived  from  Eq.\ \eqref{eq:psi_dissoc_up_down}. As stated above, in this limit natural orbitals are just the RHF orbitals, with $0,1,2$ fully occupied (core) and $3,4,5,6$ approaching all half occupation $1/2$.   This shows that the present exact GS departs significantly from a SD as $R$ increases, as already indicated by the eigenvalues $\Gamma_\nu$ of the total $\rho_{\uparrow}$ and $\rho_{\downarrow}$ RDMs.  

In the case of  $\dab$ (third panel),
just as in the previous case, its eigenvalues range from 0  and 1 at equilibrium distance, to fractional numbers in the dissociation limit, except for the largest eigenvalue $\lambda_{\rm max}[\dab]$, that is slightly greater than 1 for all $R$, indicating weak pairing effects,  and  approaches 1 when $R \rightarrow \infty$ or $R$ is low. 
It attains its maximum value of $\sim 1.021$ at $R \approx 1.5$\AA~ (as clearly seen  in the bottom panel of Fig.\ \ref{fig:dms_gs}). The associated eigenvector can be written approximately as $A_{\nu_{\rm max}}^\dagger \approx 0.5(C^\dagger_{1\bar{1}} + C^\dagger_{2\bar{2}}) + 0.45(C^\dagger_{3\bar{3}} + C^\dagger_{4\bar{4}}) - 0.16 (C^\dagger_{5\bar{5}} + C^\dagger_{6\bar{6}})$ at the maximum. 
The reason two ``core'' electrons (labelled as 1 and 2 here) are involved in this paired state is due to their occupation numbers being slightly lower than 1, i.e. they are not exactly core electrons, as seen in the bottom panel of Fig.\ \ref{fig:dms_gs}.
If the occupation numbers of these orbitals were set exactly to 1 (i.e., only the ``active'' part of the Hamiltonian diagonalized), then $\lambda_{\rm max}[\dab] = 0.9971<1$, meaning that this weak pairing effect would be lost. 

In order to  understand more clearly the eigenvalues and eigenvectors of $\dab$ (which are up-down pairs) in  the dissociation limit,  an alternative representation of the GS \eqref{eq:core_act}-\eqref{eq:psi_dissoc_up_down}  is 
\begin{equation}
|\Psi\rangle=\tfrac{1}{\sqrt{3}}(C_{3{\bar 4}_{\ps}}^\dagger C_{5\bar{6}_{\ps}}^\dagger - C_{3\bar{5}_{\ps}}^\dagger C_{4\bar{6}_{\ps}}^\dagger)\,C^\dagger_{\rm core}|0\rangle\label{eq:psi_pairs_ud}\,,\end{equation}
where \begin{equation}
        C^\dagger_{i\bar j_{\pm\!}}:= \tfrac{1}{\sqrt{2}}(c_i^\dagger c_{\bar{j}}^\dagger \pm c_j^\dagger c_{\bar{i}}^\dagger)\,,
    \label{eq:Cops}
\end{equation} 
creates Bell-type pairs. 
Here the two-particle states created by the $C^\dagger_{k\bar{k}}$ operators that form $C^\dagger_{\rm core}$ are obviously eigenvectors of  $\rho^{(2)}_{\uparrow\downarrow}$; they are  associated with nine eigenvalues 1. Besides,  $\rho^{(2)}_{\uparrow\downarrow}$ has a set of 24 eigenvalues 1/2 that come from the core-active blocks $\mathbbm 1^c_\uparrow \otimes \rho_\downarrow^{(1)}$ and $\rho^{(1)}_\uparrow \otimes \mathbbm 1^c_\downarrow$.

The remaining eigenvalues of $\dab$ stem from the Bell-type entangled pairs \eqref{eq:Cops}, with 
\begin{equation}
    \langle C^\dag_{i\bar j_s} C_{k\bar l_{s'}}\rangle=\delta_{ik}\delta_{jl}\delta_{ss'}\lambda_{ijs}\,.
\end{equation}
Explicitly, the active operators involved in the representation \eqref{eq:psi_pairs_ud}, $C^\dagger_{3\bar{4}_{\ps}}, C^\dagger_{5\bar{6}_{\ps}}, C^\dagger_{3\bar{5}_{\ps}}$, and $C^\dagger_{4\bar{6}_{\ps}}$,  lead to four eigenvalues 3/4. 
The $1/3$ eigenvalues come from the operators $C^{\dagger}_{3\bar{6}_{\ms}}$ and $C^{\dagger}_{4\bar{5}_{\ms}}$, and the 1/12 ones come from $C^\dagger_{3\bar{4}_{\ms}}, C^\dagger_{5\bar{6}_{\ms}}, C^\dagger_{3\bar{5}_{\ms}}$, and $C^\dagger_{4\bar{6}_{\ms}}$.
 
As can be shown after some algebra, they yield an alternative representation of the active part in \eqref{eq:psi_pairs_ud}: 
\begin{equation}
C^\dagger_{\rm active} = \tfrac{2}{\sqrt{3}}C^\dagger_{3\bar{6}_{\ms}}C^\dagger_{4\bar{5}_{\ms}}-\tfrac{2}{\sqrt{12}}(C^\dagger_{3\bar{4}_{\ms}} C^\dagger_{5\bar{6}_{\ms}} - C^\dagger_{3\bar{5}_{\ms}} C^\dagger_{4\bar{6}_{\ms}})\,,
\end{equation}
(the scalar coefficients are written as to show that they are proportional to the eigenvalues of $\rho^{(2)}_{\uparrow\downarrow}$). 
It becomes apparent then that the eigenvectors of $\rho^{(2)}_{\uparrow\downarrow}$ allow for a different Schmidt-like decomposition of the active part of $\ket{\Psi}$, $\ket{\Psi_{\rm active}} = C^\dagger_{\rm active} \ket{0}$, into two subsystems of $\uparrow \downarrow$ pairs,
\begin{equation}
\ket{\Psi^{\rm active}} = \frac{1}{N_\uparrow N_\downarrow} \sum_\nu \sigma_\nu A^\dagger_\nu B^\dagger_\nu \ket{0},
\label{eq:schmidt_2b}
\end{equation}
where $\sigma_\nu = \sqrt{\lambda_\nu[\dab]}$, and $A^\dagger_\nu$ and $B^\dagger_\nu$ are normal operators that create a single $\uparrow \downarrow$ electron pair, 
in agreement with the general expansions introduced in \cite{GDR.21,CR.24}. 
Their explicit forms are shown in Table \ref{tab:schmidt2ud}.

\begin{table}
\begin{center}
    \begin{footnotesize}
\begin{tabular}{crrrrrrrrrr}
\multicolumn{1}{c|}{$\sigma_\nu$}    &                                                                                    \multicolumn{4}{c|}{$\sqrt{\frac{3}{4}}$}                                                                                    &                               \multicolumn{2}{c|}{$\sqrt{\frac{1}{3}}$}                               &                                                                         \multicolumn{4}{c}{$\sqrt{\frac{1}{12}}$}\\                                                                         
\hline
\multicolumn{1}{c|}{$A^\dagger_\nu$} & \multicolumn{1}{r|}{$C^\dagger_{3\bar{4}_{\ps}}$} & \multicolumn{1}{r|}{$C^\dagger_{5\bar{6}_{\ps}}$} & \multicolumn{1}{r|}{-$C^\dagger_{3\bar{5}_{\ps}}$} & \multicolumn{1}{r|}{$C^\dagger_{4\bar{6}_{\ps}}$}  & \multicolumn{1}{r|}{$C^\dagger_{3\bar{6}_{\ms}}$} & \multicolumn{1}{r|}{$C^\dagger_{4\bar{5}_{\ms}}$} & \multicolumn{1}{r|}{$-C^\dagger_{3\bar{4}_{\ms}}$} & \multicolumn{1}{r|}{$C^\dagger_{5\bar{6}_{\ms}}$}  & \multicolumn{1}{r|}{$C^\dagger_{3\bar{5}_{\ms}}$} & $C^\dagger_{4\bar{6}_{\ms}}$\\
\multicolumn{1}{c|}{$B^\dagger_\nu$} & \multicolumn{1}{r|}{$C^\dagger_{5\bar{6}_{\ps}}$} & \multicolumn{1}{r|}{$C^\dagger_{3\bar{4}_{\ps}}$} & \multicolumn{1}{r|}{$C^\dagger_{4\bar{6}_{\ps}}$}  & \multicolumn{1}{r|}{$-C^\dagger_{3\bar{5}_{\ps}}$} & \multicolumn{1}{r|}{$C^\dagger_{4\bar{5}_{\ms}}$} & \multicolumn{1}{r|}{$C^\dagger_{3\bar{6}_{\ms}}$} & \multicolumn{1}{r|}{$C^\dagger_{5\bar{6}_{\ms}}$}  & \multicolumn{1}{r|}{$-C^\dagger_{3\bar{4}_{\ms}}$} & \multicolumn{1}{r|}{$C^\dagger_{4\bar{6}_{\ms}}$} & $C^\dagger_{3\bar{5}_{\ms}}$\\
\end{tabular}
\end{footnotesize}
\caption{Square root of the eigenvalues of $\rho^{(2)}_{\uparrow\downarrow}$ (labeled as $\sigma_\nu$), and pair creation operators that applied to the vacuum, create its eigenvectors, in the dissociation limit. They are paired in such a way that they form expansion \eqref{eq:schmidt_2b} of the active part of the GS in the dissociation limit. \label{tab:schmidt2ud}}
\end{center}
\end{table}

Finally, regarding $\daa$, we recall that, since there are three core electrons, it can be written as $\rho^{(2)}_{\uparrow\uparrow} = \rho^{(2)\,a}_{\uparrow\uparrow} \oplus (\rho_\uparrow^{(1)\,a}\otimes \mathbbm 1^c_{\downarrow}) \oplus (\mathbbm 1^c_\uparrow \otimes \rho_\downarrow^{(1)\,a}) \oplus (\mathbbm 1^c_\uparrow \otimes \mathbbm 1^c_\downarrow)$, with $a$ and $c$ standing for active and core, respectively.
Since $C^\dagger_{\rm active}$ contains only two pairs of electrons, $\rho_{\uparrow\uparrow}^{(2)\,a}$ has the same structure as $\rho_\uparrow$, with the same eigenvalues. 
Indeed, an inspection of Fig. \ref{fig:dms_gs} confirms that the eigenvalues of $\rho^{(2)}_{\uparrow\uparrow}$ are those of $\rho_\uparrow$, plus those of $\rho_\uparrow^{(1)}$, with the latter appearing twice.

\begin{figure}
    \begin{center}
        \includegraphics[width=.95\linewidth]{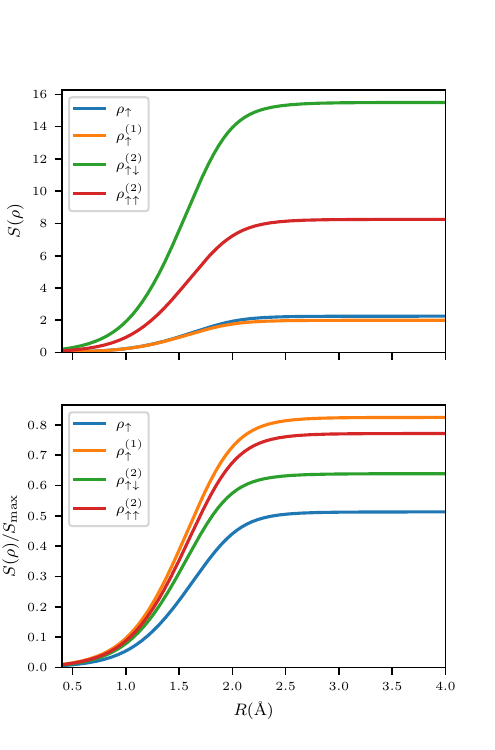}
        \vspace*{-0.5cm}        
        \caption{\label{fig:gs_entropies} Von Neumann entropies of the RDMs $\rho_{\uparrow}$, 
        and the one- and two-body blocks $\rho^{(1)}_{\uparrow}$, $\rho^{(2)}_{\uparrow\downarrow}$ and $\rho^{(2)}_{\uparrow\uparrow}$, for the GS as a function of the 
        O--H distance. Top: Plain  entropies. Bottom: Normalized entropies $S/S_{\rm max}$. We set $\log\equiv\log_2$ throughout the manuscript.}
    \end{center}
\end{figure}

The von Neumann entropies computed from the eigenvalues of the RDMs discussed above are shown in the top panel of Fig.\ \ref{fig:gs_entropies}. They measure the pertinent subsystem-rest entanglement. 
The bottom panel shows the same entropies, but divided by the maximum values they can here attain, namely,
$\log_2 \binom{d/2}{2}$ for $\rho_\uparrow$, 
$N_{\uparrow}\log_2 \frac{d}{2N_{\uparrow}}$ for $\rho^{(1)}_\uparrow$, 
$\binom{N_{\uparrow}}{2}\log_2[\binom{d/2}{2}/\binom{N_\uparrow}{2}]$ for $\rho^{(2)}_{\uparrow\uparrow}$ (similarly for $\rho^{(1)}_{\downarrow}$, $\rho^{(2)}_{\downarrow\downarrow}$) and $N_\uparrow N_{\downarrow}\log_2\frac{(d/2)^2}{N_{\uparrow}N_{\downarrow}}$ for $\rho^{(2)}_{\uparrow\downarrow}$, 
with $d=14$, $N_\uparrow=N_{\downarrow}=5$. 
They all increase monotonically with increasing $R$, 
 reaching well-defined limits that can be computed analytically from the rational eigenvalues shown in Table \ref{tab:eigenvalues_dms}.
 When compared to their saturation values, the one-body entanglement entropy $S(\rho^{(1)}_\uparrow)$ leads to the highest  ratio, 
followed by the two-body entropies $S(\rho^{(2)}_{\uparrow\uparrow})$ and $S(\rho^{(2)}_{\uparrow\downarrow})$. 

\begin{figure}
    \begin{center}
        \includegraphics[width=1\linewidth]{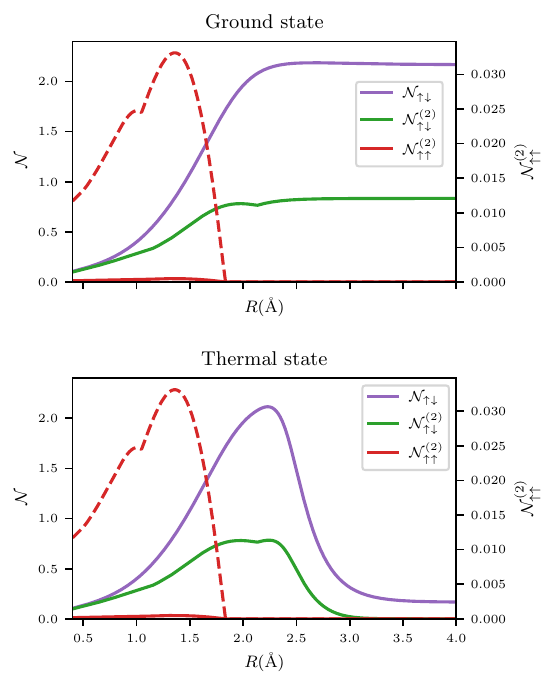}
        \caption{\label{fig:negativity} Negativities of the GS (top) and the thermal state  \eqref{eq:thermal} (bottom) at $\beta=1000 E_h^{-1}$, as a function of the O--H distance.  Both the  full, Eq.\ \eqref{neg}, and two-body, Eq.\ \eqref{neg2}, up-down negativities are depicted, together with the  reduced up-up negativity \eqref{Neg4}. The latter is also  shown in a different scale (dashed line, right axis) to improve visibility.\vspace*{-.35cm}}
    \end{center}
\end{figure}

The up-down negativities ${\cal N}_{\uparrow\downarrow}$, ${\cal N}_{\uparrow\downarrow}^{(2)}$ associated with the whole $\rho$ and $\rho^{(2)}_{\uparrow\downarrow}$ respectively, and the negativity ${\cal N}^{(2)}_{\uparrow\uparrow}$ of $\rho^{(2)}_{\uparrow\uparrow}$,  Eq.\ \eqref{Neg4}, are shown in the top panel of Fig.\ \ref{fig:negativity} and reveal a more interesting picture: Both $\mathcal{N}_{\uparrow\downarrow}$ and $\mathcal{N}^{(2)}
_{\uparrow\downarrow}$  approach a constant value for $R \rightarrow \infty$. For a pure state the total up down negativity ${\cal N}_{\uparrow\downarrow}$ takes the value \eqref{negp}  and is an alternative  measure of the total up-down entanglement previously measured by $S(\rho_{\uparrow})$, indicating the deviation from a product up-down state. Hence, for $R\rightarrow\infty$  it approaches the value $13/6$, with $\rho^{t_\downarrow}$ having 15 negative eigenvalues $\Gamma_\nu\Gamma_{\nu'}$ for $\nu<\nu'$,  i.e., $-1/3$ (1), $-1/6$ (8) and $-1/12$ (6), according to Table \ref{tab:schmidtud}. 

On the other hand, ${\cal N}^{(2)}_{\uparrow\downarrow}$ is an indicator of the deviation of $\rho^{(2)}_{\uparrow\downarrow}$ from a convex mixture of product densities $\rho^{(1)}_{\uparrow}\otimes\rho^{(1)}_{\downarrow}$, i.e.\ of its ``inner'' entanglement. 
The number of negative eigenvalues of $({\dab})^{t_\downarrow}$ ranges here from 9 for $R = 0.4$\AA, to only one for $R \rightarrow \infty$, but $\mathcal{N}_{\uparrow\downarrow}^{(2)}$ increases with $R$, since this negative eigenvalue becomes larger in magnitude, reaching $-\frac{5}{6}$ in the dissociation limit. 

In contrast,  $\mathcal{N}_{\uparrow\uparrow}^{(2)}$,  an indicator of the deviation of $\rho^{(2)}_{\uparrow\uparrow}$  from a convex mixture of real SDs, attains its maximum at around $1.25$\AA, and then falls to 0.
This striking difference with the previous negativities arises from the fact that, as seen from Eq.\  \eqref{eq:psi_pairs_ud}, the entanglement in $\ket{\Psi}$ comes mostly from entangled $\uparrow \downarrow$ pairs. Moreover, in the dissociation limit we have seen that the total up RDM $\rho_{\uparrow}$ approaches  a convex mixture of SDs, containing then just ``classical''-type correlations, thus  implying ${\cal N}^{(2)}_{\uparrow\uparrow}=0$. 
Notice that although $S(\rho_{\uparrow\uparrow}^{(2)})$ involves the same DM as $\mathcal{N}_{\uparrow\uparrow}^{(2)}$, the former represents the entanglement of an $\uparrow\uparrow$ pair of electrons with the rest of the system,  
whereas $\mathcal{N}_{\uparrow\uparrow}^{(2)}$ measures the ``inner'' entanglement of the pair. 
We also note that what could be called ``static'' correlation 
is here captured by both $\nud$ and $\nud^{(2)}$, whereas $\nuu^{(2)}$ stems here from essentially ``dynamic'' correlations near the equilibrium distance, where there are small but non-zero deviations from a SD. 

\begin{figure}
    \begin{center}
        \includegraphics[width=.95\linewidth]{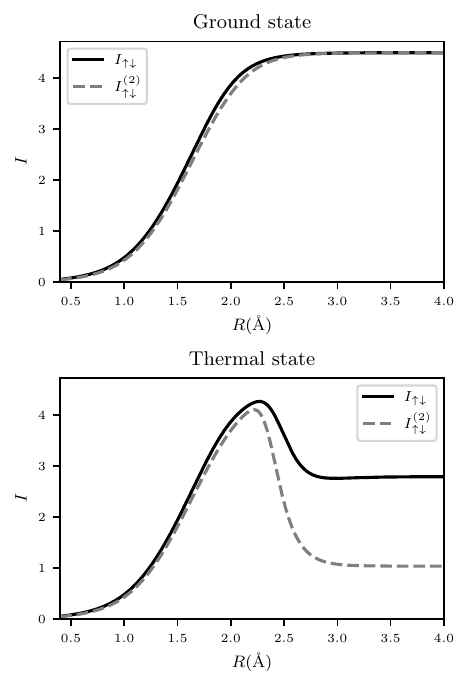}
        \vspace*{-0.5cm}
        \caption{\label{fig:IM_thermal} Total and two-body up-down mutual informations for the GS  and the thermal state \eqref{eq:thermal}  at $\beta = 1000E_h^{-1}$. The analytical limits for $R\rightarrow\infty$ are $2(\frac{2}{3}+\log_2 3)$ in the GS (for both $\iud$ and $\iud^{(2)}$) and $2 +\frac{1}{2}{\rm log}_2 3$ and $\frac{1}{2}(-37 + 10\,{\rm log}_2 15)$ respectively in the thermal state (see text).}
    \end{center}
\end{figure}

The $\uparrow\downarrow$ mutual information \eqref{Iud} and the reduced mutual information \eqref{Imud} for the GS are shown in the top panel of Fig.\ \ref{fig:IM_thermal}. In the pure case, $I_{\uparrow\downarrow}$ is just twice the total up-down entanglement entropy,  
hence reaching the value $\frac{4}{3}+2\log_2 3\approx 4.503$ in the dissociation limit, according to Table \ref{tab:schmidtud}. 
One notable feature of $I^{(2)}_{\uparrow\downarrow}$  is that it reproduces here the former almost exactly, being just slightly smaller for finite $R$ and coinciding exactly in the dissociation limit. This is not a general feature of these correlation measures, but in the present case  it is related to the fact that $\ket{\Psi}$ can be written as linear combinations of products of $\uparrow\downarrow$ electron pairs in this limit. 

\subsection{Degeneracy in the dissociation limit and thermal state correlations \label{IIIC}}

In the previous section we discussed the dissociation limit of the GS in detail.
We now focus on the space of states that become degenerate with the GS in the dissociation limit, within the $M_S=0$ subspace. This space can be smoothly captured for increasing $R$ through the thermal state $\rho_0(\beta)$ at very low temperatures, such that just these  states acquire nonvanishing weight in the dissociation limit.  
Aside from the spin interaction, when $R \rightarrow \infty$, the atoms do not interact, since $h_{ij} \rightarrow 0$ if spin orbitals $i$ and $j$ are centered around different atoms, and $R_{ij,kl} \rightarrow 0$ unless $i, j, k, $ and $l$ are centered around the same atom. 
It is reasonable then to expect that the Hamiltonian allows for eigenstates of the form $\ket{\Psi} = C^\dagger_{\rm O} C^\dagger_{\rm H_A} C^\dagger_{\rm H_B} \ket{0}$, where $C^\dagger_{\rm X}$ is a $N_{\rm X}$-particle creation operator, that creates the exact GS of the isolated atom X.
As in the previous section, we limit here the discussion to a minimal basis set, which contains one $1s$ orbital for each H atom, and $1s$, $2s$, and $2p$ orbitals for the O atom, and denote the first two orbitals as $\phi_{H_A}$ and $\phi_{H_B}$, and the orbitals centered on the O atom as $\phi_{1s}, \phi_{2s}$ and $\phi_{2p_\mu}$, $\mu=x,y,z$.  
It is essentially the limit of the same basis set used in the previous section, but with  H orbitals fully localized on each H atom.

\subsubsection{GS subspace in the dissociation limit}
Considering first the O atom, its GS is a triplet, i.e.,  states $^3P$ from a sp  configuration $1s^2 2s^2 2p^4$, hence $9$-fold degenerate (the spin-orbit coupling is here neglected). The $\phi_{1s}$ and $\phi_{2s}$ orbitals are doubly occupied, while the $2p$ orbitals are partially occupied by four electrons.
The $(N_{2p\uparrow}, N_{2p\downarrow})$ occupation numbers in the $p$ orbitals 
can be either $(1, 3)$, $(2, 2)$, or $(3, 1)$, leading to $M^O_S = -1, 0, 1$, respectively. 
In the $(3, 1)$ case, the $\downarrow$ electron  is located in one of the three $2p$ orbitals, 
leading to 3 degenerate states with $M^O_S = 1$, which are SDs. 
By interchanging $\uparrow$ and $\downarrow$, it is apparent that the $(1, 3)$ case leads to three similar degenerate states with $M^O_S = -1$, leading to a total of 6 degenerate states with $|M^O_S| = 1$.
In contrast, in the $M^O_S = 0$ case, while there are in principle 9 possible states, 
just three of them correspond to the $^3P$ subspace 
(the other ones belong to the $^1D$ and $^1S$ subspaces), in which one of the $2p$ orbitals is doubly occupied, and the other two form a Bell pair $C^\dagger_{i\bar{j}_\ms}\ket{0}$ (where we have used $i$ and $j$ to refer to any two of the three $2p$ orbitals). This leads again to three degenerate states (one for each choice of $i$ and $j$) which now are not SDs. 

Regarding the two additional electrons occupying the $\phi_{H_A}$,  $\phi_{H_B}$ orbitals, when $(N_{2p\uparrow}, N_{2p\downarrow})=(3,1)$ or $(1,3)$,  
both $\phi_{{\rm H}_A}$ and $\phi_{{\rm H}_B}$ are occupied with one $\downarrow$ ($\uparrow$) electron  in the first (second) case for total $M_S=0$, 
 leading to six $M_S=0$ SD eigenstates of  the whole system. 
On the other hand, in the $(2,2)$ configuration, 
either $\phi_{\mathrm{H}_A}$ or $\phi_{\mathrm{H}_B}$ is occupied with one $\uparrow$ electron, the other  with a $\downarrow$ one,  for total $M_S=0$. This yields two distinct H configurations for each of the three $^3P$ $M_S^O=0$ non-SD O eigenstates,  leading to six non-SD $M_S=0$ degenerate eigenstates of the whole system. 

We emphasize that the simple $M_S=0$ states described above, summarized in Eqs.\ \eqref{KS}--\eqref{Astates} below,  are eigenstates of $H$ in the dissociation limit only.
They are not total spin $S^2$ eigenstates, since they involve singly occupied spatial orbitals (having only one $\uparrow$ or one $\downarrow$ electron), but they can be linearly combined to form total spin eigenstates; three of them are singlets ($S=0$), six of them are triplets ($S=1$), and the remaining three are quintuplets ($S=2$). It is also verified in Fig.\  \ref{fig:energies} that the lowest $12$ levels of the molecule for $R\agt 1.6${\AA} comprise precisely three $S=0$, six $S=1$ and three $S=2$ nondegenerate levels (not counting their $2S+1$ degeneracy), whose energy difference becomes vanishingly small for $R\agt 3${\AA}, and which approach definite spin linear combinations of previous 12 states for $R\rightarrow\infty$. 

\begin{figure}
    \begin{center} \includegraphics[width=0.95\linewidth]{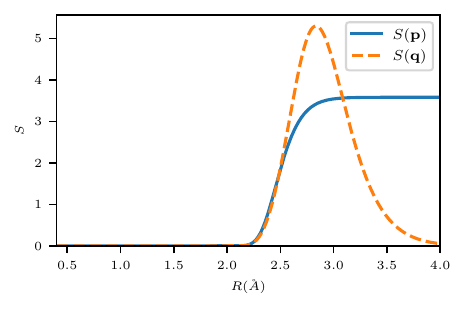}
    \vspace*{-0.5cm}
    \caption{\label{fig:thermal_entropy} Von Neumann entropy of the thermal distributions  
    $\mathbf{p} = \mathbf{q}/{\rm Tr}(\mathbf{q})$ (normalized) and $\mathbf{q}=\{e^{-\beta (E_n-E_0)}\}$, at $\beta = 1000 E_h^{-1}$, as a function of the O--H distance.\vspace*{-.5cm}}
    \end{center}
\end{figure}

\subsubsection{Thermal state}
We now consider the thermal state \eqref{eq:thermal}. 
We  set $\beta=1000 E_h^{-1}$ in all cases,  with $E_h\approx 27.211 eV$ the Hartree energy, so that only  states that become degenerate with the GS contribute to $\rho_0(\beta)$ near the dissociation limit. The GS energy in this limit is $-74.737 E_h$ and the gap between the lowest 12 states and the next band is $0.095 E_h$ (at $R=4$\AA\ the energy difference between the highest and lowest of these 12  states is just  $6\times 10^{-6}E_h$). 

For small $R$, $\rho_0(\beta)$ is essentially the same as the GS, since for $R \leq 2.2$\AA, it amounts to more than 99.9\% of the ensemble, but as $R$ increases they start to differ.
Near the dissociation limit, $\rho_0(\beta)$ becomes at this $\beta$ proportional to a projector onto the  $M_S=0$ GS subspace, i.e., that spanned by the previous $12$ $M_S=0$ eigenstates almost degenerate with the  GS at large but finite $R$.

This is verified in Fig.\ \ref{fig:thermal_entropy}, which shows the von Neumann entropy of the normalized and unnormalized distributions $\mathbf{p}=\{p_n\}$ and $\mathbf{q}=\{q_n\}$ in Eq.\ \eqref{eq:thermal}, as a function of $R$ at fixed previous $\beta$. While $S({\bf p})=S(\rho_0(\beta))$ is the standard thermal entropy, approaching $\log_2 12$ for $R\rightarrow\infty$ ($\rho_0(\beta)$ maximally mixed in the GS subspace at this $\beta$), $S({\bf q})$ 
 represents a measure of its ``complexity'',  indicating  the deviation of $\sum_n e^{-\beta (E_n-E_0)}|\Psi_n\rangle\langle\Psi_n|$ from a projector onto the $M_S=0$ GS subspace, 
 hence vanishing for both small $R$ ($q_n=\delta_{n 0}$) and  large $R$ ($q_n=1$ for $n=0,\ldots,11$ and $0$  otherwise).

Choosing now for this degenerate GS subspace for $R\rightarrow\infty$ a basis of 12 orthogonal states $|K\rangle$, $K = 1, \ldots, 12$, $\langle K|K'\rangle=\delta_{KK'}$, where   the first six $\ket{K}$ states are the SDs described 
in previous subsection and the last  six of them  the remaining states, 
we obtain a ``minimally entangled'' representation of $\rho_0(\beta)$ in this limit,
\begin{equation}        \lim_{{T \rightarrow 0^+}\atop{\!\!R\rightarrow\infty}}\rho_0(\beta) =\tfrac{1}{12}\sum_{K=1}^{12}|K\rangle\langle K|= \tfrac{1}{2} (\rho_{\rm sep} + \rho_{\rm bp})
\label{eq:rho_beta}\,,
\end{equation}
where $\rho_{\rm sep}= \frac{1}{6}\sum_{K=1}^{6}\ket{K}\bra{K}$, $\rho_{\rm bp} = \frac{1}{6}\sum_{K=7}^{12} \ket{K}\bra{K}$, with bp standing for ``Bell pair''  (Eq.\ \eqref{eq:Cops}). 
Explicitly,  
\begin{equation}    |K\rangle=C^\dag_{K_H}C^\dag_{K_{2p}}C^\dag_{\rm core}|0\rangle\,,\label{KS}
\end{equation}
where, labelling the O sp orbitals $\phi_{1s}$, $\phi_{2s}$ and $\phi_{2p_\mu}$ as $0,1$ and $2,3,6$ respectively (as in Sec. \ref{sec:GS}), and  $\phi_{H_{A(B)}}$ as $4,5$, we obtain $C^\dag_{\rm core}=C^\dag_{01\bar 0\bar 1}$ and 
\begin{subequations}\label{Astates}
\begin{equation}\!\!\!\!\!\!C^\dag_{K_{2p}}=\left\{\begin{array}{l}
C^\dag_{236\bar i}
\\ C^\dag_{i\bar 2\bar 3\bar 6}
\end{array}\right.,\;
C^\dag_{K_{H}}=\left\{\begin{array}{l}C^\dag_{\bar 4\bar5}
\\ C^\dag_{45}
\end{array}\right.,\;
\begin{array}{l}K=1,2,3\\K=4,5,6\end{array}
\end{equation}
\begin{equation}\!\!\!\!\!C^\dag_{K_{2p}}=
C^\dag_{i\bar i}
C^\dag_{j{\bar k}_-},\;
C^\dag_{K_{H}}=\left\{\begin{array}{l}C^\dag_{4\bar 5}
\\C^\dag_{5\bar 4} 
\end{array}\right.,
\begin{array}{l}K=7,8,9\\K=10,11,12\end{array}
\end{equation}
\end{subequations}
for $i\neq j\neq k\in\{2,3,6\}$. 
Here $C^\dag_{j\bar k_-}=\frac{c^\dag_jc^\dag_{\bar k}-c^\dag_k c^\dag_{\bar j}}{\sqrt{2}}$ denotes the Bell pair creation operator \eqref{eq:Cops}, which provides the sole quantum correlation in the asymptotic  $\rho_0(\beta)$. 
No linear combination of the three O states $C^\dag_{i\bar i}C^\dag_{j\bar k_-}C^\dag_{\rm core}|0\rangle$ yields a SD (all normalized combinations 
lead in fact to the same eigenvalues  $\frac{1}{2}$ ($\times 4$) and $1$ ($\times 4$)
 of $\rho^{(1)}_{\rm O}$).  
 
Previous states lead  to the effective representation  
\begin{subequations}  \begin{align}
   \rho_{\rm sep} &= \tfrac{1}{2} \rho_{\rm core} \otimes (\rho_{2p}^+\otimes \rho^{\downarrow\downarrow}_{\rm H} +\rho_{2p}^- \otimes \rho^{\uparrow\uparrow}_{\rm H}) ,\label{rsep}\\
    \rho_{\rm bp} &= \tfrac{1}{2}\rho_{\rm core} \otimes\rho_{2p}^0\otimes (\rho_{\rm H}^{\uparrow\downarrow}+\rho_{\rm H}^{\downarrow\uparrow}),
    \label{rbp}
    \end{align}
    \label{rhod}
    \end{subequations}
of the two parts  in \eqref{eq:rho_beta}, where 
\begin{equation}
    \rho_{2p}^\mu = \tfrac{1}{3}\sum_{K=K_\mu}^{K_\mu+2} C_{K_{2p}}^\dagger \ket{0}\bra{0} C_{K_{2p}}, \label{rh}
\end{equation}
are mixed states with $K_\mu=1,4,7$ for $\mu=+,-,0$,  while  $\rho_{\rm H}^{\mu\nu} = C^\dag_{K_{H}}\ket{0}\bra{0} C_{K_H}$ are pure states,  with e.g.\ $K = 1,4,7,10$ for $\mu\nu = \downarrow\downarrow$, $\uparrow\uparrow$, $\uparrow\downarrow$, $\downarrow\uparrow$ respectively. 

We also remark that the $R\rightarrow\infty$ limit \eqref{eq:psi_dissoc_up_down} of the molecule GS (nondegenerate for finite $R$)  is 
just the  $S^2=0$ superposition of the four asymptotic eigenstates $|K\rangle$ which have the same O core $C^\dag_{012\bar 0\bar 1\bar 2}|0\rangle$ as the GS, i.e., $K=1,4,7,10$: 
\begin{equation}
    \ket{\Psi}=-\tfrac{1}{\sqrt{3}}(\ket{1}+\ket{4})+\tfrac{1}{\sqrt{6}}(\ket{7}-\ket{10})\,.\label{asint}   
\end{equation}
  When expanded, this leads explicitly to  expression \eqref{eq:psi_dissoc_up_down}.

\subsubsection{Entanglement and correlation measures}
In the dissociation limit, all eigenstates $|K\rangle$ in \eqref{KS} are ``product'' states regarding the O (core + $2p$) and H atoms.  In addition, $\rho_{\rm sep}$ in \eqref{eq:rho_beta}  is clearly a convex mixture of SDs  with definite number of up and down electrons, Eq.\ \eqref{rsep}, implying that all negativities ${\cal N}_{\uparrow\downarrow}$, ${\cal N}^{(2)}_{\uparrow\downarrow}$ and ${\cal N}_{\uparrow\uparrow}$ also vanish when evaluated at $\rho_{\rm sep}$. This is not necessarily the case with the second part $\rho_{\rm bp}$ due to the Bell pair at the $O$ atom in the last six eigenstates. 

\begin{figure}
        \begin{center}
            \includegraphics[width=.95\linewidth]{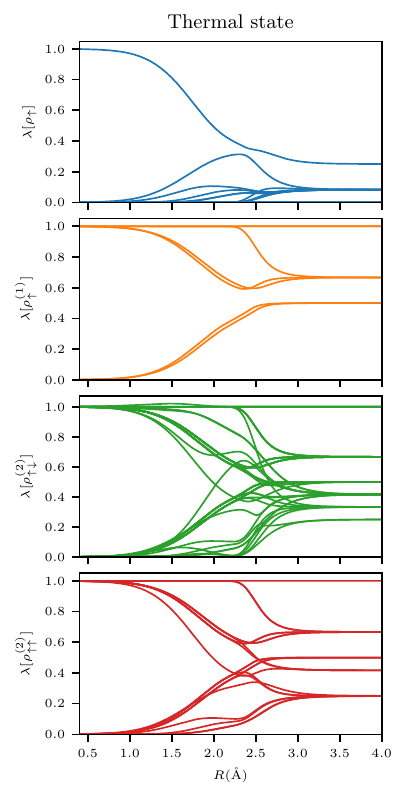}
            \vspace*{-0.5cm}
            \caption{\label{fig:dms_thermal} Eigenvalues of the total $\uparrow$-reduced state (top), and  those of the blocks of the one- and two- body reduced density matrices (remaining panels),  for the thermal state \eqref{eq:thermal} at $\beta=1000 E_h^{-1}$.  They approach the fractional values summarized in Table \ref{tab:eigenvalues_dms}.}
        \end{center}
\end{figure}

We first depict in Fig.\ \ref{fig:dms_thermal}  the eigenvalues of $\rho_{\uparrow}$, $\rho_{\uparrow}^{(1)}$, $\daa$ and $\dab$ in the full $\rho_0(\beta)$ for increasing $R$ at the same previous  $\beta$.  In the present mixed state case these eigenvalues  no longer represent an entanglement spectrum, but nonetheless they still provide a basic characterization of the main features of the thermal state. The intermediate sector $1$\AA\ $\alt R\alt 3$\AA\  is seen again to be the interval of maximum complexity of these eigenvalues. 
In the dissociation limit, and at the indicated low temperature, they approach rational numbers that are different from those of the GS, and  are listed in Table \ref{tab:eigenvalues_dms} with their degeneracy numbers. They can be readily derived from previous asymptotic expressions \eqref{eq:rho_beta}--\eqref{rhod}.

As seen in the second panel of Fig.\ \ref{fig:dms_thermal} and Table \ref{tab:eigenvalues_dms}, the eigenvalues of $\rho^{(1)}_{\uparrow}$ in $\rho_0(\beta)$ approach three distinct values in the dissociation limit. The two $1$ eigenvalues  obviously correspond to the core orbitals that are fully occupied in all  12 states $\ket{K}$ ($\phi_{1s}$ and $\phi_{2s}$ orbitals). 
The three 2/3 eigenvalues correspond to the $2p$ orbitals: three of the $\ket{K}$ SDs have all three of them fully occupied, the next three have only one of them occupied, and the last six of them have two out of three occupied, leading to 
a total probability $2/3$ of each of the $2p_\mu$ orbitals being occupied. 
Finally, the two $1/2$ eigenvalues stem from the two orbitals centered around the H atoms.  

Regarding the blocks of $\rho^{(2)}$, the eigenvalues of $\rho^{(2)}_{\uparrow\uparrow}$ and $\rho^{(2)}_{\uparrow\downarrow}$ approach five  and six distinct asymptotic values respectively, as seen in the fourth and third panels of Fig.\ \ref{fig:dms_thermal}. In the following discussion we set $i, j \in \{2, 3, 6\}$, and $k \in \{4, 5\}$ and omit the  three obvious upper eigenvalues $1/2$, $2/3$ and $1$ arising from one or two core electrons in the pair.    
$\daa$ has three additional eigenvalues 5/12, whose eigenvectors are $C^\dag_{ij}\ket{0}$, and seven eigenvalues 1/4, whose eigenvectors are $C_{45}^\dag\ket{0}$, and $C_{ik}^\dag\ket{0}$.
It is clear from them that $\daa$ is a convex mixture of SDs.
$\dab$ has 3 additional eigenvalues 1/2, corresponding to eigenvectors $C_{i \bar j_-}^\dag\ket{0}$, 12 eigenvalues 5/12, whose eigenvectors are $C_{i\bar k}^\dag \ket{0}$ and $C_{k \bar i}^\dag \ket{0}$, 6 eigenvalues 1/3, with eigenvectors $C^\dag_{i\bar j_+}\ket{0}$ and $C_{i \bar i}^\dag\ket{0}$, and two eigenvalues 1/4, with eigenvectors $C_{4 \bar 5}^\dag \ket{0}$ and $C_{5 \bar 4}^\dag \ket{0}$. 

Finally, the eigenvalues of $\rho_\uparrow$ (top panel) approach two distinct asymptotic values in the dissociation limit:  1/4, with eigenvector $C_{10236}^\dag\ket{0}$, arising from $\rho_{\rm sep}$, and nine eigenvalues 1/12, three of them from $\rho_{\rm sep}$ (eigenvectors $C^{\dag}_{10i45}\ket{0}$) and the rest from $\rho_{\rm bp}$ (eigenvectors $C^\dag_{10ijk}\ket{0}$).
This implies  $\rho_\uparrow$ is a convex mixture of SDs, as is also apparent from Eqs.\ \eqref{eq:rho_beta}-\eqref{rhod}: $\rho_{\rm sep}$ can only lead to  convex mixtures of SDs when $\downarrow$ modes are traced out, while each two-electron up-down Bell pair in  $\rho_{\rm bp}$ leads to a convex mixture of two up sp states 
after tracing out the $\downarrow$ modes, and hence to a convex mixture of SDs in the full $\rho_\uparrow$. This already implies that $\rho^{(2)}_{\uparrow\uparrow}$ is also a convex mixture of SDs, in agreement with previous result. Similar results hold of course for $\rho_{\downarrow}$. 

It is then apparent that the classical mixing of such 12 asymptotic eigenstates destroys most (but not all) of the fermionic entanglement present in the dissociation limit of the pure GS \eqref{eq:psi_dissoc_up_down}--\eqref{asint}. This can be seen, for example, in the bottom panel of Fig.\ \ref{fig:negativity},  where the up-down two-body negativity $\mathcal{N}_{\uparrow\downarrow}^{(2)} \rightarrow 0$, while the total up-down negativity $\mathcal{N}_{\uparrow\downarrow} \rightarrow 1/6$ (as compared with 5/6 and 13/6 for the GS, respectively). 
The nonzero asymptotic total negativity arises from two negative eigenvalues $-1/12$ of $\rho^{t_\downarrow}$, and certifies that the up-down entanglement stemming from the Bell pairs in $\rho_{\rm bp}$ is not fully destroyed in the mixture \eqref{eq:rho_beta} of the 12 asymptotic GSs. It is also nonzero $\forall\,R$, meaning that  at this $\beta$,  $\rho_0(\beta)$ is never a convex mixture of up-down product states  

Explicitly, the asymptotic total up-down negativity ${\cal N}_{\uparrow\downarrow}$  comes  from $\rho_{2p}^0$ in $\rho_{\rm bp}$, as remaining terms are separable: 
$\rho^{0 t_\downarrow}_{2p}$ has a diagonal block with elements $C^\dagger_{i\bar{i}j\bar{k}}\ket{0}\bra{0}C_{i\bar i j\bar k}$, and a separate non-diagonal block,
{\small\begin{equation}
    \frac{1}{6}\begin{pmatrix}
        0 & -1 & -1\\
        -1 & 0 & -1\\
        -1 & -1 & 0
    \end{pmatrix},
    \label{eq:matrix}
\end{equation}}
in the space spanned by the pair states $\{C^\dagger_{i\bar i j\bar j}\ket{0}, \,i \neq j\}$, which leads to a single negative eigenvalue $-1/3$. 
Since it appears twice in Eq. \eqref{rbp}, $\rho_{\rm bp}^{t_\downarrow}$ presents two $-1/6$ negative eigenvalues. 

Besides, recalling Eq.\ \eqref{eq:rho_beta}, we note that mixing $\rho_{\rm bp}$ with $\rho_{\rm sep}$ does not destroy these negative eigenvalues, since the terms in $\rho_{\rm sep}$ contain three (one) $\uparrow$ and one (three) $\downarrow$ electrons occupying these orbitals, which live in an orthogonal subspace. The final value is then ${\cal N}_{\uparrow\downarrow}(\rho_0(\beta))=1/6$ in this limit. 

Now, focusing on $\dab$, it is sufficient to note that the block that may lead to a negative partial transpose can be obtained by tracing $\rho_{2p}^0$ over one $\uparrow \downarrow$ pair of electrons.
This produces non-diagonal terms $-\frac{1}{6}\, C_{i \bar j}^\dagger \ket{0}\bra{0} C_{j \bar i}$, but also diagonal terms $\frac{1}{3}\,C_{i\bar i}^\dagger \ket{0}\bra{0} C_{i \bar i}$, which lead in $\rho^{(2)t_{\downarrow}}_{\uparrow\downarrow}$ to a block  similar to \eqref{eq:matrix} but with $+2/6$ in the diagonal. This just makes its lowest eigenvalue vanish, leading to $\mathcal{N}_{\uparrow\downarrow}^{{(2)}} = 0$.
We also mention that any non-uniform mixture of these three Bell pairs in $\rho^0_{2p}$ leads instead to ${\cal N}^{(2)}_{\uparrow\downarrow}>0$.

Finally, the up-down mutual informations $\iud$ and $\iud^{(2)}$ in $\rho_0(\beta)$ are depicted in the bottom panel of Fig.\ \ref{fig:IM_thermal}. In contrast with the GS case, they now exhibit a maximum value at $R \approx 2.2$\AA, the point where the excited states begin to contribute to $\rho_0(\beta)$.
This maximum  marks the transition from the pure GS, where the $\uparrow$ and $\downarrow$ subsystems are entangled, to an ensemble of separable and entangled states. 
In order to understand their limit values (indicated in the caption),  
we first note that $\rho_{{\rm sep}_\uparrow}$ (and similarly $\rho_{{\rm sep}_\downarrow}$) is not uniform, i.e., it has an eigenvalue $1/2$ with eigenvector $C^\dag_{01236}|0\rangle$ and three eigenvalues $1/6$.  
This part then has $\iud(\rho_{\rm sep})=1$ (for $\log=\log_2$). On the other hand, $\rho_{{\rm bp}\uparrow}$ (and similarly $\rho_{{\rm bp}\downarrow}$) is clearly uniform (a uniform mixture of six SDs, as stated before) so that $\iud(\rho_{\rm bp})=\log_2 6$ (which, though coinciding with the value for 
maximum classical-like correlation for a rank $6$ state,  it contains here quantum correlations in the  Bell pairs,  confirmed by the nonzero total negativity $\nud(\rho_0(\beta))$). These results then lead to the 
total value $\iud=2+\frac{1}{2}\log_2 3$, which is lower than in the pure GS. In the same way, the limit value of $\iud^{(2)}$ can be obtained from the corresponding limit spectrum  of $\rho^{(2)}_{\uparrow\downarrow}$ and $\rho^{(1)}_{\uparrow(\downarrow)}$ (table \ref{tab:eigenvalues_dms}), and lead now to a finite but much lower value in comparison with those of the pure GS or the total $\iud$.  

All calculations performed (here and in Appendix \ref{apB}) conserved    exactly the spin symmetry and did not involve any spin contamination in the system eigenstates (all had fixed $N_\uparrow$, $N_{\downarrow}$), implying that the up-up, up-down  and down-down sectors of the two-body DM were indeed  proper blocks  with no elements connecting them. Present measures could in principle  be formally  applied  to more general states after proper extension, as these blocks remain positive semidefinite, although $N_{\uparrow}$, $N_{\downarrow}$  and hence the trace of these blocks  will no longer be fixed in general, and the effects of elements linking these blocks will be omitted. 

\section{Conclusions}

We have examined various entanglement and fermionic correlation measures in the GS of the water molecule along the dissociation curve,  
including the dissociation limit.  
Due to the GS degeneracy emerging in this limit  we have also considered the $M_S=0$ thermal state for very low temperatures ($T\rightarrow 0^+$ limit) in order to obtain a consistent description of the latter. 

From the theoretical side, 
we have introduced some new correlation measures for fermionic systems having a fixed number of up and down  particles. In the first place, the  total up-down mutual information $\iud$ and negativity $\nud$. The first one 
provides a measure of the total (classical + quantum) correlations  between the spin up and spin down subsystems,  whereas $\nud$ measures only quantum correlations between them. 
In the case of pure states, $\iud$ is just twice the up-down entanglement entropy, while $\nud$ 
 becomes just another entropic measure of the total up-down entanglement, being determined  by the singular values $\Gamma_\nu$ of the up-down Schmidt decomposition of the state, but differ for mixed  states (i.e., in the thermal case here considered), where a positive $\nud>0$ ensures  a nonzero up-down entanglement of formation of the mixed state,  and can obviously vanish even if $\iud>0$. 

We have also presented equivalent measures based  on the reduced two-body DM, namely, the two-body up-down mutual information and negativity, $\iud^{(2)}$ and ${\cal N}_{\uparrow\downarrow}^{(2)}$,  the former here appropriately rescaled in order to be core independent. 
Essentially these quantities are analogous to the previous ones but at the two-body level, requiring thus less information (and becoming equivalent for two-particle  states with $N_{\uparrow}=N_{\downarrow}=1$). In particular, 
$\nud^{(2)}>0$ already ensures that the whole state is up-down entangled, i.e.,  not a convex mixture of product up-down states.

Additionally, a new two-body quantum correlation measure was introduced, the two-body fermionic negativity ${\cal N}^{(2)}$, which is applicable to any real fermionic state and is not based on any a priori partition of the sp space, being a measure of the ``internal'' fermionic entanglement of $\rho^{(2)}$.  It is based on an antisymmetrized partial transpose and vanishes if $\rho^{(2)}$ is a convex mixture of two-fermion SDs. On the other hand, it is positive for real entangled pure two-fermion states, taking the proper value determined by their Schmidt decomposition. In general ${\cal N}^{(2)}>0$  already ensures that the full 
many-fermion state cannot be a convex  mixture of real SDs. Here we used this negativity for measuring the inner entanglement of the up-up sector  $\rho^{(2)}_{\uparrow\uparrow}$, leading to ${\cal N}^{(2)}_{\uparrow\uparrow}$. 
 
 Besides, we have also explored other  measures, like the one-body entanglement entropy determined by the blocks of $\rho^{(1)}$  and the two-body entanglement entropies  determined by the blocks of $\rho^{(2)}$, which   essentially measure the entanglement between one fermion or a fermion pair with the rest and are  mode independent (though here adapted to systems with a fixed number of up and down electrons).

The dissociation curve of both the ground and the $M_S=0$ thermal states of the water molecule as discussed above were analyzed in detail, including analytical solutions for the dissociation limit.
The behavior of the entanglement measures presented here is fully compatible with the physical description of the system: in particular, towards the dissociation limit, $\nud$ is larger for the asymptotic GS than for the thermal state, since the only source of up-down entanglement for the latter are the local Bell pairs occupying the $2p$ orbitals of the O atom, whereas the former also has up-down entanglement between electrons located in the orbitals centered around the H atoms and the $2p$ orbitals in the O ones, due the total $S=0$ constraint, and is just a linear combination of the ``product'' asymptotic eigenstates. 
$\nuu^{(2)}$ is only nonzero around  the equilibrium geometry, where the GS is a slight deviation from a SD. On the other hand,
$\iud$ and $\iud^{(2)}$  are always $> 0$, since for the pure GS they reflect the up-down entanglement, whereas for the thermal state they also capture the classical correlation between subsystems.
These computations were performed using a minimal basis set, which may not be sufficiently large for an accurate quantitative description of the water molecule, but is ideal for a qualitative approach, since the dissociation limit is analytic. 
We have also performed  computations using the larger 6-31G basis set, which provides a better quantitative picture, 
although the shapes of the curves obtained, shown in Appendix \ref{apB},  are similar to the present ones. Illustrative results for the ground and low temperature thermal state of the H$_2$ molecule in bases of increasing dimension are also depicted there. 

All of the above entails that these correlation measures are useful tools for the characterization of molecules obtained by exact or approximate QC methods.
For example, some implementations of the configuration interaction (CI) method allow the user to access the wavefunction directly, 
and the $\iud$ and $\nud$ quantities defined here are readily applicable. 
Additionally, many implementations of CI, or other approximate correlated methods, such as M{\o}ller-Plesset perturbation theory methods or Coupled Cluster approaches \cite{MP},
provide the one- and two- body DMs, already separated in spin blocks (and in some cases  even the three- and four- body DMs are available as well).
The two-body quantities defined in the present work can be used to characterize these systems directly and with ease. But also the full-body $\iud$ and $\nud$ quantities can be applied to the blocks of the three- and four- body DMs, if available.

We finally remark that the correlation measures here defined are also directly applicable to systems with fixed $N_\uparrow \neq N_\downarrow$, such as in the case of triplet, quintuplet, or higher multiplicity eigenstates, and also convex mixtures of them. Application to other molecules and  use of larger basis sets are currently under investigation. 

\acknowledgments 
Authors acknowledge support from CONICET (J.G. and J.A.C.) and CIC (R.R.) of Argentina. Work supported by CONICET PIP Grant No. 11220200101877CO. Discussions with M. Cerezo, N.L. Diaz (Los Alamos National Lab) and M. Fonseca are gratefully acknowledged. 

\appendix
\section{Two-body entanglement witnesses and pairing example\label{apA}}

We discuss here  entanglement witnesses of up-down and fermionic entanglement, with emphasis placed on those based on the two-body DM. 
  \subsection{Up-down entanglement}
We start with the total up-down entanglement of the many-electron state $\rho$ (pure or mixed). As the set of up-down separable states, defined as convex mixtures $\rho_s=\sum_\nu p_\nu |\Phi_\nu\rangle\langle\Phi_\nu|$ ($p_\nu\in[0,1]$, $\sum_\nu p_\nu=1$)  of product states $|\Phi_\nu\rangle=A^\dag_\nu B^\dag_{\bar\nu}|0\rangle$, is convex,
for any up-down entangled state $\rho$ an entanglement witness in the form of an observable  $O$ (hermitian operator) always exists such that ${\rm Tr}\,\rho_s\, O\geq 0$ for any up-down separable state $\rho_s$, while ${\rm Tr}\,\rho\,O<0$. This follows from the general result for distinguishable bipartite systems \cite{GT.09,HHHH.09,LKCH.00}. 

For example,  an entanglement witness of an up-down entangled pure state  $\rho=|\Psi\rangle\langle\Psi|$ with definite $N_{\uparrow},N_{\downarrow}$   is
\begin{equation}O=-\rho+\langle \rho\rangle_{p_{\rm max}}\,,\label{A1}\end{equation}
where $\langle \rho\rangle_{p_{\rm max}}={\rm Max}_{\{|\Phi_p\rangle\}}|\langle \Phi_p|\Psi\rangle|^2$ is its maximum average  in all up-down product states $|\Phi_p\rangle=A^\dag_{\alpha}B^\dag_{\bar \beta}|0\rangle$, i.e.\ its  maximum squared fidelity with an up-down product state: Clearly $\langle O\rangle_{\rho_s}={\rm Tr}\,\rho_s O\geq 0$ $\forall$ up-down separable  $\rho_s$ 
whereas $\langle\Psi| O|\Psi\rangle=\langle \rho\rangle_{p_{\rm max}}-1<0$. Moreover, from the up-down Schmidt expansion \eqref{Scd} of $|\Psi\rangle$ it follows that $\langle\rho\rangle_{p_{\rm max}}=\Gamma_{\rm max}^2$, where $\Gamma_{\rm max}$ denotes the maximum singular value $\Gamma_\nu$ in \eqref{Scd} \cite{GT.09,BG.04}. 

More generally,  given an hermitian two-body  operator preserving the number of up and down electrons, i.e.\ 
\begin{equation}V=\sum_{i,j,k,l} V_{ij,kl}c^\dag_{i}c^\dag_{\bar j} c_{\bar l}c_k\,,\label{V}\end{equation} assumed to have a nondegenerate up-down entangled GS 
for given values of $N_{\uparrow},N_{\downarrow}$,  let $\langle V\rangle_{p_{\rm min}}$ be its  minimum average among all up-down product 
states $|\Phi_p\rangle$ with the same values of $N_{\uparrow}, N_{\downarrow}$. Then 
\begin{equation}\tilde V=V-\langle V\rangle_{p_{\rm min}}\label{Vt}\end{equation}
is also a witness of up-down entanglement: $\langle \tilde V\rangle\geq 0$ in any up-down product  state or separable mixed state $\rho_s$ for these $N_{\uparrow}$, $N_{\downarrow}$, whereas $\langle \tilde V\rangle<0$ in its GS. 

Similar arguments hold for the two-body DM up-down block $\rho^{(2)}_{\uparrow\downarrow}$, since the set of separable  $\rho^{(2)}_{\uparrow\downarrow}$  is also convex.  
Moreover,  since    $\langle V\rangle=\sum_{i,j,k,l}\!V_{ij,kl}\rho^{(2)}_{\uparrow\downarrow_{kl,ij}}={\rm Tr}[\rho^{(2)}_{\uparrow\downarrow}{\bf V}]$ depends just on $\rho^{(2)}_{\uparrow\downarrow}$ (${\bf V}$ is  the  matrix of elements $V_{ij,kl}$)   
and product states $|\Phi_p\rangle$ lead to a product $\rho^{(2)}_{\uparrow\downarrow}$, 
$\tilde V$ is also {\it an entanglement witness for} $\rho^{(2)}_{\uparrow\downarrow}$: its GS leads to an up-down entangled $\rho^{(2)}_{\uparrow\downarrow}$ since otherwise  a separable state providing the same average would exist. Then 
$\langle\tilde V\rangle\geq 0$
in all  separable $\rho^{(2)}_{\uparrow\downarrow}$ but $\langle \tilde V\rangle<0$ in the entangled $\rho^{(2)}_{\uparrow\downarrow}$ of its GS. Notice that  \eqref{Vt} leads to $\tilde {\bf V}={\bf V}-\frac{\langle V\rangle_{p_{\rm min}}}{N_\uparrow N_\downarrow}\mathbbm 1$,    such that  $\langle\tilde V\rangle={\rm Tr}[\rho^{(2)}_{\uparrow\downarrow}\tilde {\bf V}]$ is also  linear in $\rho^{(2)}_{\uparrow\downarrow}$. 

As a simple example, for a  plain pairing coupling
\begin{equation}V=-A^\dag A\,,\;\;A^\dag=\tfrac{1}{\sqrt{n}}\sum_{k=1}^nc^\dag_{k}c^\dag_{\bar k}\,,\label{VP}\end{equation}
($V_{ij,kl}=-\delta_{ij}\delta_{kl}/n$ in \eqref{V}) for $n$ up (down) sp levels $k$ ($\bar k$), 
its exact and nondegenerate GS for $N_{\uparrow}=N_{\downarrow}=m$, $1\leq m\leq n-1$, is the associated $m$-pair condensate \cite{GDR.21}:
\begin{equation}V|\Psi_m\rangle=-m(1-\tfrac{m-1}{n})|\Psi_m\rangle\,,\;\; |\Psi_m\rangle\propto (A^\dag)^m|0\rangle\,.\label{cond}\end{equation}
Such $|\Psi_m\rangle$ is {\it maximally up-down entangled} at fixed $m\in[1,n-1]$, with plain   up-down Schmidt representation  \cite{CR.24}  $|\Psi_m\rangle=\binom{n}{m}^{-\frac{1}{2}}\sum_{\bm\alpha}C^\dag_{\bm\alpha}C^\dag_{\bar{\bm\alpha}}|0\rangle$ for  
$C^\dag_{\bm\alpha(\bar{\bm\alpha})}=
\prod_k (c^\dag_{k(\bar k)})^{n_k}$,  $n_k=0,1$, $\bm\alpha=(n_1,\ldots,n_n)$ and $\sum_k n_k=m$, implying  maximum up down entanglement entropy $E_{\uparrow\downarrow}=\log\binom{n}{m}$ and negativity   ${\cal  N}_{\uparrow\downarrow}=\frac{1}{2}(\binom{n}{m}-1)$ (Eq.\ \eqref{negp}). 

In contrast, its minimum average in up-down product states with $N_{\uparrow}=N_{\downarrow}=m$ is just $\langle V\rangle_{p_{\rm min}}=-\frac{m}{n}$, attained e.g.\  in a SD $(\prod_{k=1}^m c^\dag_k c^\dag_{\bar k})|0\rangle$. Hence $\tilde V=V+\frac{m}{n}$ is an up-down {\it entanglement witness}:  $\langle \tilde V\rangle\geq 0$ in any such product state (or convex mixture of them)  while  $\langle \Psi_m|\tilde V|\Psi_m\rangle=-\frac{m(n-m)}{n}<0$ according to \eqref{cond},  vanishing just in the trivial cases $m=0$ or $n$   where $|\Psi_m\rangle$ becomes a vacuum or SD. In fact $\langle \tilde V\rangle<0$ also in many other up-down entangled eigenstates of $V$.

The nonzero elements of $\rho^{(2)}_{\uparrow\downarrow}$ in  \eqref{cond} are $\langle c^\dag_k c^\dag_{\bar k}c_{\bar k'}c_{k'}\rangle=\frac{m}{n}[\delta_{kk'}+(1\!-\!\delta_{kk'})\frac{n-m}{n-1}]$ and  $\langle c^\dag_k c^\dag_{\bar k'}c_{\bar k'}c_{k}\rangle=\frac{m(m-1)}{n(n-1)}$ (for $k\neq k'$) \cite{CR.24}. It is up-down entangled for $m\in[1,n-1]$ as implied by 
$\langle \Psi_m|\tilde V|\Psi_m\rangle={\rm Tr}\,\rho^{(2)}_{\uparrow\downarrow}\tilde{\bf V}<0$.  The ensuing up-down  two-body negativity \eqref{neg2} becomes 
  ${\cal N}^{(2)}_{\uparrow\downarrow}=\frac{m(n+1-2m)}{2}>0$ for $1\leq m\leq \frac{n}{2}$  (and $0$ otherwise),  stemming from $\frac{n(n-1)}{2}$ negative eigenvalues $-\frac{m(n+1-2m)}{n(n-1)}$ of $\rho^{(2)t_{\downarrow}}_{\uparrow\downarrow}$ for $1\leq m\leq \frac{n}{2}$, hence detecting  entanglement up to the half-filled case (though the equivalent  hole representation $m\rightarrow n-m$  of \eqref{cond} \cite{PR.25}  
   can be used if $m>\frac{n}{2}$). 

Other witnesses are also feasible. As the maximum eigenvalue of a product  $\rho^{(2)}_{\uparrow\downarrow}=\rho^{(1)}_{\uparrow}\otimes\rho^{(1)}_{\downarrow}$ is always $\leq 1$  
(since those of $\rho^{(1)}_{\uparrow}$,   $\rho^{(1)}_{\downarrow}$ are  $\leq 1$), the same holds for a separable $\rho^{(2)}_{\uparrow\downarrow}$. Then 
the {\it maximum eigenvalue}  $\lambda^{(2)}_{\uparrow\downarrow_{\rm max}}$ of $\rho^{(2)}_{\uparrow\downarrow}$  {\it is an up-down entanglement witness if exceeding $1$:}  
\begin{equation}\lambda^{(2)}_{\uparrow\downarrow_{\rm max}}>1\implies  \rho^{(2)}_{\uparrow\downarrow}\; {\rm entangled}\label{l1}\,,\end{equation}  (and hence the full state it comes from is also up-down entangled). For instance, from Eq.\ \eqref{cond} it follows that in the plain pair condensate $|\Psi_m\rangle$, 
$\lambda^{(2)}_{\rm \uparrow\downarrow_{\rm max}}=m(1-\frac{m-1}{n})>1$ for $2\leq m
\leq n-1$, reaching its highest possible value for $m$ up (down)  electrons in $n$ up (down) sp states \cite{Yang.62,SS.21,CR.24}.   Moreover, $\lambda^{(2)}_{ \uparrow\downarrow_{\rm max}}\geq 1$ in any pair condensate $\propto (A^\dag)^m|0\rangle$ for 
general $A^\dag=\sum_k \sigma_kc^\dag_kc^\dag_{\bar k}$ \cite{CR.24} and $2\leq m\leq 1$, reflecting the up-down pairing correlations. 

Nonetheless, Eq.\ \eqref{l1} is of course a sufficient but  not necessary condition for up-down entanglement. Since   $\lambda^{(2)}_{\uparrow\downarrow_{\rm max}}=\langle  A^\dag_{\rm max} A_{\rm max}\rangle$  with $A_{\rm max}^\dag$ the associated eigenvector (normalized pair creation operator) of $\rho^{(2)}_{\uparrow\downarrow}$, this witness  corresponds to $V=-A^\dag_{\rm max} A_{\rm max}$ in \eqref{Vt} with  $\langle V\rangle_{p_{\rm min}}$ replaced by the lower bound $-1$. 

Other witnesses can be derived from conditions on the elements of $\rho^{(2)}_{\uparrow\downarrow}$ which ensure its partial transpose is not positive: For instance, if the $\rho^{(2)t_\downarrow}_{\uparrow\downarrow}$ subdeterminant 
\begin{equation} D_{kk'}=\langle c^\dag_kc^\dag_{\bar k'}c_{\bar k'}c_k\rangle\langle c^\dag_{k'}c^\dag_{\bar k}c_{\bar k}c_{k'}\rangle-|\langle c^\dag_k c^\dag_{\bar k}c_{\bar k'}c_{k'}\rangle|^2\,,\label{Dkk}\end{equation} fulfills $D_{kk'}<0$, then $\rho^{(2)t_\downarrow}_{\uparrow\downarrow}$ cannot be positive, ensuring a negative eigenvalue and hence up-down entanglement.  

\subsection{Fermionic entanglement}

A similar discussion applies to witnesses of  fermionic entanglement, except that no a priori partition of the sp space is assumed or chosen. 
A positive value ${\cal N}^{(2)}>0$ of  the two-body negativity of  Eqs.\ \eqref{rtp}--\eqref{Neg2},  applicable to real states (in a given  basis of real SDs) with definite particle number,  indicates  that the full state $\rho$ is `entangled', in the sense that it is not a convex mixture $\rho_s=\sum_\nu p_\nu|\Phi_\nu\rangle\langle\Phi_\nu|$  ($p_\nu\in[0,1]$, $\sum_\nu p_\nu=1$) of real SDs $|\Phi_\nu\rangle$ (or equivalently, of real fermionic gaussian states with definite particle number), since for each term of such sum the defined partial transpose \eqref{rtp} of $\rho^{(2)}$ is positive 
semidefinite, as shown in main text, implying ${\cal N}^{(2)}=0$. 
Hence, ${\cal N}^{(2)}$ is a witness of this fermionic entanglement, providing a sufficient (but not necessary) condition. 
We also remark that such mode-independent fermionic entanglement \cite{GDR.20,GDR.21} is relevant  as a resource in quantum information theory, as a quantum computation  involving just convex mixtures of SDs  
(or in general  `convex gaussian' states) and non-interacting fermionic gates can be  classically efficiently simulated \cite{MCT.13}.

Again, since the set of such convex mixtures is convex,  for any entangled state $\rho$ an entanglement witness in the form an hermitian operator $O$  exists such that ${\rm Tr}\,\rho_s O\geq 0$ for any such convex mixture $\rho_s$ but ${\rm Tr}\,\rho\, O<0$. For example, for a non SD pure state  $|\Psi\rangle$ with  $\rho=|\Psi\rangle\langle\Psi|$, 
an entanglement witness similar to Eq.\ \eqref{A1} is 
\begin{equation}
O=-\rho+\langle\rho\rangle_{s_{\rm max}}\label{A1s}\,,\end{equation}
where $\langle\rho\rangle_{s_{\rm max}}={\rm Max}_{\{|\Phi_s\rangle\}}|
\langle\Phi_s|\Psi\rangle|^2$ is the maximum squared fidelity between $|\Psi\rangle$ and a 
SD $|\Phi_s\rangle$:  $\langle O\rangle\geq 0$ $\forall$ $|\Phi_s\rangle$ or convex mixture $\rho_s$, but $\langle\Psi|O|\Psi\rangle=\langle\rho\rangle_{s_{\rm max}}-1<0$.   

 Similarly,  given an hermitian two-body operator $V=\sum_{i,j,k,l} V_{ijkl}c^\dag_ic^\dag_jc_lc_k$ with an entangled (non SD) nondegenerate GS,  let $\langle V\rangle_{s_{\rm min}}\equiv\langle V\rangle_{_{\rm HF}}$ be its minimum average   over all  SDs (HF approximation). 
Then \begin{equation}\tilde V=V-\langle V\rangle_{_{\rm HF}}\,,\label{VHF}\end{equation} is a witness  of this entanglement: $\langle \tilde V\rangle\geq 0$ in any  SD or   convex mixture of  SDs, 
 but $\langle \tilde V\rangle<0$ in its GS. Since $\langle V\rangle$ is fully determined by $\rho^{(2)}$, such GS will lead to an {\it entangled} $\rho^{(2)}$, in the sense of not being a convex mixture of two-body DMs of SDs (which are diagonal in a basis of two-fermion SDs \cite{GDR.21,CR.24}),   since otherwise a separable state or SD leading to the same minimum average would exist. Then $\tilde V$  is  also an entanglement witness of $\rho^{(2)}$.
 And if $V$ is real, its eigenstates are (or can be chosen as) all real, such that a real HF approximation is suitable (at least if complex conjugation symmetry is not broken). 
 
  For example, considering again the pairing coupling  \eqref{VP} for $2m$ fermions in $2n$ levels, $\tilde V=V+\frac{m}{n}$ is also a witness of this fermionic entanglement, 
since here $\langle V\rangle_{_{\rm HF}}=\langle V\rangle_{p_{\rm min}}=-\frac{m}{n}$  while its GS, the pair condensate \eqref{cond}, is obviously not a SD. The associated one-body DM is in fact {\it maximally mixed}: $\rho^{(1)}_{\uparrow}=\rho^{(1)}_{\downarrow}=\frac{m}{n}\mathbbm{1}$ \cite{GDR.21,CR.24}, leading to  maximum one-body entanglement entropy  $E^{(1)}=2m\log\frac{n}{m}$ for fixed  $m,n$. Then $\langle \tilde V\rangle=-\frac{m(n-m)}{n}<0$ in its GS  for $1\leq m\leq n-1$ while $\langle\tilde V\rangle\geq 0$ in any SD or convex mixture of SDs.   We also obtain   ${\cal N}^{(2)}=m(n+1-2m)=2{\cal N}^{(2)}_{\uparrow\downarrow}>0$ for $1\leq m\leq n/2$ in \eqref{cond}, thus detecting this entanglement {\it without} a priori  knowledge  of the underlying up-down partition.  

In the present work we may  also consider for $V$ in \eqref{VHF} the full electronic  Hamiltonian \eqref{eq:hamiltonian}, such that $\tilde H=H-\langle H\rangle_{_{\rm HF}}$ is a witness  of the fermionic entanglement of both its GS (when  nondegenerate) and  the ensuing full  $\rho^{(2)}$   (for  fixed  particle number $N\geq 2$,  $\rho^{(1)}$ is a partial trace of $\rho^{(2)}$,  
 $\langle c^\dag_{j}c_i\rangle=\frac{1}{N-1}\sum_k\langle c^\dag_j c^\dag_k c_kc_i\rangle$, such that $\langle H\rangle$ is a function of $\rho^{(2)}$ only). Accordingly, $\langle \tilde H\rangle$ will be most negative in the intermediate sector where its GS 
 deviates most from a SD and is nondegenerate,  while in the minimal basis it vanishes for large $R$ as the GS becomes degenerate with SD states (see \ref{IIIC} and also App.\  \ref{apB}). 
 
 On the other hand, the behavior of the full two-body ${\cal N}^{(2)}$ of Eq.\  \eqref{Neg2}, also a witness of this fermionic entanglement, is quite similar to that of ${\cal N}^{(2)}_{\uparrow\downarrow}$ (Fig.\ \ref{fig:negativity}) in both the GS and the thermal state considered, since the up-down block $\rho^{(2)}_{\uparrow\downarrow}$ contains the main quantum correlations at the two-body level. Accordingly, in the dissociation limit ${\cal N}^{(2)}$  remains finite in the GS, while vanishing in the thermal state (in minimal basis). 
 We have then  applied  this negativity to  detect entanglement in the 
 up-up sector  $\rho^{(2)}_{\uparrow\uparrow}$ (here identical with $\rho^{(2)}_{\downarrow\downarrow}$),  which has no a priori preferred partition, through ${\cal N}^{(2)}_{\uparrow\uparrow}$. Though much smaller than ${\cal N}^{(2)}_{\uparrow\downarrow}$  (Fig.\ \ref{fig:negativity}) it is still nonzero for small $R$, reflecting ``dynamic'' correlations.  Witnesses similar to \eqref{VHF} for this entanglement can also be used, using couplings involving just the up electrons and hence $\rho^{(2)}_{\uparrow\uparrow}$ only.  
 
The extension of   \eqref{l1}, $\lambda^{(2)}_{\rm max}>1\Rightarrow\rho^{(2)}$ {\it entangled},  
with $\lambda^{(2)}_{\rm max}$  the maximum eigenvalue of the full $\rho^{(2)}$, remains valid as witness of fermionic entanglement (and pairing correlations \cite{Yang.62,SS.21,CR.24}) at the two-body level, implying entanglement in $\rho^{(2)}$ and hence in the full $\rho$, since in SDs the eigenvalues of $\rho^{(2)}$ are just $1$ or $0$ \cite{GDR.21,CR.24}  and hence  never exceed $1$ in a convex mixture of SDs. In fact, as previously mentioned $\lambda^{(2)}_{\rm max}\geq 1$ in any $m$-pair condensate $\propto (A^\dag)^m|0\rangle$ for a general pair creation operator $A^\dag$, with $\lambda^{(2)}_{\rm max}>1$ if $A^\dag$ has rank $\geq 2$ and $2\leq m\leq n-1$ in a $2n$ dimensional sp space \cite{CR.24}, and 
\eqref{cond} providing the upper limit for $2m$ fermions in $2n$ sp states \cite{Yang.62,SS.21,CR.24}.  In the present work an eigenvalue  slightly larger than $1$  arises just in the $\rho^{(2)}_{\uparrow\downarrow}$ block for moderate $R$ (Fig.\ \ref{fig:dms_gs}, bottom panel), reflecting weak pairing-like correlations.    Finally,  witnesses based on subdeterminants of $\rho^{(2)t_p}$ similar to \eqref{Dkk}, which imply it is not positive semidefinite, are in principle also available, though the optimum sp elements and partition are in general not a priori known.

\section{Results for other basis sets \label{apB}}

Here we show how the information measures defined in this work change with bases of increasing dimension. 

\subsection*{H$_2$O molecule}
\begin{table}[!ht]
    \begin{center}
        \begin{tabular}{c|c|c}
            Basis set $\;$ & $\;d/2\;$ & $N_{\rm str}$ \\
            \hline
            STO-3G  &  7 & 441 \\\hline
            6-31G   & 13 & 1656369  
        \end{tabular}
    \end{center}
    \caption{Dimension of the sp basis sets and total Hilbert spaces ($N_{\rm str}$) for the water molecule.}
    \label{t4}
\end{table}

\begin{figure}[!ht]
    \centering
    \includegraphics[width=0.9\linewidth]{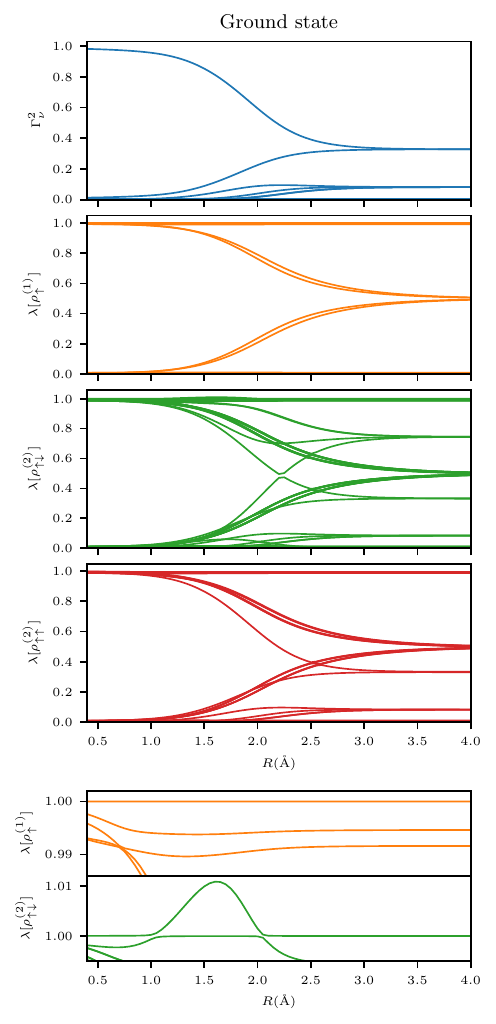}
    \caption{Eigenvalues of $\rho_\uparrow$, $\rho^{(1)}_\uparrow$, $\dab$ and $\daa$ for the GS of the H$_2$O molecule computed with the 6-31G basis set, as a function of the O-H  distance $R$.}
    \label{fig:eigenvalues_dms_631g}
\end{figure}

We resort here to the 6-31G basis set \cite{ditchfield1971a,hehre1972a} which, although considered not fully sufficient for quantitative Quantum Chemistry, it does add a measurable amount of dynamic correlation which serves to illustrate the 
behavior  of the correlation measures introduced here. This set comprises 13 sp orbitals for each spin, implying a basis  of $D=\binom{13}{5}^2=1656369$ SDs (Table \ref{t4}).   We used native PySCF algorithms for the diagonalization and the determination of the DMs.

\begin{figure}[!ht]
    \centering    \includegraphics[width=0.85\linewidth]{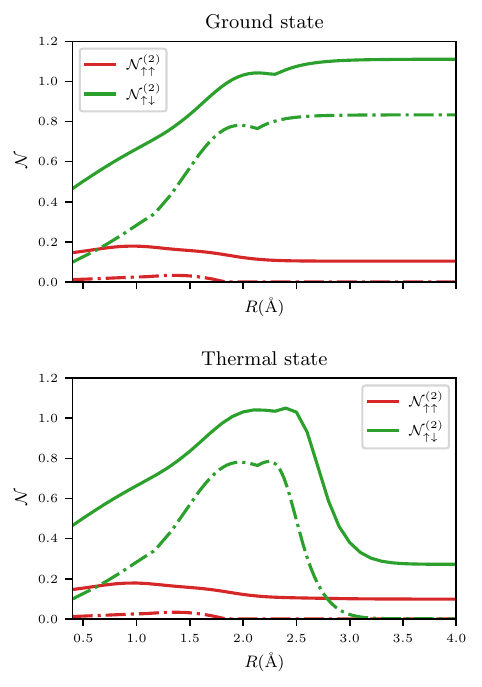}
    \caption{Two-body negativities of the H$_2$O molecule as a function of the distance  $R$, computed with the STO-3G (dashed-dotted  lines) and 6-31G (solid lines) basis sets. Results for both the GS and the thermal state in the $T\rightarrow 0^+$ limit are depicted.}
    \label{fig:neg_h2o_631g}
\end{figure}

First, we show in Fig.\  \ref{fig:eigenvalues_dms_631g} the entanglement spectra of the one and two-body DMs computed for the GS.
A comparison with the corresponding results of Fig.\ \ref{fig:dms_gs} (obtained with the minimal set) shows that they are quite similar, 
with a few minor differences:\\
i) A few very small, non-zero eigenvalues appear in every $\rho$, replacing the exact $0$s for the minimal basis set. Similarly,  
the blowup in the bottom panel shows that $\rho_{\uparrow}^{(1)}$ now behaves less like a three-electron core and more like a one-electron core: Two of the previous three eigenvalues $\approx 1$ for $R\agt 2.5 \AA$ now have values between $0.99$ and $0.995$ in the dissociation limit.  \\
ii) The eigenvalues in the dissociation limit are not exact fractional numbers, but lie  very close to previous values. These features reflect essentially the dynamic correlations now present  in the lone O atom. \\
iii) The previous pairing effect in  $\rho^{(2)}_{\uparrow\downarrow}$ is diminished: the maximum eigenvalue of $\dab$ at $R\approx 1.6$ goes from $\sim 1.022$ for the STO-3G basis set to $\sim 1.012$ for the 6-31G one. 

Next, we show in Fig.\  \ref{fig:neg_h2o_631g} the corresponding two-body negativities.  
We also include those previously computed in the minimal basis set for comparison. As can be seen, both $\nud^{(2)}$ and $\nuu^{(2)}$ show appreciable  increments  when the basis set is enlarged.
In particular, $\nuu^{(2)}$
approaches  now a finite value in the dissociation limit, 
 both for the GS as well as $\rho_0(\beta)$.
It is therefore safe to assume that both correlation measures are able to capture the added dynamic correlation.
We also expect them to increase further when employing larger basis sets. 
These measures could then be used as an additional criterion to detect when the complete basis set limit is reached.

\subsection*{H$_2$ molecule}

\begin{figure}[!ht]
    \centering
    \includegraphics[width=0.7\linewidth]{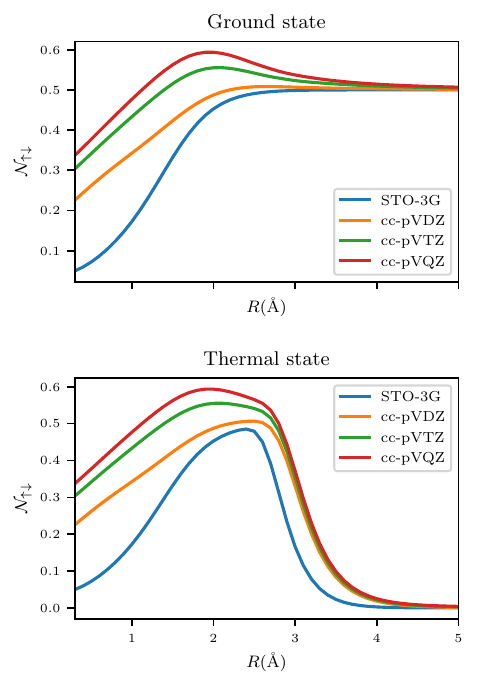}
    \caption{The negativity $\nud$ for the H$_2$ molecule as a function of the internuclear distance for different basis sets. Top: GS. Bottom: Thermal state $\rho_0(\beta)$.}
    \label{fig:neg_h2}
\end{figure}

In order to show the behavior   
of entanglement measures for increasingly larger basis sets, we depict here illustrative results  
for the  H$_2$ molecule in both its GS and the thermal state $\rho_0(\beta)$ for $\beta\rightarrow 0^+$, for the minimal basis set STO-3G, as well as the correlation consistent cc-pV$X$Z basis sets with $X = D, T, Q$ \cite{dunning1989a}, 
indicated in Table \ref{tab}.
Considering only the $M_S = 0$ subspace, the H$_2$ molecule has just one up  and one down electron, implying $\daa = 0$ 
and $\dab = \rho$ (full DM). Hence,  $\nud \propto \nud^{(2)}$ whereas $\nuu = \nuu^{(2)} = 0$.
In Fig.\ \ref{fig:neg_h2} we show results for the  negativity $\nud$ in  the previous basis sets. It can be seen that as the basis set size increases, the correlation measure increases, though all approaching  essentially the same limit value for $R\rightarrow\infty$. 
In the case of the GS, $\nud \rightarrow 1/2$, reflecting the exact  GS limit
$\frac{1}{\sqrt{2}} (C_{0\bar0}^\dag - C_{1 \bar{1}}^\dag) \ket{0}$.
The thermal state is an equally weighted mixture of the GS and the triplet state 
$C_{0\bar{1}_-}^\dag \ket{0}$, which becomes degenerate with previous state  for $R \rightarrow \infty$. Here $0$ ($1$) denote the $\pm$ orthogonal superpositions of the orbitals localized in each atom.

\begin{table}[!ht]
\begin{center}
    \begin{tabular}{c|c|c}
        Basis set $\;$& $\;d/2\;$ & $D$ \\
        \hline
        STO-3G  & 2 & 4 \\\hline
        cc-pVDZ & 10 & 100 \\\hline
        cc-pVTZ & 28 & 784 \\\hline
        cc-pVQZ & 60 & 3600 
    \end{tabular}
\end{center}
    \caption{Dimension of the sp basis sets and the associated  Hilbert space ($D=(d/2)^2$)   for the H$_2$ molecule.}
    \label{tab}
\end{table}

\newpage

%\bibliography{biblio}

\begin{thebibliography}{75}%
\makeatletter
\providecommand \@ifxundefined [1]{%
 \@ifx{#1\undefined}
}%
\providecommand \@ifnum [1]{%
 \ifnum #1\expandafter \@firstoftwo
 \else \expandafter \@secondoftwo
 \fi
}%
\providecommand \@ifx [1]{%
 \ifx #1\expandafter \@firstoftwo
 \else \expandafter \@secondoftwo
 \fi
}%
\providecommand \natexlab [1]{#1}%
\providecommand \enquote  [1]{``#1''}%
\providecommand \bibnamefont  [1]{#1}%
\providecommand \bibfnamefont [1]{#1}%
\providecommand \citenamefont [1]{#1}%
\providecommand \href@noop [0]{\@secondoftwo}%
\providecommand \href [0]{\begingroup \@sanitize@url \@href}%
\providecommand \@href[1]{\@@startlink{#1}\@@href}%
\providecommand \@@href[1]{\endgroup#1\@@endlink}%
\providecommand \@sanitize@url [0]{\catcode `\\12\catcode `\$12\catcode `\&12\catcode `\#12\catcode `\^12\catcode `\_12\catcode `\%12\relax}%
\providecommand \@@startlink[1]{}%
\providecommand \@@endlink[0]{}%
\providecommand \url  [0]{\begingroup\@sanitize@url \@url }%
\providecommand \@url [1]{\endgroup\@href {#1}{\urlprefix }}%
\providecommand \urlprefix  [0]{URL }%
\providecommand \Eprint [0]{\href }%
\providecommand \doibase [0]{http://dx.doi.org/}%
\providecommand \selectlanguage [0]{\@gobble}%
\providecommand \bibinfo  [0]{\@secondoftwo}%
\providecommand \bibfield  [0]{\@secondoftwo}%
\providecommand \translation [1]{[#1]}%
\providecommand \BibitemOpen [0]{}%
\providecommand \bibitemStop [0]{}%
\providecommand \bibitemNoStop [0]{.\EOS\space}%
\providecommand \EOS [0]{\spacefactor3000\relax}%
\providecommand \BibitemShut  [1]{\csname bibitem#1\endcsname}%
\let\auto@bib@innerbib\@empty
%</preamble>
\bibitem [{\citenamefont {Wigner}(1934)}]{Wigner_1934}%
  \BibitemOpen
  \bibfield  {author} {\bibinfo {author} {\bibfnamefont {E.}~\bibnamefont {Wigner}},\ }\bibfield  {title} {\enquote {\bibinfo {title} {On the interaction of electrons in metals},}\ }\href {\doibase 10.1103/physrev.46.1002} {\bibfield  {journal} {\bibinfo  {journal} {Phys. Rev.}\ }\textbf {\bibinfo {volume} {46}},\ \bibinfo {pages} {1002} (\bibinfo {year} {1934})}\BibitemShut {NoStop}%
\bibitem [{\citenamefont {Löwdin}(1955)}]{L_wdin_1955_b}%
  \BibitemOpen
  \bibfield  {author} {\bibinfo {author} {\bibfnamefont {P.O.}\ \bibnamefont {Löwdin}},\ }\bibfield  {title} {\enquote {\bibinfo {title} {Quantum theory of many-particle systems. {III}. {E}xtension of the {H}artree-{F}ock scheme to include degenerate systems and correlation effects},}\ }\href {\doibase 10.1103/physrev.97.1509} {\bibfield  {journal} {\bibinfo  {journal} {Phys. Rev.}\ }\textbf {\bibinfo {volume} {97}},\ \bibinfo {pages} {1509} (\bibinfo {year} {1955})}\BibitemShut {NoStop}%
\bibitem [{\citenamefont {Nielsen}\ and\ \citenamefont {Chuang}(2010)}]{NielsenBook}%
  \BibitemOpen
  \bibfield  {author} {\bibinfo {author} {\bibfnamefont {M.~A.}\ \bibnamefont {Nielsen}}\ and\ \bibinfo {author} {\bibfnamefont {I.~L.}\ \bibnamefont {Chuang}},\ }\href@noop {} {\emph {\bibinfo {title} {Quantum Computation and Quantum Information}}},\ \bibinfo {edition} {2nd}\ ed.\ (\bibinfo  {publisher} {Cambridge University Press},\ \bibinfo {year} {2010})\BibitemShut {NoStop}%
\bibitem [{\citenamefont {Horodecki}\ \emph {et~al.}(2009)\citenamefont {Horodecki}, \citenamefont {Horodecki}, \citenamefont {Horodecki},\ and\ \citenamefont {Horodecki}}]{HHHH.09}%
  \BibitemOpen
  \bibfield  {author} {\bibinfo {author} {\bibfnamefont {R.}~\bibnamefont {Horodecki}}, \bibinfo {author} {\bibfnamefont {P.}~\bibnamefont {Horodecki}}, \bibinfo {author} {\bibfnamefont {M.}~\bibnamefont {Horodecki}}, \ and\ \bibinfo {author} {\bibfnamefont {K.}~\bibnamefont {Horodecki}},\ }\bibfield  {title} {\enquote {\bibinfo {title} {Quantum entanglement},}\ }\href {\doibase 10.1103/RevModPhys.81.865} {\bibfield  {journal} {\bibinfo  {journal} {Rev. Mod. Phys.}\ }\textbf {\bibinfo {volume} {81}},\ \bibinfo {pages} {865} (\bibinfo {year} {2009})}\BibitemShut {NoStop}%
\bibitem [{\citenamefont {Amico}\ \emph {et~al.}(2008)\citenamefont {Amico}, \citenamefont {Fazio}, \citenamefont {Osterloh},\ and\ \citenamefont {Vedral}}]{A.08}%
  \BibitemOpen
  \bibfield  {author} {\bibinfo {author} {\bibfnamefont {L.}~\bibnamefont {Amico}}, \bibinfo {author} {\bibfnamefont {R.}~\bibnamefont {Fazio}}, \bibinfo {author} {\bibfnamefont {A.}~\bibnamefont {Osterloh}}, \ and\ \bibinfo {author} {\bibfnamefont {V.}~\bibnamefont {Vedral}},\ }\bibfield  {title} {\enquote {\bibinfo {title} {Entanglement in many-body systems},}\ }\href {\doibase 10.1103/RevModPhys.80.517} {\bibfield  {journal} {\bibinfo  {journal} {Rev. Mod. Phys.}\ }\textbf {\bibinfo {volume} {80}},\ \bibinfo {pages} {517} (\bibinfo {year} {2008})}\BibitemShut {NoStop}%
\bibitem [{\citenamefont {Aliverti-Piuri}\ \emph {et~al.}(2024)\citenamefont {Aliverti-Piuri}, \citenamefont {Chatterjee}, \citenamefont {Ding}, \citenamefont {Liao}, \citenamefont {Liebert},\ and\ \citenamefont {Schilling}}]{Aliverti.24}%
  \BibitemOpen
  \bibfield  {author} {\bibinfo {author} {\bibfnamefont {D.}~\bibnamefont {Aliverti-Piuri}}, \bibinfo {author} {\bibfnamefont {K.}~\bibnamefont {Chatterjee}}, \bibinfo {author} {\bibfnamefont {L.}~\bibnamefont {Ding}}, \bibinfo {author} {\bibfnamefont {K.}~\bibnamefont {Liao}}, \bibinfo {author} {\bibfnamefont {J.}~\bibnamefont {Liebert}}, \ and\ \bibinfo {author} {\bibfnamefont {C.}~\bibnamefont {Schilling}},\ }\bibfield  {title} {\enquote {\bibinfo {title} {What can quantum information theory offer to quantum chemistry?}}\ }\href {\doibase 10.1039/D4FD00059E} {\bibfield  {journal} {\bibinfo  {journal} {Faraday Discuss.}\ }\textbf {\bibinfo {volume} {254}},\ \bibinfo {pages} {76} (\bibinfo {year} {2024})}\BibitemShut {NoStop}%
\bibitem [{\citenamefont {Benatti}\ \emph {et~al.}(2020)\citenamefont {Benatti}, \citenamefont {Floreanini}, \citenamefont {Franchini},\ and\ \citenamefont {Marzolino}}]{BFFM.20}%
  \BibitemOpen
  \bibfield  {author} {\bibinfo {author} {\bibfnamefont {F.}~\bibnamefont {Benatti}}, \bibinfo {author} {\bibfnamefont {R.}~\bibnamefont {Floreanini}}, \bibinfo {author} {\bibfnamefont {F.}~\bibnamefont {Franchini}}, \ and\ \bibinfo {author} {\bibfnamefont {U.}~\bibnamefont {Marzolino}},\ }\bibfield  {title} {\enquote {\bibinfo {title} {Entanglement in indistinguishable particle systems},}\ }\href {\doibase https://doi.org/10.1016/j.physrep.2020.07.003} {\bibfield  {journal} {\bibinfo  {journal} {Phys. Rep.}\ }\textbf {\bibinfo {volume} {878}},\ \bibinfo {pages} {1} (\bibinfo {year} {2020})}\BibitemShut {NoStop}%
\bibitem [{\citenamefont {Schliemann}\ \emph {et~al.}(2001)\citenamefont {Schliemann}, \citenamefont {Cirac}, \citenamefont {Ku{\'s}}, \citenamefont {Lewenstein},\ and\ \citenamefont {Loss}}]{SC.01}%
  \BibitemOpen
  \bibfield  {author} {\bibinfo {author} {\bibfnamefont {J.}~\bibnamefont {Schliemann}}, \bibinfo {author} {\bibfnamefont {J.~I.}\ \bibnamefont {Cirac}}, \bibinfo {author} {\bibfnamefont {M.}~\bibnamefont {Ku{\'s}}}, \bibinfo {author} {\bibfnamefont {M.}~\bibnamefont {Lewenstein}}, \ and\ \bibinfo {author} {\bibfnamefont {D.}~\bibnamefont {Loss}},\ }\bibfield  {title} {\enquote {\bibinfo {title} {Quantum correlations in two-fermion systems},}\ }\href@noop {} {\bibfield  {journal} {\bibinfo  {journal} {Phys. Rev. A}\ }\textbf {\bibinfo {volume} {64}},\ \bibinfo {pages} {022303} (\bibinfo {year} {2001})}\BibitemShut {NoStop}%
\bibitem [{\citenamefont {Eckert}\ \emph {et~al.}(2002)\citenamefont {Eckert}, \citenamefont {Schliemann}, \citenamefont {Bru{\ss}},\ and\ \citenamefont {Lewenstein}}]{ES.02}%
  \BibitemOpen
  \bibfield  {author} {\bibinfo {author} {\bibfnamefont {K.}~\bibnamefont {Eckert}}, \bibinfo {author} {\bibfnamefont {J.}~\bibnamefont {Schliemann}}, \bibinfo {author} {\bibfnamefont {D.}~\bibnamefont {Bru{\ss}}}, \ and\ \bibinfo {author} {\bibfnamefont {M.}~\bibnamefont {Lewenstein}},\ }\bibfield  {title} {\enquote {\bibinfo {title} {Quantum correlations in systems of indistinguishable particles},}\ }\href@noop {} {\bibfield  {journal} {\bibinfo  {journal} {Ann. Phys.}\ }\textbf {\bibinfo {volume} {299}},\ \bibinfo {pages} {88} (\bibinfo {year} {2002})}\BibitemShut {NoStop}%
\bibitem [{\citenamefont {Zanardi}(2002)}]{Za.02}%
  \BibitemOpen
  \bibfield  {author} {\bibinfo {author} {\bibfnamefont {P.}~\bibnamefont {Zanardi}},\ }\bibfield  {title} {\enquote {\bibinfo {title} {Quantum entanglement in fermionic lattices},}\ }\href@noop {} {\bibfield  {journal} {\bibinfo  {journal} {Phys. Rev. A}\ }\textbf {\bibinfo {volume} {65}},\ \bibinfo {pages} {042101} (\bibinfo {year} {2002})}\BibitemShut {NoStop}%
\bibitem [{\citenamefont {Friis}\ \emph {et~al.}(2013)\citenamefont {Friis}, \citenamefont {Lee},\ and\ \citenamefont {Bruschi}}]{FL.13}%
  \BibitemOpen
  \bibfield  {author} {\bibinfo {author} {\bibfnamefont {N.}~\bibnamefont {Friis}}, \bibinfo {author} {\bibfnamefont {A.R.}\ \bibnamefont {Lee}}, \ and\ \bibinfo {author} {\bibfnamefont {D.E.}\ \bibnamefont {Bruschi}},\ }\bibfield  {title} {\enquote {\bibinfo {title} {Fermionic-mode entanglement in quantum information},}\ }\href@noop {} {\bibfield  {journal} {\bibinfo  {journal} {Phys. Rev. A}\ }\textbf {\bibinfo {volume} {87}},\ \bibinfo {pages} {022338} (\bibinfo {year} {2013})}\BibitemShut {NoStop}%
\bibitem [{\citenamefont {Spee}\ \emph {et~al.}(2018)\citenamefont {Spee}, \citenamefont {Schwaiger}, \citenamefont {Giedke},\ and\ \citenamefont {Kraus}}]{SSG.18}%
  \BibitemOpen
  \bibfield  {author} {\bibinfo {author} {\bibfnamefont {C.}~\bibnamefont {Spee}}, \bibinfo {author} {\bibfnamefont {K.}~\bibnamefont {Schwaiger}}, \bibinfo {author} {\bibfnamefont {G.}~\bibnamefont {Giedke}}, \ and\ \bibinfo {author} {\bibfnamefont {B.}~\bibnamefont {Kraus}},\ }\bibfield  {title} {\enquote {\bibinfo {title} {Mode entanglement of gaussian fermionic states},}\ }\href {\doibase 10.1103/PhysRevA.97.042325} {\bibfield  {journal} {\bibinfo  {journal} {Phys. Rev. A}\ }\textbf {\bibinfo {volume} {97}},\ \bibinfo {pages} {042325} (\bibinfo {year} {2018})}\BibitemShut {NoStop}%
\bibitem [{\citenamefont {Gigena}\ and\ \citenamefont {Rossignoli}(2015)}]{GR.15}%
  \BibitemOpen
  \bibfield  {author} {\bibinfo {author} {\bibfnamefont {N.}~\bibnamefont {Gigena}}\ and\ \bibinfo {author} {\bibfnamefont {R.}~\bibnamefont {Rossignoli}},\ }\bibfield  {title} {\enquote {\bibinfo {title} {Entanglement in fermion systems},}\ }\href {\doibase 10.1103/PhysRevA.92.042326} {\bibfield  {journal} {\bibinfo  {journal} {Phys. Rev. A}\ }\textbf {\bibinfo {volume} {92}},\ \bibinfo {pages} {042326} (\bibinfo {year} {2015})}\BibitemShut {NoStop}%
\bibitem [{\citenamefont {Gigena}\ \emph {et~al.}(2020)\citenamefont {Gigena}, \citenamefont {Tullio},\ and\ \citenamefont {Rossignoli}}]{GDR.20}%
  \BibitemOpen
  \bibfield  {author} {\bibinfo {author} {\bibfnamefont {N.}~\bibnamefont {Gigena}}, \bibinfo {author} {\bibfnamefont {M.~Di}\ \bibnamefont {Tullio}}, \ and\ \bibinfo {author} {\bibfnamefont {R.}~\bibnamefont {Rossignoli}},\ }\bibfield  {title} {\enquote {\bibinfo {title} {One-body entanglement as a quantum resource in fermionic systems},}\ }\href {\doibase 10.1103/PhysRevA.102.042410} {\bibfield  {journal} {\bibinfo  {journal} {Phys. Rev. A}\ }\textbf {\bibinfo {volume} {102}},\ \bibinfo {pages} {042410} (\bibinfo {year} {2020})}\BibitemShut {NoStop}%
\bibitem [{\citenamefont {Gigena}\ \emph {et~al.}(2021)\citenamefont {Gigena}, \citenamefont {Tullio},\ and\ \citenamefont {Rossignoli}}]{GDR.21}%
  \BibitemOpen
  \bibfield  {author} {\bibinfo {author} {\bibfnamefont {N.}~\bibnamefont {Gigena}}, \bibinfo {author} {\bibfnamefont {M.~Di}\ \bibnamefont {Tullio}}, \ and\ \bibinfo {author} {\bibfnamefont {R.}~\bibnamefont {Rossignoli}},\ }\bibfield  {title} {\enquote {\bibinfo {title} {Many-body entanglement in fermion systems},}\ }\href {\doibase 10.1103/PhysRevA.103.052424} {\bibfield  {journal} {\bibinfo  {journal} {Phys. Rev. A}\ }\textbf {\bibinfo {volume} {103}},\ \bibinfo {pages} {052424} (\bibinfo {year} {2021})}\BibitemShut {NoStop}%
\bibitem [{\citenamefont {Iemini}\ \emph {et~al.}(2014)\citenamefont {Iemini}, \citenamefont {Debarba},\ and\ \citenamefont {Vianna}}]{Iem.14}%
  \BibitemOpen
  \bibfield  {author} {\bibinfo {author} {\bibfnamefont {F.}~\bibnamefont {Iemini}}, \bibinfo {author} {\bibfnamefont {T.}~\bibnamefont {Debarba}}, \ and\ \bibinfo {author} {\bibfnamefont {R.O.}\ \bibnamefont {Vianna}},\ }\bibfield  {title} {\enquote {\bibinfo {title} {Quantumness of correlations in indistinguishable particles},}\ }\href@noop {} {\bibfield  {journal} {\bibinfo  {journal} {Phys. Rev. A}\ }\textbf {\bibinfo {volume} {89}},\ \bibinfo {pages} {032324} (\bibinfo {year} {2014})}\BibitemShut {NoStop}%
\bibitem [{\citenamefont {Majtey}\ \emph {et~al.}(2016)\citenamefont {Majtey}, \citenamefont {Bouvrie}, \citenamefont {Vald{\'e}s-Hern{\'a}ndez},\ and\ \citenamefont {Plastino}}]{MB.16}%
  \BibitemOpen
  \bibfield  {author} {\bibinfo {author} {\bibfnamefont {A.P.}\ \bibnamefont {Majtey}}, \bibinfo {author} {\bibfnamefont {P.A.}\ \bibnamefont {Bouvrie}}, \bibinfo {author} {\bibfnamefont {A.}~\bibnamefont {Vald{\'e}s-Hern{\'a}ndez}}, \ and\ \bibinfo {author} {\bibfnamefont {A.R.}\ \bibnamefont {Plastino}},\ }\bibfield  {title} {\enquote {\bibinfo {title} {Multipartite concurrence for identical-fermion systems},}\ }\href@noop {} {\bibfield  {journal} {\bibinfo  {journal} {Phys. Rev. A}\ }\textbf {\bibinfo {volume} {93}},\ \bibinfo {pages} {032335} (\bibinfo {year} {2016})}\BibitemShut {NoStop}%
\bibitem [{\citenamefont {Gigena}\ and\ \citenamefont {Rossignoli}(2017)}]{GR.17}%
  \BibitemOpen
  \bibfield  {author} {\bibinfo {author} {\bibfnamefont {N.}~\bibnamefont {Gigena}}\ and\ \bibinfo {author} {\bibfnamefont {R.}~\bibnamefont {Rossignoli}},\ }\bibfield  {title} {\enquote {\bibinfo {title} {Bipartite entanglement in fermion systems},}\ }\href@noop {} {\bibfield  {journal} {\bibinfo  {journal} {Phys. Rev. A}\ }\textbf {\bibinfo {volume} {95}},\ \bibinfo {pages} {062320} (\bibinfo {year} {2017})}\BibitemShut {NoStop}%
\bibitem [{\citenamefont {da~Silva~Souza}\ \emph {et~al.}(2018)\citenamefont {da~Silva~Souza}, \citenamefont {Debarba}, \citenamefont {Braga-Ferreira}, \citenamefont {Iemini},\ and\ \citenamefont {Vianna}}]{SD.18}%
  \BibitemOpen
  \bibfield  {author} {\bibinfo {author} {\bibfnamefont {L.}~\bibnamefont {da~Silva~Souza}}, \bibinfo {author} {\bibfnamefont {T.}~\bibnamefont {Debarba}}, \bibinfo {author} {\bibfnamefont {D.~L.}\ \bibnamefont {Braga-Ferreira}}, \bibinfo {author} {\bibfnamefont {F.}~\bibnamefont {Iemini}}, \ and\ \bibinfo {author} {\bibfnamefont {R.~O.}\ \bibnamefont {Vianna}},\ }\bibfield  {title} {\enquote {\bibinfo {title} {Completely positive maps for reduced states of indistinguishable particles},}\ }\href@noop {} {\bibfield  {journal} {\bibinfo  {journal} {Phys. Rev. A}\ }\textbf {\bibinfo {volume} {98}},\ \bibinfo {pages} {052135} (\bibinfo {year} {2018})}\BibitemShut {NoStop}%
\bibitem [{\citenamefont {Tullio}\ \emph {et~al.}(2018)\citenamefont {Tullio}, \citenamefont {Gigena},\ and\ \citenamefont {Rossignoli}}]{DGR.18}%
  \BibitemOpen
  \bibfield  {author} {\bibinfo {author} {\bibfnamefont {M.~Di}\ \bibnamefont {Tullio}}, \bibinfo {author} {\bibfnamefont {N.}~\bibnamefont {Gigena}}, \ and\ \bibinfo {author} {\bibfnamefont {R.}~\bibnamefont {Rossignoli}},\ }\bibfield  {title} {\enquote {\bibinfo {title} {Fermionic entanglement in superconducting systems},}\ }\href@noop {} {\bibfield  {journal} {\bibinfo  {journal} {Phys. Rev. A}\ }\textbf {\bibinfo {volume} {97}},\ \bibinfo {pages} {062109} (\bibinfo {year} {2018})}\BibitemShut {NoStop}%
\bibitem [{\citenamefont {Tullio}\ \emph {et~al.}(2019)\citenamefont {Tullio}, \citenamefont {Rossignoli}, \citenamefont {Cerezo},\ and\ \citenamefont {Gigena}}]{DRCG.19}%
  \BibitemOpen
  \bibfield  {author} {\bibinfo {author} {\bibfnamefont {M.~Di}\ \bibnamefont {Tullio}}, \bibinfo {author} {\bibfnamefont {R.}~\bibnamefont {Rossignoli}}, \bibinfo {author} {\bibfnamefont {M.}~\bibnamefont {Cerezo}}, \ and\ \bibinfo {author} {\bibfnamefont {N.}~\bibnamefont {Gigena}},\ }\bibfield  {title} {\enquote {\bibinfo {title} {Fermionic entanglement in the {L}ipkin model},}\ }\href@noop {} {\bibfield  {journal} {\bibinfo  {journal} {Phys. Rev. A}\ }\textbf {\bibinfo {volume} {100}},\ \bibinfo {pages} {062104} (\bibinfo {year} {2019})}\BibitemShut {NoStop}%
\bibitem [{\citenamefont {Cianciulli}\ \emph {et~al.}(2024)\citenamefont {Cianciulli}, \citenamefont {Rossignoli}, \citenamefont {Tullio}, \citenamefont {Gigena},\ and\ \citenamefont {Petrovich}}]{CR.24}%
  \BibitemOpen
  \bibfield  {author} {\bibinfo {author} {\bibfnamefont {J.~A.}\ \bibnamefont {Cianciulli}}, \bibinfo {author} {\bibfnamefont {R.}~\bibnamefont {Rossignoli}}, \bibinfo {author} {\bibfnamefont {M.~Di}\ \bibnamefont {Tullio}}, \bibinfo {author} {\bibfnamefont {N.}~\bibnamefont {Gigena}}, \ and\ \bibinfo {author} {\bibfnamefont {F.}~\bibnamefont {Petrovich}},\ }\bibfield  {title} {\enquote {\bibinfo {title} {Bipartite representations and many-body entanglement of pure states of {$N$} indistinguishable particles},}\ }\href {\doibase 10.1103/PhysRevA.110.032414} {\bibfield  {journal} {\bibinfo  {journal} {Phys. Rev. A}\ }\textbf {\bibinfo {volume} {110}},\ \bibinfo {pages} {032414} (\bibinfo {year} {2024})}\BibitemShut {NoStop}%
\bibitem [{\citenamefont {Wang}\ and\ \citenamefont {Kais}(2007)}]{Wang_2007}%
  \BibitemOpen
  \bibfield  {author} {\bibinfo {author} {\bibfnamefont {H.}~\bibnamefont {Wang}}\ and\ \bibinfo {author} {\bibfnamefont {S.}~\bibnamefont {Kais}},\ }\bibfield  {title} {\enquote {\bibinfo {title} {Quantum entanglement and electron correlation in molecular systems},}\ }\href {\doibase 10.1560/ijc.47.1.59} {\bibfield  {journal} {\bibinfo  {journal} {Isr. J. Chem.}\ }\textbf {\bibinfo {volume} {47}},\ \bibinfo {pages} {59} (\bibinfo {year} {2007})}\BibitemShut {NoStop}%
\bibitem [{\citenamefont {Luzanov}\ and\ \citenamefont {Prezhdo}(2007)}]{Luzanov_2007}%
  \BibitemOpen
  \bibfield  {author} {\bibinfo {author} {\bibfnamefont {A.~V.}\ \bibnamefont {Luzanov}}\ and\ \bibinfo {author} {\bibfnamefont {O.~V.}\ \bibnamefont {Prezhdo}},\ }\bibfield  {title} {\enquote {\bibinfo {title} {High-order entropy measures and spin-free quantum entanglement for molecular problems},}\ }\href {\doibase 10.1080/00268970701725039} {\bibfield  {journal} {\bibinfo  {journal} {Mol. Phys.}\ }\textbf {\bibinfo {volume} {105}},\ \bibinfo {pages} {2879} (\bibinfo {year} {2007})}\BibitemShut {NoStop}%
\bibitem [{\citenamefont {Ivanov}\ \emph {et~al.}(2005)\citenamefont {Ivanov}, \citenamefont {Lyakh},\ and\ \citenamefont {Adamowicz}}]{Ivanov_2005}%
  \BibitemOpen
  \bibfield  {author} {\bibinfo {author} {\bibfnamefont {V.V.}\ \bibnamefont {Ivanov}}, \bibinfo {author} {\bibfnamefont {D.I.}\ \bibnamefont {Lyakh}}, \ %and\ 
  \bibinfo {author} {\bibfnamefont {L.}~\bibnamefont {Adamowicz}},\ }\bibfield  {title} {\enquote {\bibinfo {title} {New indices for describing the multi-configurational nature of the coupled cluster wave function},}\ }\href {\doibase 10.1080/00268970500083283} {\bibfield  {journal} {\bibinfo  {journal} {Mol. Phys.}\ }\textbf {\bibinfo {volume} {103}},\ \bibinfo {pages} {2131} (\bibinfo {year} {2005})}\BibitemShut {NoStop}%
\bibitem [{\citenamefont {Alcoba}\ \emph {et~al.}(2016)\citenamefont {Alcoba}, \citenamefont {Torre}, \citenamefont {Lain}, \citenamefont {Massaccesi}, \citenamefont {{O\~na}}, \citenamefont {Ayers}, \citenamefont {Raemdonck}, \citenamefont {Bultinck},\ and\ \citenamefont {Neck}}]{Alcoba_2016}%
  \BibitemOpen
  \bibfield  {author} {\bibinfo {author} {\bibfnamefont {D.~R.}\ \bibnamefont {Alcoba}}, \bibinfo {author} {\bibfnamefont {A.}~\bibnamefont {Torre}}, \bibinfo {author} {\bibfnamefont {L.}~\bibnamefont {Lain}}, \bibinfo {author} {\bibfnamefont {G.~E.}\ \bibnamefont {Massaccesi}}, \bibinfo {author} {\bibfnamefont {O.~B.}\ \bibnamefont {{O\~na}}}, \bibinfo {author} {\bibfnamefont {P.~W.}\ \bibnamefont {Ayers}}, \bibinfo {author} {\bibfnamefont {M.~Van}\ \bibnamefont {Raemdonck}}, \bibinfo {author} {\bibfnamefont {P.}~\bibnamefont {Bultinck}}, \ and\ \bibinfo {author} {\bibfnamefont {D.~Van}\ \bibnamefont {Neck}},\ }\bibfield  {title} {\enquote {\bibinfo {title} {Performance of {S}hannon-entropy compacted {$N$}-electron wave functions for configuration interaction methods},}\ }\href {\doibase 10.1007/s00214-016-1905-x} {\bibfield  {journal} {\bibinfo  {journal} {Theor. Chem. Acc.}\ }\textbf {\bibinfo {volume} {135}},\ \bibinfo {pages} {153} (\bibinfo {year} {2016})}\BibitemShut {NoStop}%
\bibitem [{\citenamefont {Juhász}\ and\ \citenamefont {Mazziotti}(2006)}]{Juh_sz_2006}%
  \BibitemOpen
  \bibfield  {author} {\bibinfo {author} {\bibfnamefont {T.}~\bibnamefont {Juhász}}\ and\ \bibinfo {author} {\bibfnamefont {D.A.}\ \bibnamefont {Mazziotti}},\ }\bibfield  {title} {\enquote {\bibinfo {title} {The cumulant two-particle reduced density matrix as a measure of electron correlation and entanglement},}\ }\href {\doibase 10.1063/1.2378768} {\bibfield  {journal} {\bibinfo  {journal} {J. Chem. Phys.}\ }\textbf {\bibinfo {volume} {125}},\ \bibinfo {pages} {174105} (\bibinfo {year} {2006})}\BibitemShut {NoStop}%
\bibitem [{\citenamefont {Alcoba}\ \emph {et~al.}(2010)\citenamefont {Alcoba}, \citenamefont {Bochicchio}, \citenamefont {Lain},\ and\ \citenamefont {Torre}}]{Alcoba_2010}%
  \BibitemOpen
  \bibfield  {author} {\bibinfo {author} {\bibfnamefont {D.~R.}\ \bibnamefont {Alcoba}}, \bibinfo {author} {\bibfnamefont {R.~C.}\ \bibnamefont {Bochicchio}}, \bibinfo {author} {\bibfnamefont {L.}~\bibnamefont {Lain}}, \ and\ \bibinfo {author} {\bibfnamefont {A.}~\bibnamefont {Torre}},\ }\bibfield  {title} {\enquote {\bibinfo {title} {On the measure of electron correlation and entanglement in quantum chemistry based on the cumulant of the second-order reduced density matrix},}\ }\href {\doibase 10.1063/1.3503766} {\bibfield  {journal} {\bibinfo  {journal} {J. Chem. Phys.}\ }\textbf {\bibinfo {volume} {133}},\ \bibinfo {pages} {144104} (\bibinfo {year} {2010})}\BibitemShut {NoStop}%
\bibitem [{\citenamefont {Li}\ \emph {et~al.}(2021)\citenamefont {Li}, \citenamefont {Liebenthal},\ and\ \citenamefont {DePrince}}]{Li_2021}%
  \BibitemOpen
  \bibfield  {author} {\bibinfo {author} {\bibfnamefont {R.~R.}\ \bibnamefont {Li}}, \bibinfo {author} {\bibfnamefont {M.~D.}\ \bibnamefont {Liebenthal}}, \ and\ \bibinfo {author} {\bibfnamefont {A.~E.}\ \bibnamefont {DePrince}},\ }\bibfield  {title} {\enquote {\bibinfo {title} {Challenges for variational reduced-density-matrix theory with three-particle {$N$}-representability conditions},}\ }\href {\doibase 10.1063/5.0066404} {\bibfield  {journal} {\bibinfo  {journal} {J. Chem. Phys.}\ }\textbf {\bibinfo {volume} {155}},\ \bibinfo {pages} {174110} (\bibinfo {year} {2021})}\BibitemShut {NoStop}%
\bibitem [{\citenamefont {Benavides-Riveros}\ \emph {et~al.}(2017)\citenamefont {Benavides-Riveros}, \citenamefont {Lathiotakis},\ and\ \citenamefont {Marques}}]{Benavides_Riveros_2017}%
  \BibitemOpen
  \bibfield  {author} {\bibinfo {author} {\bibfnamefont {C.~L.}\ \bibnamefont {Benavides-Riveros}}, \bibinfo {author} {\bibfnamefont {N.~N.}\ \bibnamefont {Lathiotakis}}, \ and\ \bibinfo {author} {\bibfnamefont {M.~A.~L.}\ \bibnamefont {Marques}},\ }\bibfield  {title} {\enquote {\bibinfo {title} {Towards a formal definition of static and dynamic electronic correlations},}\ }\href {\doibase 10.1039/c7cp01137g} {\bibfield  {journal} {\bibinfo  {journal} {Phys. Chem. Chem. Phys.}\ }\textbf {\bibinfo {volume} {19}},\ \bibinfo {pages} {12655} (\bibinfo {year} {2017})}\BibitemShut {NoStop}%
\bibitem [{\citenamefont {Izs\'ak}\ \emph {et~al.}(2023)\citenamefont {Izs\'ak}, \citenamefont {Ivanov}, \citenamefont {Blunt}, \citenamefont {Holzmann},\ and\ \citenamefont {Neese}}]{Izs_k_2023}%
  \BibitemOpen
  \bibfield  {author} {\bibinfo {author} {\bibfnamefont {R.}~\bibnamefont {Izs\'ak}}, \bibinfo {author} {\bibfnamefont {A.~V.}\ \bibnamefont {Ivanov}}, \bibinfo {author} {\bibfnamefont {N.~S.}\ \bibnamefont {Blunt}}, \bibinfo {author} {\bibfnamefont {N.}~\bibnamefont {Holzmann}}, \ and\ \bibinfo {author} {\bibfnamefont {F.}~\bibnamefont {Neese}},\ }\bibfield  {title} {\enquote {\bibinfo {title} {Measuring electron correlation: The impact of symmetry and orbital transformations},}\ }\href {\doibase 10.1021/acs.jctc.3c00122} {\bibfield  {journal} {\bibinfo  {journal} {J. Chem. Theory Comput.}\ }\textbf {\bibinfo {volume} {19}},\ \bibinfo {pages} {2703} (\bibinfo {year} {2023})}\BibitemShut {NoStop}%
\bibitem [{\citenamefont {\v{S}ulka}\ \emph {et~al.}(2023)\citenamefont {\v{S}ulka}, \citenamefont {\v{S}ulkov\'a}, \citenamefont {Jure\v{c}ka},\ and\ \citenamefont {Dubeck\'y}}]{_ulka_2023}%
  \BibitemOpen
  \bibfield  {author} {\bibinfo {author} {\bibfnamefont {M.}~\bibnamefont {\v{S}ulka}}, \bibinfo {author} {\bibfnamefont {K.}~\bibnamefont {\v{S}ulkov\'a}}, \bibinfo {author} {\bibfnamefont {P.}~\bibnamefont {Jure\v{c}ka}}, \ and\ \bibinfo {author} {\bibfnamefont {M.}~\bibnamefont {Dubeck\'y}},\ }\bibfield  {title} {\enquote {\bibinfo {title} {Dynamic and nondynamic electron correlation energy decomposition based on the node of the {H}artree–{F}ock {S}later determinant},}\ }\href {\doibase 10.1021/acs.jctc.3c00828} {\bibfield  {journal} {\bibinfo  {journal} {J. Chem. Theory Comput.}\ }\textbf {\bibinfo {volume} {19}},\ \bibinfo {pages} {8147} (\bibinfo {year} {2023})}\BibitemShut {NoStop}%
\bibitem [{\citenamefont {Ganoe}\ and\ \citenamefont {Shee}(2024)}]{Ganoe_2024}%
  \BibitemOpen
  \bibfield  {author} {\bibinfo {author} {\bibfnamefont {B.}~\bibnamefont {Ganoe}}\ and\ \bibinfo {author} {\bibfnamefont {J.}~\bibnamefont {Shee}},\ }\bibfield  {title} {\enquote {\bibinfo {title} {On the notion of strong correlation in electronic structure theory},}\ }\href {\doibase 10.1039/d4fd00066h} {\bibfield  {journal} {\bibinfo  {journal} {Faraday Discuss.}\ }\textbf {\bibinfo {volume} {254}},\ \bibinfo {pages} {53} (\bibinfo {year} {2024})}\BibitemShut {NoStop}%
\bibitem [{\citenamefont {Boguslawski}\ \emph {et~al.}(2012)\citenamefont {Boguslawski}, \citenamefont {Tecmer}, \citenamefont {Legeza},\ and\ \citenamefont {Reiher}}]{Boguslawski_2012}%
  \BibitemOpen
  \bibfield  {author} {\bibinfo {author} {\bibfnamefont {K.}~\bibnamefont {Boguslawski}}, \bibinfo {author} {\bibfnamefont {P.}~\bibnamefont {Tecmer}}, \bibinfo {author} {\bibfnamefont {\"O.}\ \bibnamefont {Legeza}}, \ and\ \bibinfo {author} {\bibfnamefont {M.}~\bibnamefont {Reiher}},\ }\bibfield  {title} {\enquote {\bibinfo {title} {Entanglement measures for single- and multireference correlation effects},}\ }\href {\doibase 10.1021/jz301319v} {\bibfield  {journal} {\bibinfo  {journal} {J. Phys. Chem. Lett.}\ }\textbf {\bibinfo {volume} {3}},\ \bibinfo {pages} {3129} (\bibinfo {year} {2012})}\BibitemShut {NoStop}%
\bibitem [{\citenamefont {Boguslawski}\ \emph {et~al.}(2013)\citenamefont {Boguslawski}, \citenamefont {Tecmer}, \citenamefont {Barcza}, \citenamefont {Legeza},\ and\ \citenamefont {Reiher}}]{Boguslawski_2013}%
  \BibitemOpen
  \bibfield  {author} {\bibinfo {author} {\bibfnamefont {K.}~\bibnamefont {Boguslawski}}, \bibinfo {author} {\bibfnamefont {P.}~\bibnamefont {Tecmer}}, \bibinfo {author} {\bibfnamefont {G.}~\bibnamefont {Barcza}}, \bibinfo {author} {\bibfnamefont {\"O.}\ \bibnamefont {Legeza}}, \ and\ \bibinfo {author} {\bibfnamefont {M.}~\bibnamefont {Reiher}},\ }\bibfield  {title} {\enquote {\bibinfo {title} {Orbital entanglement in bond-formation processes},}\ }\href {\doibase 10.1021/ct400247p} {\bibfield  {journal} {\bibinfo  {journal} {J. Chem. Theory Comput.}\ }\textbf {\bibinfo {volume} {9}},\ \bibinfo {pages} {2959} (\bibinfo {year} {2013})}\BibitemShut {NoStop}%
\bibitem [{\citenamefont {Boguslawski}\ and\ \citenamefont {Tecmer}(2014)}]{Boguslawski_2014_a}%
  \BibitemOpen
  \bibfield  {author} {\bibinfo {author} {\bibfnamefont {K.}~\bibnamefont {Boguslawski}}\ and\ \bibinfo {author} {\bibfnamefont {P.}~\bibnamefont {Tecmer}},\ }\bibfield  {title} {\enquote {\bibinfo {title} {Orbital entanglement in quantum chemistry},}\ }\href {\doibase 10.1002/qua.24832} {\bibfield  {journal} {\bibinfo  {journal} {Int. J. Q. Chem.}\ }\textbf {\bibinfo {volume} {115}},\ \bibinfo {pages} {1289} (\bibinfo {year} {2014})}\BibitemShut {NoStop}%
\bibitem [{\citenamefont {Ding}\ \emph {et~al.}(2021)\citenamefont {Ding}, \citenamefont {Mardazad}, \citenamefont {Das}, \citenamefont {Szalay}, \citenamefont {Schollwo\"ck}, \citenamefont {Zimbor\'as},\ and\ \citenamefont {Schilling}}]{DS.21}%
  \BibitemOpen
  \bibfield  {author} {\bibinfo {author} {\bibfnamefont {L.}~\bibnamefont {Ding}}, \bibinfo {author} {\bibfnamefont {S.}~\bibnamefont {Mardazad}}, \bibinfo {author} {\bibfnamefont {S.}~\bibnamefont {Das}}, \bibinfo {author} {\bibfnamefont {S.}~\bibnamefont {Szalay}}, \bibinfo {author} {\bibfnamefont {U.}~\bibnamefont {Schollwo\"ck}}, \bibinfo {author} {\bibfnamefont {Z.}~\bibnamefont {Zimbor\'as}}, \ and\ \bibinfo {author} {\bibfnamefont {C.}~\bibnamefont {Schilling}},\ }\bibfield  {title} {\enquote {\bibinfo {title} {Concept of orbital entanglement and correlation in quantum chemistry},}\ }\href {\doibase 10.1021/acs.jctc.0c00559} {\bibfield  {journal} {\bibinfo  {journal} {J. Chem. Theory Comput.}\ }\textbf {\bibinfo {volume} {17}},\ \bibinfo {pages} {79} (\bibinfo {year} {2021})}\BibitemShut {NoStop}%
\bibitem [{\citenamefont {Ding}\ \emph {et~al.}(2022)\citenamefont {Ding}, \citenamefont {Knecht}, \citenamefont {Zimborás},\ and\ \citenamefont {Schilling}}]{Ding_2022}%
  \BibitemOpen
  \bibfield  {author} {\bibinfo {author} {\bibfnamefont {L.}~\bibnamefont {Ding}}, \bibinfo {author} {\bibfnamefont {S.}~\bibnamefont {Knecht}}, \bibinfo {author} {\bibfnamefont {Z.}~\bibnamefont {Zimborás}}, \ and\ \bibinfo {author} {\bibfnamefont {C.}~\bibnamefont {Schilling}},\ }\bibfield  {title} {\enquote {\bibinfo {title} {Quantum correlations in molecules: from quantum resourcing to chemical bonding},}\ }\href {\doibase 10.1088/2058-9565/aca4ee} {\bibfield  {journal} {\bibinfo  {journal} {Quantum Sci. Technol.}\ }\textbf {\bibinfo {volume} {8}},\ \bibinfo {pages} {015015} (\bibinfo {year} {2022})}\BibitemShut {NoStop}%
\bibitem [{\citenamefont {Ding}\ \emph {et~al.}(2023{\natexlab{a}})\citenamefont {Ding}, \citenamefont {D\"unnweber},\ and\ \citenamefont {Schilling}}]{Ding_2023}%
  \BibitemOpen
  \bibfield  {author} {\bibinfo {author} {\bibfnamefont {L.}~\bibnamefont {Ding}}, \bibinfo {author} {\bibfnamefont {G.}~\bibnamefont {D\"unnweber}}, \ and\ \bibinfo {author} {\bibfnamefont {C.}~\bibnamefont {Schilling}},\ }\bibfield  {title} {\enquote {\bibinfo {title} {Physical entanglement between localized orbitals},}\ }\href {\doibase 10.1088/2058-9565/ad00d9} {\bibfield  {journal} {\bibinfo  {journal} {Quantum Sci. Technol.}\ }\textbf {\bibinfo {volume} {9}},\ \bibinfo {pages} {015005} (\bibinfo {year} {2023}{\natexlab{a}})}\BibitemShut {NoStop}%
\bibitem [{\citenamefont {Ding}\ \emph {et~al.}(2023{\natexlab{b}})\citenamefont {Ding}, \citenamefont {Knecht},\ and\ \citenamefont {Schilling}}]{Ding_2023_a}%
  \BibitemOpen
  \bibfield  {author} {\bibinfo {author} {\bibfnamefont {L.}~\bibnamefont {Ding}}, \bibinfo {author} {\bibfnamefont {S.}~\bibnamefont {Knecht}}, \ and\ \bibinfo {author} {\bibfnamefont {C.}~\bibnamefont {Schilling}},\ }\bibfield  {title} {\enquote {\bibinfo {title} {Quantum information-assisted complete active space optimization {(QICAS)}},}\ }\href {\doibase 10.1021/acs.jpclett.3c02536} {\bibfield  {journal} {\bibinfo  {journal} {J. Phys. Chem. Lett.}\ }\textbf {\bibinfo {volume} {14}},\ \bibinfo {pages} {11022} (\bibinfo {year} {2023}{\natexlab{b}})}\BibitemShut {NoStop}%
\bibitem [{\citenamefont {Ratini}\ \emph {et~al.}(2024)\citenamefont {Ratini}, \citenamefont {Capecci},\ and\ \citenamefont {Guidoni}}]{Ratini_2024}%
  \BibitemOpen
  \bibfield  {author} {\bibinfo {author} {\bibfnamefont {L.}~\bibnamefont {Ratini}}, \bibinfo {author} {\bibfnamefont {C.}~\bibnamefont {Capecci}}, \ and\ \bibinfo {author} {\bibfnamefont {L.}~\bibnamefont {Guidoni}},\ }\bibfield  {title} {\enquote {\bibinfo {title} {Natural orbitals and sparsity of quantum mutual information},}\ }\href {\doibase 10.1021/acs.jctc.3c01325} {\bibfield  {journal} {\bibinfo  {journal} {J. Chem. Theory Comput.}\ }\textbf {\bibinfo {volume} {20}},\ \bibinfo {pages} {3535} (\bibinfo {year} {2024})}\BibitemShut {NoStop}%
\bibitem [{\citenamefont {Vidal}\ and\ \citenamefont {Werner}(2002)}]{VW.02}%
  \BibitemOpen
  \bibfield  {author} {\bibinfo {author} {\bibfnamefont {G.}~\bibnamefont {Vidal}}\ and\ \bibinfo {author} {\bibfnamefont {R.~F.}\ \bibnamefont {Werner}},\ }\bibfield  {title} {\enquote {\bibinfo {title} {Computable measure of entanglement},}\ }\href {\doibase 10.1103/PhysRevA.65.032314} {\bibfield  {journal} {\bibinfo  {journal} {Phys. Rev. A}\ }\textbf {\bibinfo {volume} {65}},\ \bibinfo {pages} {032314} (\bibinfo {year} {2002})}\BibitemShut {NoStop}%
\bibitem [{\citenamefont {Plenio}(2005)}]{P.05}%
  \BibitemOpen
  \bibfield  {author} {\bibinfo {author} {\bibfnamefont {M.~B.}\ \bibnamefont {Plenio}},\ }\bibfield  {title} {\enquote {\bibinfo {title} {Logarithmic negativity: A full entanglement monotone that is not convex},}\ }\href {\doibase 10.1103/PhysRevLett.95.090503} {\bibfield  {journal} {\bibinfo  {journal} {Phys. Rev. Lett.}\ }\textbf {\bibinfo {volume} {95}},\ \bibinfo {pages} {090503} (\bibinfo {year} {2005})}\BibitemShut {NoStop}%
\bibitem [{\citenamefont {Ding}\ and\ \citenamefont {Schilling}(2020)}]{DS.20}%
  \BibitemOpen
  \bibfield  {author} {\bibinfo {author} {\bibfnamefont {L.}~\bibnamefont {Ding}}\ and\ \bibinfo {author} {\bibfnamefont {C.}~\bibnamefont {Schilling}},\ }\bibfield  {title} {\enquote {\bibinfo {title} {Correlation paradox of the dissociation limit: A quantum information perspective},}\ }\href {\doibase 10.1021/acs.jctc.0c00054} {\bibfield  {journal} {\bibinfo  {journal} {J. Chem. Theory Comput.}\ }\textbf {\bibinfo {volume} {16}},\ \bibinfo {pages} {4159} (\bibinfo {year} {2020})}\BibitemShut {NoStop}%
\bibitem [{\citenamefont {Graves}\ \emph {et~al.}(2025)\citenamefont {Graves}, \citenamefont {S\"underhauf}, \citenamefont {Blunt}, \citenamefont {Izsák},\ and\ \citenamefont {Szőri}}]{GSB.25}%
  \BibitemOpen
  \bibfield  {author} {\bibinfo {author} {\bibfnamefont {V.}~\bibnamefont {Graves}}, \bibinfo {author} {\bibfnamefont {C.}~\bibnamefont {S\"underhauf}}, \bibinfo {author} {\bibfnamefont {N.S.}\ \bibnamefont {Blunt}}, \bibinfo {author} {\bibfnamefont {R.}~\bibnamefont {Izsák}}, \ and\ \bibinfo {author} {\bibfnamefont {M.}~\bibnamefont {Szőri}},\ }\bibfield  {title} {\enquote {\bibinfo {title} {The electronic structure of the hydrogen molecule: A tutorial exercise in classical and quantum computation},}\ }\href {\doibase doi.org/10.1021/acsphyschemau.5c0003010.1088/0264-9381/12/5/011} {\bibfield  {journal} {\bibinfo  {journal} {ACS Phys. Chem Au}\ }\textbf {\bibinfo {volume} {5}},\ \bibinfo {pages} {435} (\bibinfo {year} {2025})}\BibitemShut {NoStop}%
\bibitem [{\citenamefont {Nielsen}\ and\ \citenamefont {Kempe}(2001)}]{NK.01}%
  \BibitemOpen
  \bibfield  {author} {\bibinfo {author} {\bibfnamefont {M.~A.}\ \bibnamefont {Nielsen}}\ and\ \bibinfo {author} {\bibfnamefont {J.}~\bibnamefont {Kempe}},\ }\bibfield  {title} {\enquote {\bibinfo {title} {Separable states are more disordered globally than locally},}\ }\href {\doibase 10.1103/PhysRevLett.86.5184} {\bibfield  {journal} {\bibinfo  {journal} {Phys. Rev. Lett.}\ }\textbf {\bibinfo {volume} {86}},\ \bibinfo {pages} {5184} (\bibinfo {year} {2001})}\BibitemShut {NoStop}%
\bibitem [{\citenamefont {Rossignoli}\ and\ \citenamefont {Canosa}(2002)}]{RC.02}%
  \BibitemOpen
  \bibfield  {author} {\bibinfo {author} {\bibfnamefont {R.}~\bibnamefont {Rossignoli}}\ and\ \bibinfo {author} {\bibfnamefont {N.}~\bibnamefont {Canosa}},\ }\bibfield  {title} {\enquote {\bibinfo {title} {Generalized entropic criterion for separability},}\ }\href@noop {} {\bibfield  {journal} {\bibinfo  {journal} {Phys.\ Rev.\ A}\ }\textbf {\bibinfo {volume} {66}},\ \bibinfo {pages} {423061} (\bibinfo {year} {2002})}\BibitemShut {NoStop}%
\bibitem [{\citenamefont {H.~Araki}(1970)}]{AL.70}%
  \BibitemOpen
  \bibfield  {author} {\bibinfo {author} {\bibfnamefont {E.H.~Lieb}\ \bibnamefont {H.~Araki}},\ }\bibfield  {title} {\enquote {\bibinfo {title} {Entropy inequalities},}\ }\href {https://link.springer.com/article/10.1007/BF01646092} {\bibfield  {journal} {\bibinfo  {journal} {Commun. Math. Phys.}\ }\textbf {\bibinfo {volume} {18}},\ \bibinfo {pages} {160} (\bibinfo {year} {1970})}\BibitemShut {NoStop}%
\bibitem [{\citenamefont {Yang}(1962)}]{Yang.62}%
  \BibitemOpen
  \bibfield  {author} {\bibinfo {author} {\bibfnamefont {C.N.}\ \bibnamefont {Yang}},\ }\bibfield  {title} {\enquote {\bibinfo {title} {Concept of off-diagonal long range order and the quantum phases of liquid he and of superconductors},}\ }\href@noop {} {\bibfield  {journal} {\bibinfo  {journal} {Rev. Mod. Phys.}\ }\textbf {\bibinfo {volume} {34}},\ \bibinfo {pages} {694} (\bibinfo {year} {1962})}\BibitemShut {NoStop}%
\bibitem [{\citenamefont {Sambataro}\ and\ \citenamefont {Sandulescu}(2021)}]{SS.21}%
  \BibitemOpen
  \bibfield  {author} {\bibinfo {author} {\bibfnamefont {M.}~\bibnamefont {Sambataro}}\ and\ \bibinfo {author} {\bibfnamefont {N.}~\bibnamefont {Sandulescu}},\ }\bibfield  {title} {\enquote {\bibinfo {title} {Pairing correlations and eigenvalues of two-body density matrix in atomic nuclei},}\ }\href@noop {} {\bibfield  {journal} {\bibinfo  {journal} {Ann. Phys.}\ }\textbf {\bibinfo {volume} {413}},\ \bibinfo {pages} {168061} (\bibinfo {year} {2021})}\BibitemShut {NoStop}%
\bibitem [{\citenamefont {Peres}(1996)}]{P.96}%
  \BibitemOpen
  \bibfield  {author} {\bibinfo {author} {\bibfnamefont {A.}~\bibnamefont {Peres}},\ }\bibfield  {title} {\enquote {\bibinfo {title} {Separability criterion for density matrices},}\ }\href {\doibase 10.1103/PhysRevLett.77.1413} {\bibfield  {journal} {\bibinfo  {journal} {Phys. Rev. Lett.}\ }\textbf {\bibinfo {volume} {77}},\ \bibinfo {pages} {1413} (\bibinfo {year} {1996})}\BibitemShut {NoStop}%
\bibitem [{\citenamefont {Horodecki}\ \emph {et~al.}(1996)\citenamefont {Horodecki}, \citenamefont {Horodecki},\ and\ \citenamefont {Horodecki}}]{HHH.96}%
  \BibitemOpen
  \bibfield  {author} {\bibinfo {author} {\bibfnamefont {M.}~\bibnamefont {Horodecki}}, \bibinfo {author} {\bibfnamefont {P.}~\bibnamefont {Horodecki}}, \ and\ \bibinfo {author} {\bibfnamefont {R.}~\bibnamefont {Horodecki}},\ }\bibfield  {title} {\enquote {\bibinfo {title} {Separability of mixed states: necessary and sufficient conditions},}\ }\href {\doibase https://doi.org/10.1016/S0375-9601(96)00706-2} {\bibfield  {journal} {\bibinfo  {journal} {Phys. Lett. A}\ }\textbf {\bibinfo {volume} {223}},\ \bibinfo {pages} {1} (\bibinfo {year} {1996})}\BibitemShut {NoStop}%
\bibitem [{\citenamefont {Werner}(1989)}]{W.89}%
  \BibitemOpen
  \bibfield  {author} {\bibinfo {author} {\bibfnamefont {R.~F.}\ \bibnamefont {Werner}},\ }\bibfield  {title} {\enquote {\bibinfo {title} {Quantum states with einstein-podolsky-rosen correlations admitting a hidden-variable model},}\ }\href {\doibase 10.1103/PhysRevA.40.4277} {\bibfield  {journal} {\bibinfo  {journal} {Phys. Rev. A}\ }\textbf {\bibinfo {volume} {40}},\ \bibinfo {pages} {4277} (\bibinfo {year} {1989})}\BibitemShut {NoStop}%
\bibitem [{\citenamefont {Brown}\ \emph {et~al.}(1984)\citenamefont {Brown}, \citenamefont {Shavitt},\ and\ \citenamefont {Shepard}}]{Brown_1984}%
  \BibitemOpen
  \bibfield  {author} {\bibinfo {author} {\bibfnamefont {F.~B.}\ \bibnamefont {Brown}}, \bibinfo {author} {\bibfnamefont {I.}~\bibnamefont {Shavitt}}, \ and\ \bibinfo {author} {\bibfnamefont {R.}~\bibnamefont {Shepard}},\ }\bibfield  {title} {\enquote {\bibinfo {title} {Multireference configuration interaction treatment of potential energy surfaces: symmetric dissociation of {H}$_2${O} in a double-zeta basis},}\ }\href {\doibase 10.1016/0009-2614(84)80042-1} {\bibfield  {journal} {\bibinfo  {journal} {Chem. Phys. Lett.}\ }\textbf {\bibinfo {volume} {105}},\ \bibinfo {pages} {363} (\bibinfo {year} {1984})}\BibitemShut {NoStop}%
\bibitem [{\citenamefont {Olsen}\ \emph {et~al.}(1996)\citenamefont {Olsen}, \citenamefont {J{\o}rgensen}, \citenamefont {Koch}, \citenamefont {Balkova},\ and\ \citenamefont {Bartlett}}]{Olsen_1996}%
  \BibitemOpen
  \bibfield  {author} {\bibinfo {author} {\bibfnamefont {J.}~\bibnamefont {Olsen}}, \bibinfo {author} {\bibfnamefont {P.}~\bibnamefont {J{\o}rgensen}}, \bibinfo {author} {\bibfnamefont {H.}~\bibnamefont {Koch}}, \bibinfo {author} {\bibfnamefont {A.}~\bibnamefont {Balkova}}, \ and\ \bibinfo {author} {\bibfnamefont {R.~J.}\ \bibnamefont {Bartlett}},\ }\bibfield  {title} {\enquote {\bibinfo {title} {Full configuration–interaction and state of the art correlation calculations on water in a valence double-zeta basis with polarization functions},}\ }\href {\doibase 10.1063/1.471518} {\bibfield  {journal} {\bibinfo  {journal} {J. Chem. Phys.}\ }\textbf {\bibinfo {volume} {104}},\ \bibinfo {pages} {8007} (\bibinfo {year} {1996})}\BibitemShut {NoStop}%
\bibitem [{\citenamefont {Li}\ and\ \citenamefont {Paldus}(1998)}]{Li_1998}%
  \BibitemOpen
  \bibfield  {author} {\bibinfo {author} {\bibfnamefont {X.}~\bibnamefont {Li}}\ and\ \bibinfo {author} {\bibfnamefont {J.}~\bibnamefont {Paldus}},\ }\bibfield  {title} {\enquote {\bibinfo {title} {Reduced multireference couple cluster method. {I}{I}. {A}pplication to potential energy surfaces of {H}{F}, {F}$_2$, and {H}$_2${O}},}\ }\href {\doibase 10.1063/1.475425} {\bibfield  {journal} {\bibinfo  {journal} {J. Chem. Phys.}\ }\textbf {\bibinfo {volume} {108}},\ \bibinfo {pages} {637} (\bibinfo {year} {1998})}\BibitemShut {NoStop}%
\bibitem [{\citenamefont {Ma}\ \emph {et~al.}(2005)\citenamefont {Ma}, \citenamefont {Li},\ and\ \citenamefont {Li}}]{Ma_2005}%
  \BibitemOpen
  \bibfield  {author} {\bibinfo {author} {\bibfnamefont {J.}~\bibnamefont {Ma}}, \bibinfo {author} {\bibfnamefont {S.}~\bibnamefont {Li}}, \ and\ \bibinfo {author} {\bibfnamefont {W.}~\bibnamefont {Li}},\ }\bibfield  {title} {\enquote {\bibinfo {title} {A multireference configuration interaction method based on the separated electron pair wave functions},}\ }\href {\doibase 10.1002/jcc.20319} {\bibfield  {journal} {\bibinfo  {journal} {J. Comp. Chem.}\ }\textbf {\bibinfo {volume} {27}},\ \bibinfo {pages} {39} (\bibinfo {year} {2005})}\BibitemShut {NoStop}%
\bibitem [{\citenamefont {Lee}\ \emph {et~al.}(2018)\citenamefont {Lee}, \citenamefont {Huggins}, \citenamefont {Head-Gordon},\ and\ \citenamefont {Whaley}}]{Lee_2018}%
  \BibitemOpen
  \bibfield  {author} {\bibinfo {author} {\bibfnamefont {J.}~\bibnamefont {Lee}}, \bibinfo {author} {\bibfnamefont {W.~J.}\ \bibnamefont {Huggins}}, \bibinfo {author} {\bibfnamefont {M.}~\bibnamefont {Head-Gordon}}, \ and\ \bibinfo {author} {\bibfnamefont {K.~B.}\ \bibnamefont {Whaley}},\ }\bibfield  {title} {\enquote {\bibinfo {title} {Generalized unitary coupled cluster wave functions for quantum computation},}\ }\href {\doibase 10.1021/acs.jctc.8b01004} {\bibfield  {journal} {\bibinfo  {journal} {J. Chem. Theory Comput.}\ }\textbf {\bibinfo {volume} {15}},\ \bibinfo {pages} {311} (\bibinfo {year} {2018})}\BibitemShut {NoStop}%
\bibitem [{\citenamefont {McClean}\ \emph {et~al.}(2019)\citenamefont {McClean}, \citenamefont {Sung}, \citenamefont {Kivlichan}, \citenamefont {Cao}, \citenamefont {Dai}, \citenamefont {Fried}, \citenamefont {Gidney}, \citenamefont {Gimby}, \citenamefont {Gokhale}, \citenamefont {Häner} \emph {et~al.}}]{mcclean2019openfermionelectronicstructurepackage}%
  \BibitemOpen
  \bibfield  {author} {\bibinfo {author} {\bibfnamefont {J.~R.}\ \bibnamefont {McClean}}, \bibinfo {author} {\bibfnamefont {K.~J.}\ \bibnamefont {Sung}}, \bibinfo {author} {\bibfnamefont {I.~D.}\ \bibnamefont {Kivlichan}}, \bibinfo {author} {\bibfnamefont {Y.}~\bibnamefont {Cao}}, \bibinfo {author} {\bibfnamefont {C.}~\bibnamefont {Dai}}, \bibinfo {author} {\bibfnamefont {E.~Schuyler}\ \bibnamefont {Fried}}, \bibinfo {author} {\bibfnamefont {C.}~\bibnamefont {Gidney}}, \bibinfo {author} {\bibfnamefont {B.}~\bibnamefont {Gimby}}, \bibinfo {author} {\bibfnamefont {P.}~\bibnamefont {Gokhale}}, \bibinfo {author} {\bibfnamefont {T.}~\bibnamefont {Häner}},  \emph {et~al.},\ }\href {https://arxiv.org/abs/1710.07629} {\enquote {\bibinfo {title} {{O}pen{F}ermion: The electronic structure package for quantum computers},}\ } (\bibinfo {year} {2019}),\ \Eprint {http://arxiv.org/abs/1710.07629} {arXiv:1710.07629 [quant-ph]; J.R. McClean et al, Quantum Sci. Technol. {\bf 5}, 034014 (2020)} \BibitemShut {NoStop}%
\bibitem [{\citenamefont {Sun}(2015)}]{Sun_2015}%
  \BibitemOpen
  \bibfield  {author} {\bibinfo {author} {\bibfnamefont {Q.}~\bibnamefont {Sun}},\ }\bibfield  {title} {\enquote {\bibinfo {title} {Libcint: An efficient general integral library for gaussian basis functions},}\ }\href {\doibase 10.1002/jcc.23981} {\bibfield  {journal} {\bibinfo  {journal} {J. Comput. Chem.}\ }\textbf {\bibinfo {volume} {36}},\ \bibinfo {pages} {1664} (\bibinfo {year} {2015})}\BibitemShut {NoStop}%
\bibitem [{\citenamefont {Sun}\ \emph {et~al.}(2017)\citenamefont {Sun}, \citenamefont {Berkelbach}, \citenamefont {Blunt}, \citenamefont {Booth}, \citenamefont {Guo}, \citenamefont {Li}, \citenamefont {Liu}, \citenamefont {McClain}, \citenamefont {Sayfutyarova}, \citenamefont {Sharma}, \citenamefont {Wouters},\ and\ \citenamefont {Chan}}]{Sun_2017}%
  \BibitemOpen
  \bibfield  {author} {\bibinfo {author} {\bibfnamefont {Q.}~\bibnamefont {Sun}}, \bibinfo {author} {\bibfnamefont {T.~C.}\ \bibnamefont {Berkelbach}}, \bibinfo {author} {\bibfnamefont {N.~S.}\ \bibnamefont {Blunt}}, \bibinfo {author} {\bibfnamefont {G.~H.}\ \bibnamefont {Booth}}, \bibinfo {author} {\bibfnamefont {S.}~\bibnamefont {Guo}}, \bibinfo {author} {\bibfnamefont {Z.}~\bibnamefont {Li}}, \bibinfo {author} {\bibfnamefont {J.}~\bibnamefont {Liu}}, \bibinfo {author} {\bibfnamefont {J.~D.}\ \bibnamefont {McClain}}, \bibinfo {author} {\bibfnamefont {E.~R.}\ \bibnamefont {Sayfutyarova}}, \bibinfo {author} {\bibfnamefont {S.}~\bibnamefont {Sharma}}, \bibinfo {author} {\bibfnamefont {S.}~\bibnamefont {Wouters}}, \ and\ \bibinfo {author} {\bibfnamefont {G.~K.-L.}\ \bibnamefont {Chan}},\ }\bibfield  {title} {\enquote {\bibinfo {title} {{P}y{S}{C}{F}: the {P}ython‐based simulations of chemistry framework},}\ }\href {\doibase 10.1002/wcms.1340} {\bibfield  {journal} {\bibinfo  {journal} {WIREs Comput. Mol.
  Sci.}\ }\textbf {\bibinfo {volume} {8}},\ \bibinfo {pages} {e1340} (\bibinfo {year} {2017})}\BibitemShut {NoStop}%
\bibitem [{\citenamefont {Sun}\ \emph {et~al.}(2020)\citenamefont {Sun}, \citenamefont {Zhang}, \citenamefont {Banerjee}, \citenamefont {Bao}, \citenamefont {Barbry}, \citenamefont {Blunt}, \citenamefont {Bogdanov}, \citenamefont {Booth}, \citenamefont {Chen}, \citenamefont {Cui} \emph {et~al.}}]{Sun_2020}%
  \BibitemOpen
  \bibfield  {author} {\bibinfo {author} {\bibfnamefont {Q.}~\bibnamefont {Sun}}, \bibinfo {author} {\bibfnamefont {X.}~\bibnamefont {Zhang}}, \bibinfo {author} {\bibfnamefont {S.}~\bibnamefont {Banerjee}}, \bibinfo {author} {\bibfnamefont {P.}~\bibnamefont {Bao}}, \bibinfo {author} {\bibfnamefont {M.}~\bibnamefont {Barbry}}, \bibinfo {author} {\bibfnamefont {N.~S.}\ \bibnamefont {Blunt}}, \bibinfo {author} {\bibfnamefont {N.~A.}\ \bibnamefont {Bogdanov}}, \bibinfo {author} {\bibfnamefont {G.~H.}\ \bibnamefont {Booth}}, \bibinfo {author} {\bibfnamefont {J.}~\bibnamefont {Chen}}, \bibinfo {author} {\bibfnamefont {Z.-H.}\ \bibnamefont {Cui}},  \emph {et~al.},\ }\bibfield  {title} {\enquote {\bibinfo {title} {Recent developments in the {P}y{S}{C}{F} program package},}\ }\href {\doibase 10.1063/5.0006074} {\bibfield  {journal} {\bibinfo  {journal} {J. Chem. Phys.}\ }\textbf {\bibinfo {volume} {153}},\ \bibinfo {pages} {024109} (\bibinfo {year} {2020})}\BibitemShut {NoStop}%
\bibitem [{\citenamefont {Hehre}\ \emph {et~al.}(1969)\citenamefont {Hehre}, \citenamefont {Stewart},\ and\ \citenamefont {Pople}}]{hehre1969a}%
  \BibitemOpen
  \bibfield  {author} {\bibinfo {author} {\bibfnamefont {W.~J.}\ \bibnamefont {Hehre}}, \bibinfo {author} {\bibfnamefont {R.~F.}\ \bibnamefont {Stewart}}, \ and\ \bibinfo {author} {\bibfnamefont {J.~A.}\ \bibnamefont {Pople}},\ }\bibfield  {title} {\enquote {\bibinfo {title} {Self-consistent molecular-orbital methods. {I}. {U}se of {G}aussian expansions of {S}later-type atomic orbitals},}\ }\href {\doibase 10.1063/1.1672392} {\bibfield  {journal} {\bibinfo  {journal} {J. Chem. Phys.}\ }\textbf {\bibinfo {volume} {51}},\ \bibinfo {pages} {2657} (\bibinfo {year} {1969})}\BibitemShut {NoStop}%
\bibitem [{Note1()}]{Note1}%
  \BibitemOpen
  \bibinfo {note} {All entanglement measures were computed using our own programs \cite {fermionic-mbody, q-chemistry}, which rely on numpy and scipy for the linear-algebraic operations, and are made available online.}\BibitemShut {Stop}%
\bibitem [{\citenamefont {Shavitt}\ and\ \citenamefont {Bartlett}(2009)}]{MP}%
  \BibitemOpen
  \bibfield  {author} {\bibinfo {author} {\bibfnamefont {I.}~\bibnamefont {Shavitt}}\ and\ \bibinfo {author} {\bibfnamefont {R.J.}\ \bibnamefont {Bartlett}},\ }\href@noop {} {\emph {\bibinfo {title} {Many-body methods in chemistry and physics: MBPT and coupled-cluster theory}}}\ (\bibinfo  {publisher} {Cambridge {U}niversity press},\ \bibinfo {year} {2009})\BibitemShut {NoStop}%
\bibitem [{\citenamefont {G\"uhne}\ and\ \citenamefont {T\'oth}(2009)}]{GT.09}%
  \BibitemOpen
  \bibfield  {author} {\bibinfo {author} {\bibfnamefont {O.}~\bibnamefont {G\"uhne}}\ and\ \bibinfo {author} {\bibfnamefont {G.}~\bibnamefont {T\'oth}},\ }\bibfield  {title} {\enquote {\bibinfo {title} {Entanglement detection},}\ }\href {\doibase https://doi.org/10.1016/j.physrep.2009.02.004} {\bibfield  {journal} {\bibinfo  {journal} {Phys.\ Rep.}\ }\textbf {\bibinfo {volume} {474}},\ \bibinfo {pages} {1} (\bibinfo {year} {2009})}\BibitemShut {NoStop}%
\bibitem [{\citenamefont {Lewenstein}\ \emph {et~al.}(2000)\citenamefont {Lewenstein}, \citenamefont {Kraus}, \citenamefont {Cirac},\ and\ \citenamefont {Horodecki}}]{LKCH.00}%
  \BibitemOpen
  \bibfield  {author} {\bibinfo {author} {\bibfnamefont {M.}~\bibnamefont {Lewenstein}}, \bibinfo {author} {\bibfnamefont {B.}~\bibnamefont {Kraus}}, \bibinfo {author} {\bibfnamefont {J.I.}\ \bibnamefont {Cirac}}, \ and\ \bibinfo {author} {\bibfnamefont {P.}~\bibnamefont {Horodecki}},\ }\bibfield  {title} {\enquote {\bibinfo {title} {Optimization of entanglement witnesses},}\ }\href {\doibase 10.1103/PhysRevA.62.052310} {\bibfield  {journal} {\bibinfo  {journal} {Phys. Rev. A}\ }\textbf {\bibinfo {volume} {62}},\ \bibinfo {pages} {052310} (\bibinfo {year} {2000})}\BibitemShut {NoStop}%
\bibitem [{\citenamefont {Bourennane}\ \emph {et~al.}(2004)\citenamefont {Bourennane}, \citenamefont {Eibl}, \citenamefont {Kurtsiefer}, \citenamefont {Gaertner}, \citenamefont {Weinfurter}, \citenamefont {G\"uhne}, \citenamefont {Hyllus}, \citenamefont {Bru\ss{}}, \citenamefont {Lewenstein},\ and\ \citenamefont {Sanpera}}]{BG.04}%
  \BibitemOpen
  \bibfield  {author} {\bibinfo {author} {\bibfnamefont {M.}~\bibnamefont {Bourennane}}, \bibinfo {author} {\bibfnamefont {M.}~\bibnamefont {Eibl}}, \bibinfo {author} {\bibfnamefont {C.}~\bibnamefont {Kurtsiefer}}, \bibinfo {author} {\bibfnamefont {S.}~\bibnamefont {Gaertner}}, \bibinfo {author} {\bibfnamefont {H.}~\bibnamefont {Weinfurter}}, \bibinfo {author} {\bibfnamefont {O.}~\bibnamefont {G\"uhne}}, \bibinfo {author} {\bibfnamefont {P.}~\bibnamefont {Hyllus}}, \bibinfo {author} {\bibfnamefont {D.}~\bibnamefont {Bru\ss{}}}, \bibinfo {author} {\bibfnamefont {M.}~\bibnamefont {Lewenstein}}, \ and\ \bibinfo {author} {\bibfnamefont {A.}~\bibnamefont {Sanpera}},\ }\bibfield  {title} {\enquote {\bibinfo {title} {Experimental detection of multipartite entanglement using witness operators},}\ }\href {\doibase 10.1103/PhysRevLett.92.087902} {\bibfield  {journal} {\bibinfo  {journal} {Phys. Rev. Lett.}\ }\textbf {\bibinfo {volume} {92}},\ \bibinfo {pages} {087902} (\bibinfo {year} {2004})}\BibitemShut {NoStop}%
\bibitem [{\citenamefont {Petrovich}\ and\ \citenamefont {Rossignoli}(2025)}]{PR.25}%
  \BibitemOpen
  \bibfield  {author} {\bibinfo {author} {\bibfnamefont {F.}~\bibnamefont {Petrovich}}\ and\ \bibinfo {author} {\bibfnamefont {R.}~\bibnamefont {Rossignoli}},\ }\bibfield  {title} {\enquote {\bibinfo {title} {Conserved operators and exact conditions for pair condensation},}\ }\href {\doibase 10.1103/qbwh-c3gp} {\bibfield  {journal} {\bibinfo  {journal} {Phys. Rev. A}\ }\textbf {\bibinfo {volume} {112}},\ \bibinfo {pages} {032214} (\bibinfo {year} {2025})}\BibitemShut {NoStop}%
\bibitem [{\citenamefont {de~Melo}\ \emph {et~al.}(2013)\citenamefont {de~Melo}, \citenamefont {\'Cwikli\'nski},\ and\ \citenamefont {Terhal}}]{MCT.13}%
  \BibitemOpen
  \bibfield  {author} {\bibinfo {author} {\bibfnamefont {F.}~\bibnamefont {de~Melo}}, \bibinfo {author} {\bibfnamefont {P.}~\bibnamefont {\'Cwikli\'nski}}, \ and\ \bibinfo {author} {\bibfnamefont {B.M.}\ \bibnamefont {Terhal}},\ }\bibfield  {title} {\enquote {\bibinfo {title} {The power of noisy fermionic quantum computation},}\ }\href@noop {} {\bibfield  {journal} {\bibinfo  {journal} {New J.\ Phys.}\ }\textbf {\bibinfo {volume} {15}},\ \bibinfo {pages} {013015} (\bibinfo {year} {2013})}\BibitemShut {NoStop}%
\bibitem [{\citenamefont {Ditchfield}\ \emph {et~al.}(1971)\citenamefont {Ditchfield}, \citenamefont {Hehre},\ and\ \citenamefont {Pople}}]{ditchfield1971a}%
  \BibitemOpen
  \bibfield  {author} {\bibinfo {author} {\bibfnamefont {R.}~\bibnamefont {Ditchfield}}, \bibinfo {author} {\bibfnamefont {W.~J.}\ \bibnamefont {Hehre}}, \ and\ \bibinfo {author} {\bibfnamefont {J.~A.}\ \bibnamefont {Pople}},\ }\bibfield  {title} {\enquote {\bibinfo {title} {Self-consistent molecular-orbital methods. {IX}. {A}n extended gaussian-type basis for molecular-orbital studies of organic molecules},}\ }\href {\doibase 10.1063/1.1674902} {\bibfield  {journal} {\bibinfo  {journal} {J. Chem. Phys.}\ }\textbf {\bibinfo {volume} {54}},\ \bibinfo {pages} {724} (\bibinfo {year} {1971})}\BibitemShut {NoStop}%
\bibitem [{\citenamefont {Hehre}\ \emph {et~al.}(1972)\citenamefont {Hehre}, \citenamefont {Ditchfield},\ and\ \citenamefont {Pople}}]{hehre1972a}%
  \BibitemOpen
  \bibfield  {author} {\bibinfo {author} {\bibfnamefont {W.~J.}\ \bibnamefont {Hehre}}, \bibinfo {author} {\bibfnamefont {R.}~\bibnamefont {Ditchfield}}, \ and\ \bibinfo {author} {\bibfnamefont {J.~A.}\ \bibnamefont {Pople}},\ }\bibfield  {title} {\enquote {\bibinfo {title} {Self-consistent molecular orbital methods. {XII}. {F}urther extensions of gaussian-type basis sets for use in molecular orbital studies of organic molecules},}\ }\href {\doibase 10.1063/1.1677527} {\bibfield  {journal} {\bibinfo  {journal} {J. Chem. Phys.}\ }\textbf {\bibinfo {volume} {56}},\ \bibinfo {pages} {2257} (\bibinfo {year} {1972})}\BibitemShut {NoStop}%
\bibitem [{\citenamefont {Dunning}(1989)}]{dunning1989a}%
  \BibitemOpen
  \bibfield  {author} {\bibinfo {author} {\bibfnamefont {T.~H.}\ \bibnamefont {Dunning}},\ }\bibfield  {title} {\enquote {\bibinfo {title} {Gaussian basis sets for use in correlated molecular calculations. {I}. {T}he atoms boron through neon and hydrogen},}\ }\href {\doibase 10.1063/1.456153} {\bibfield  {journal} {\bibinfo  {journal} {J. Chem. Phys.}\ }\textbf {\bibinfo {volume} {90}},\ \bibinfo {pages} {1007} (\bibinfo {year} {1989})}\BibitemShut {NoStop}%
\bibitem [{\citenamefont {Cianciulli}(2026)}]{fermionic-mbody}%
  \BibitemOpen
  \bibfield  {author} {\bibinfo {author} {\bibfnamefont {J.~A.}\ \bibnamefont {Cianciulli}},\ }\href@noop {} {\enquote {\bibinfo {title} {fermionic-mbody},}\ }\bibinfo {howpublished} {\url{https://github.com/aguschanchu/fermionic-mbody}} (\bibinfo {year} {2026})\BibitemShut {NoStop}%
\bibitem [{\citenamefont {Garcia}\ \emph {et~al.}(2026)\citenamefont {Garcia}, \citenamefont {Rossignoli},\ and\ \citenamefont {Cianciulli}}]{q-chemistry}%
  \BibitemOpen
  \bibfield  {author} {\bibinfo {author} {\bibfnamefont {J.}~\bibnamefont {Garcia}}, \bibinfo {author} {\bibfnamefont {R.}~\bibnamefont {Rossignoli}}, \ and\ \bibinfo {author} {\bibfnamefont {J.~A.}\ \bibnamefont {Cianciulli}},\ }\href@noop {} {\enquote {\bibinfo {title} {q-chemistry},}\ }\bibinfo {howpublished} {\url{https://github.com/aguschanchu/q-chemistry}} (\bibinfo {year} {2026})\BibitemShut {NoStop}%
\end{thebibliography}
\end{document}